\documentclass{aa}
\usepackage[varg]{txfonts}
\usepackage{graphicx}
\usepackage{natbib}
\bibpunct{(}{)}{;}{a}{}{,}

\begin{document}

\title{SALT observations of southern post-novae\thanks{based on observations made with the Southern African Large Telescope (SALT)}}

\author{T. Tomov
\and E. Swierczynski
\and M. Mikolajewski
\and K. Ilkiewicz
}

\institute{Centre for Astronomy, Faculty of Physics, Astronomy and Informatics, Nicolaus Copernicus University, Grudziadzka 5, 87-100 Torun, Poland\label{inst1}}

\date{Received \ldots / Accepted \ldots}

\abstract{}
{We report on recent optical observations of the stellar and the nebular remnants of 22 southern post-novae.}
{In this study, for each of our targets, we obtained and analysed long-slit spectra in the spectral range 3500--6600\,\AA\ and in H$\alpha$+[\ion{N}{II}] narrow-band images.}
{The changes in the emission lines' equivalent widths with the time since the outburst agree with earlier published results of other authors. We estimated an average value $\alpha=2.37$ for the exponent of the power law fitted to the post-novae continua. Our observations clearly show the two-component structure of the \object{V842~Cen} expanding nebulae, owing to the different velocities of the ejected matter. We discovered an expanding shell around \object{V382~Vel} with an outer diameter of about $12\arcsec$.}
{}

\keywords{novae, cataclysmic variables}

\maketitle

\section{Introduction}
\label{intro}

Classical novae are cataclysmic variable (CV) systems in which a thermonuclear runaway occurs on the surface of a white dwarf accreting matter from a red dwarf companion. Exceptions are several recurrent novae in which the donor star is a red giant. Now it is generally accepted that the classical nova eruptions are also recurrent events but with much longer recurrent times $>10^\mathrm{3}$\,yr \citep{2012ApJ...756..107S}. Because of its transient character, the nova phenomenon attracts the observers' attention mainly during the eruption. Consequently, the vast majority of the novae observational data is collected during the outburst. Remarkably less data is secured for the post-outburst phase, when the star is approaching its quiescence brightness and during the subsequent years and decades. Of course, the scarcest observational data were collected for the time  before the nova explosions. 

During recent years, classical nova shells were discovered around two dwarf novae, \object{Z~Cam} and \object{AT~Cnc} \citep{2007Natur.446..159S, 2012ApJ...758..121S}. These shells are observational evidence that at least some dwarf novae have undergone classical nova eruptions. \citet{2014ApJ...780L..25J} suggest that the observed fading of \object{V603~Aql} confirms the prediction of the hibernation hypothesis \citep[see, for instance,][]{1986ApJ...311..163S} that old novae should be fading significantly in the century after their eruption is over. \citet{2013MNRAS.434.1902P} report a transition from nova-like to dwarf nova for \object{BK~Lyn}, the possible remnant of the possible classical nova \object{Nova Lyn 101}. But, a crucial example that confirms the hibernation hypothesis, an old nova in hibernation state, is still missing. Detailed studies of as many old novae as possible are needed to expand our knowledge of the nova systems, the individual components and their evolution as 
CVs.

Most old novae are rather faint and difficult to observe. Because of this, only a small fraction among the known $\sim$340 galactic novae \citep{2006yCat.5123....0D} have been studied in detail at quiescence. For most of the old novae, photometric and spectroscopic observations in quiescence are missing. Many of them are ambiguously identified, and in the catalogue of \citet{2006yCat.5123....0D}, only the field where they should be located is marked. 

Attempts to survey the stellar and nebular novae remnants, among others, were made by \citet{1983ApJS...53..523W}, \citet{1987Ap&SS.131..467D}, \citet{1996MNRAS.281..192R}, \citet{1998MNRAS.300..221G, 2000MNRAS.314..175G}, and \citet{2000AJ....120.2007D}. Recently, \citet[and references therein]{2012MNRAS.423.2476T,2013MNRAS.436.2412T} have started a program to identify, confirm spectroscopically, and determine the orbital periods of a large number of old novae, whose explosions took place before 1980.

In this paper we present long-slit spectroscopy and H$\alpha$+[\ion{N}{II}] narrow-band imaging survey of 22 southern novae remnants. Our targets were selected from the \citet{2006yCat.5123....0D} catalogue of CVs among the well-identified old novae.

\begin{table*}
\caption{Summary of SALT observations.}
\label{log}
\centering
 \begin{tabular}{llrcrrl}
\hline\hline
Object & \multicolumn{2}{c}{Outburst} & Date& Spectra  & H$\alpha$+[\ion{N}{II}] Imaging & Known  \\
\cline{2-3}
name & Year & Time since\tablefootmark{a} && exposure & exposure & Nebular  \\
 & & (yr) & & (sec) & (sec) & Remnant \\
\hline
\object{T\,Pyx} &2011\tablefootmark{b} &1& 27.02.2012 & &3$\times$60 &Yes\tablefootmark{1,2} \\
 &  && 21.03.2012 & 2$\times$300+40 &  &  \\
\object{U\,Sco} & 2010\tablefootmark{b} &2& 29.04.2012 & 250  &4$\times$60 & \\
\object{V445\,Pup} & 2000 &12& 13.04.2012 & 2$\times$35  & 3$\times$40& Yes\tablefootmark{3} \\
\object{V382\,Vel} & 1999 &13& 30.11.2011 & 2$\times$130  &1$\times$60 &Yes\tablefootmark{4} \\
\object{CP\,Cru} & 1996 &16& 27.02.2012 &  &3$\times$60 & Yes\tablefootmark{5,6} \\
 & && 01.03.2012 & 900 &  & \\
\object{BY\,Cir} & 1995 &17& 25.02.2012 & 180  &3$\times$60 & \\
\object{V888\,Cen} & 1995 &17& 13.02.2012 & 130  &6$\times$60 & \\
\object{V868\,Cen} & 1991 &21& 26.01.2012 & 2015  &2$\times$60 & \\
\object{V842\,Cen} & 1986 &26& 13.02.2012 & 100  &2$\times$60 & Yes\tablefootmark{7} \\
\object{GQ\,Mus} & 1983 &29& 01.04.2012 & 2$\times$300  &3$\times$60 & \\
\object{Nova\,Car\,1972} & 1972 &40& 17.12.2011 & 180  &3$\times$60 & \\
\object{HS\,Pup} & 1963 &49& 11.03.2012 & 380  &3$\times$60 & Yes\tablefootmark{7} \\
\object{HZ\,Pup} & 1963 &49& 11.03.2012 & 130  &3$\times$60 & \\
\object{V365\,Car} & 1948 &64& 12.02.2012 & 300  & 1$\times$60& \\
\object{CP\,Pup} & 1942 &70& 02.01.2012 & 3$\times$60  &  & Yes\tablefootmark{8} \\
            &          && 01.04.2012 & 40  &2$\times$60 & \\
\object{BT\,Mon} & 1939 &73& 15.12.2011 & 130  &3$\times$60 & Yes\tablefootmark{9} \\
\object{RR\,Pic} & 1925 &87& 07.11.2011 & 6  &3$\times$10 & Yes\tablefootmark{8} \\
 \object{GI\,Mon} & 1918 &94& 16.12.2011 &  &3$\times$60 & \\
 & && 01.01.2012 & 2$\times$130  & & \\
\object{OY\,Ara} & 1910 &102& 22.09.2011 & 150  &10$\times$60 & \\
\object{CN\,Vel} & 1905 &107& 16.01.2012 & 300  &2$\times$60 & \\
\object{X\,Ser} & 1903 &109& 02.03.2012 &  &2$\times$60 & \\
 & && 12.03.2012 & 300  & & \\
\object{DY\,Pup} & 1902 &110& 13.03.2012 & 950  &1$\times$60 & Yes\tablefootmark{7} \\
\hline
 \end{tabular}
\tablefoot{
\tablefoottext{a}{In all tables, figures, and Sec.~\ref{obj} the objects are ordered by the time since the outburst.}
\tablefoottext{b}{Recurrent nova: only the most recent outburst is shown in the table.}
}
\tablebib{(1)~\citet{1982ApJ...261..170W}; (2)~\citet{2010ApJ...708..381S}; (3)~\citet{2009ApJ...706..738W};
 (4)~this paper; (5)~\citet{1998AAS...192.5304R}; (6)~\citet{2000AJ....120.2007D}; (7)~\citet{1998MNRAS.300..221G}; (8)~\citet{1987SSRv...45....1D}; (9)~\citet{1983MNRAS.205P..33M}}
\end{table*}

\section{Observations and data reduction}
\label{obs}

The observations were obtained at the SAAO Observatory with the Robert Stobie Spectrograph \citep[RSS;][]{2003SPIE.4841.1463B, 2003SPIE.4841.1634K} and the 10\,m Southern African Large Telescope \citep[SALT;][]{2006SPIE.6267E..32B, 2006MNRAS.372..151O}. The narrow-band imaging and the long-slit spectroscopy modes of the RSS were used. The diameter of the RSS effective field of view is 8\,arcmin. We applied a binning factor of 2, which gives a spatial resolution of 0.254\,arcsec\,pixel$^\mathrm{-1}$. In the RSS narrow-band imaging mode, we used a Fabry-Perot interference filter pi06530, centered close to H$\alpha$+[\ion{N}{II}] at 6530\,\AA\ and with a FWHM=156\,\AA. In the RSS long-slit spectroscopy mode, the volume phase holographic (VPH) grating PG0900 was used in the spectral range 3500--6600\,\AA. The slit width was 1\farcs5, which gives a reciprocal dispersion of $\sim$0.98\,\AA\,pixel$^\mathrm{-1}$ and a resolving power $\sim$800. Spectra of ThAr, Ar, and Xe comparison arcs were obtained to calibrate 
the wavelength scale. For relative flux calibration, we used spectra of the spectrophotometric standard stars Hilt\,600, EG\,21, and LTT\,4364, obtained in the default calibration framework. A summary of our observations is provided in Table~\ref{log}. 

\begin{figure*}
\centering
\includegraphics[height=0.98\textheight]{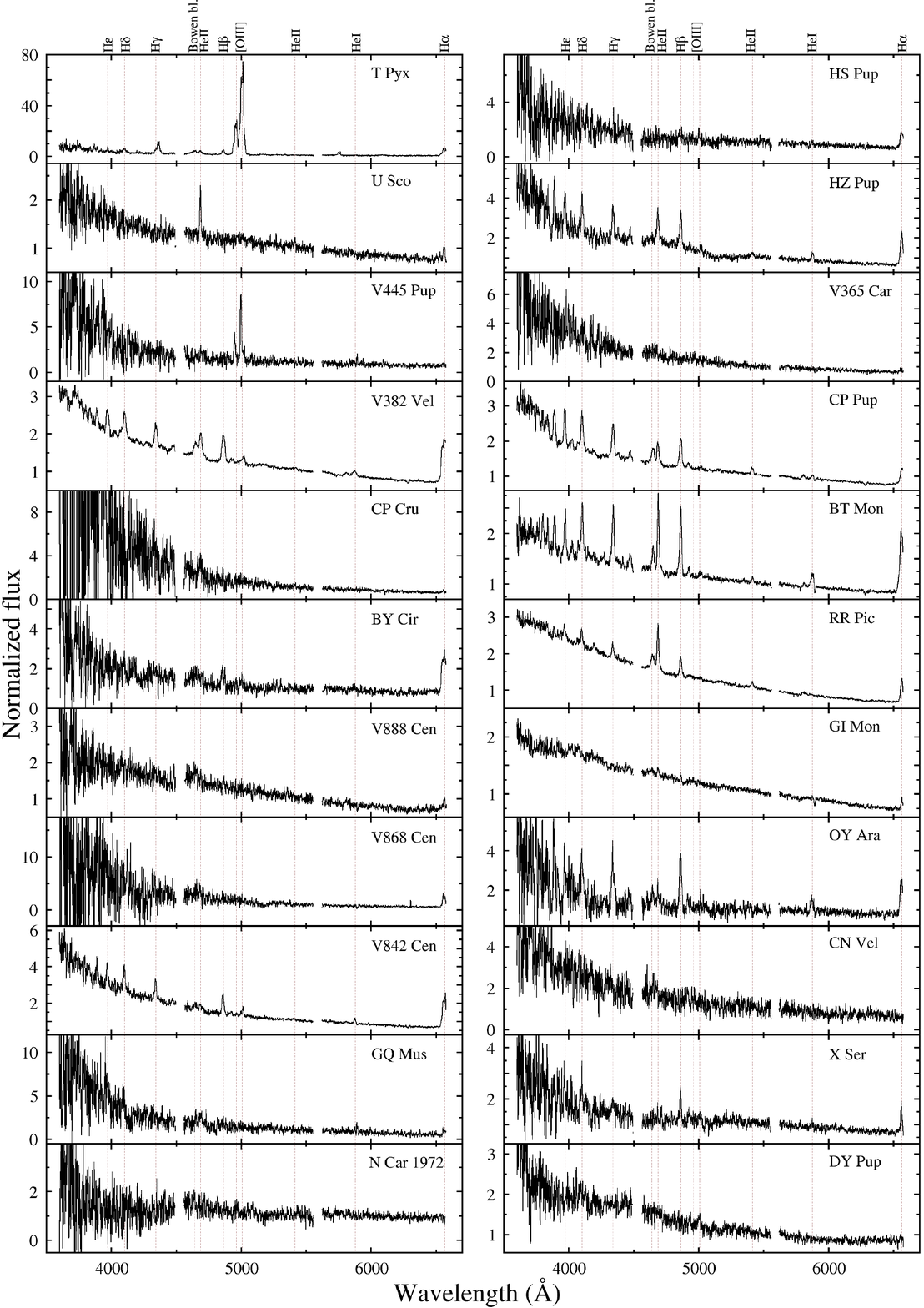}
\caption{Normalized at 5500\AA\ flux spectra of the targets . The spectra are dereddened and have been smoothed with a $3\times3$ box filter. The gaps in the spectra reflect the RSS interchip gaps.}
\label{spectra}
\end{figure*}

The RSS detector is a three-CCD-chip mosaic. The first observations revealed that our attempt to find an optimal positioning of the spectrum on the CCD mosaic using PIPT (Principal Investigator Proposal Tool, \citet{2010SPIE.7737E..30H}) was not very successful. It turned out that the H$\alpha$ line is located at the very edge of the chip (see Figs.~\ref{spectra} and \ref{map_sp}).  To avoid problems with losing other lines in the gaps between the CCD detectors, we decided to do not change the RSS long-slit setup during the observation semester.

The initial reduction of the data, including bias and overscan subtraction, gain and cross-talk corrections, trimming and mosaicking, was done with the SALT science pipeline \citep{2010SPIE.7737E..54C}. 

 We used the \textsc{dcr} software written by W.~Pych. Its algorithm is described in \citet{2004PASP..116..148P}, and it removes the cosmic rays from the long-slit spectral observations. To perform the flat field corrections and the wavelength calibration and to correct the distortion and the tilt of the frames, we used the standard \textsc{iraf}\footnote{\textsc{iraf} is distributed by the National Optical Astronomy Observatories, which are operated by the Association of Universities for Research in Astronomy, Inc., under cooperative agreement with the National Science Foundation.} tasks in the \textsc{twodspec} package. Most of the nights in which our targets were observed were not photometric. Moreover, SALT is a telescope with a variable pupil, which makes the absolute flux calibration impossible. Because of this we calibrated the spectra in relative flux using an average sensitivity curve, only to derive some info about the relative spectral energy distribution.

From the H$\alpha$+[\ion{N}{II}] narrow-band imaging, we reduced and used the central only $4\arcmin \times 4\arcmin$ field. The data were processed in a standard way, using the appropriate \textsc{iraf} tasks. Because flat field calibration images were not obtained every night, in some cases the task \textsc{imsurfit} was used for a final flattening of the field. The reduced images of each of our targets were combined to improve the S/N.

The resulting 1D dereddened (see Sect.~\ref{ebmv_sed} and Table~\ref{redd}) spectra are shown in Fig.~\ref{spectra}. The H$\alpha$+[\ion{N}{II}] narrow-band imaging in a field of $4\arcmin \times 4\arcmin$ around each nova, as well as the extracted spectrum before reddening correction, are presented in Fig.~\ref{map_sp} as online material. Profiles of individual spectral lines are shown in online figures Figs.~\ref{hb_prof}--19.

The equivalent widths (EW) and the FWHM of the strongest emission lines measured in the target spectra are provided in Table~\ref{em}. The measurements were done with the \textsc{iraf} task \textsc{splot} fitting Gaussians to the emission lines. The error estimates were computed directly in \textsc{splot} by running a number of Monte Carlo simulations based on preset instrumental parameters.

\begin{table}
\caption{Interstellar extinction and the slope ($F=\lambda^{-\alpha}$) of the continuum.}
\label{redd}
\centering
 \begin{tabular}{llll}
\hline\hline
Object & $E_\mathrm{B-V}$  & $\alpha$ & Ref. for  \\
name & (mag) & &  $E_\mathrm{B-V}$ \\
\hline
\object{T\,Pyx} & 0.25 & $3.32\pm0.21$ & 1 \\
\object{U\,Sco} & 0.20 & $1.37\pm0.06$ & 2 \\
\object{V445\,Pup} & 0.60 & $1.98\pm0.20$ & 3 \\
\object{V382\,Vel} & 0.10 & $2.11\pm0.03$ & 4 \\
\object{CP\,Cru} & 1.90 & $3.71\pm0.13$ & 5 \\
\object{BY\,Cir} & 0.55 & $1.37\pm0.13$ & 6 \\
\object{V888\,Cen} & 0.40 & $2.04\pm0.08$ & 7 \\
\object{V868\,Cen} & 1.75 & $3.54\pm0.19$ & 8 \\
\object{V842\,Cen} & 0.55 & $2.90\pm0.06$ & 9 \\
\object{GQ\,Mus} & 0.55 & $2.98\pm0.24$ & 10\\
\object{Nova\,Car\,1972} & 0.48 & $0.42\pm0.14$ & 11\tablefootmark{a} \\
\object{HS\,Pup} & 0.52 & $1.91\pm0.17$ & 12 \\
\object{HZ\,Pup} & 0.35 & $2.65\pm0.09$ & 12 \\
\object{V365\,Car} & 0.91 & $2.85\pm0.10$ & 13 \\
\object{CP\,Pup} & 0.20 & $2.09\pm0.02$ & 14 \\
\object{BT\,Mon} & 0.24  & $1.27\pm0.02$ & 14 \\
\object{RR\,Pic} & 0.00  & $2.61\pm0.02$ & 14 \\
\object{GI\,Mon} & 0.10 & $1.70\pm0.03$ & 14 \\
\object{OY\,Ara} & 0.32 & $0.95\pm0.15$ & 15 \\
\object{CN\,Vel} & 0.20 & $2.63\pm0.23$ & 11\tablefootmark{a} \\
\object{X\,Ser} & 0.25 & $1.46\pm0.12$ & 16 \\
\object{DY\,Pup} & 0.19 & $1.90\pm0.09$ & 12 \\
\hline
 \end{tabular}
\tablefoot{
\tablefoottext{a}{From NASA's IPAC Infrared Science Archive (IRSA), web interface.}
}
\tablebib{(1)~\citet{2007A&A...461..593G}; (2)~\citet{2010ApJS..187..275S}; (3)~\citet{2009ApJ...706..738W}; (4)~\citet{2002A&A...390..155D}; (5)~\citet{2003AJ....126..993L}; (6)~\citet{2002A&A...384..504E}; (7)~\citet{2001MNRAS.324..553Y}; (8)~\citet{1994ApJ...426..279W}; (9)~\citet{1994A&A...291..869A};  (10)~\citet{2008ApJ...687.1236H};  (11)~\citet{2011ApJ...737..103S}; (12)~\citet{1997ApJ...487..226S};   (13)~\citet{1975ApJ...200..694H}; (14)~\citet{2013A&A...560A..49S}; (15)~\citet{1997ApJ...483..899Z}; (16)~\citet{2004BaltA..13...93S}}
\end{table}

\section{Interstellar extinction and spectral energy distribution (SED)}
\label{ebmv_sed}

To get an idea about the SED of the observed old novae, we first corrected their spectra for the interstellar extinction. We searched for published estimations of the extinction $A_\mathrm{V}$ and/or the color excess $E_\mathrm{B-V}$ for the particular novae. The relation $A_\mathrm{V}=3.1E_\mathrm{B-V}$ was used to compare them. In cases where more than one estimation was found, we compared them very carefully, adopting the best one for further use. For two objects, \object{Nova\,Car\,1972} and \object{CN\,Vel}, we did not find any estimations of the interstellar extinction in the literature. For them we adopted the minimum value of the interstellar extinction, corresponding to the direction to these targets \citep{2011ApJ...737..103S}, taken from NASA's IPAC Infrared Science Archive (IRSA). The adopted values of $E_\mathrm{B-V}$ and the corresponding references are presented in Table~\ref{redd}. The spectra were dereddened using the appropriate  \textsc{iraf} task and applying the empirical 
 selective  extinction  function  of \cite{1989ApJ...345..245C} with $R_\mathrm{V}=3.1$. 

The exponents of the power law $F=\lambda^{-\alpha}$ fitted to the continuum of the dereddened spectra are also shown in Table~\ref{redd}. For all the objects, a single power law was fitted in the wavelength range 4000--6500\,\AA. The errors of the exponents represent only the standard deviation from the fitted power law. The average value of the exponent $\alpha$ for our sample is $2.25\pm0.78$ (excluding \object{Nova\,Car\,1972}, see Sect.~\ref{ncar1972}). For the same number of old novae \citet{1996MNRAS.281..192R} estimated an average $\alpha=2.68\pm0.82$. For a sample of eight post-novae \citet{2012MNRAS.423.2476T} found ``that $\alpha$ for most systems falls well below the value of 2.33 for a steady-state accretion disk \citep{1969Natur.223..690L}''. Excluding the objects fitted with two power laws and \object{V529 Ori}, which shows an atypical SED, 13 post-novae published by  \citet{2005A&A...432..199S} and \citet{2012MNRAS.423.2476T,2014arXiv1405.3635T} remain for which an average $\alpha=2.13\pm0.
67$ can be estimated.

\begin{figure}
\centering
\resizebox{\hsize}{!}{\includegraphics{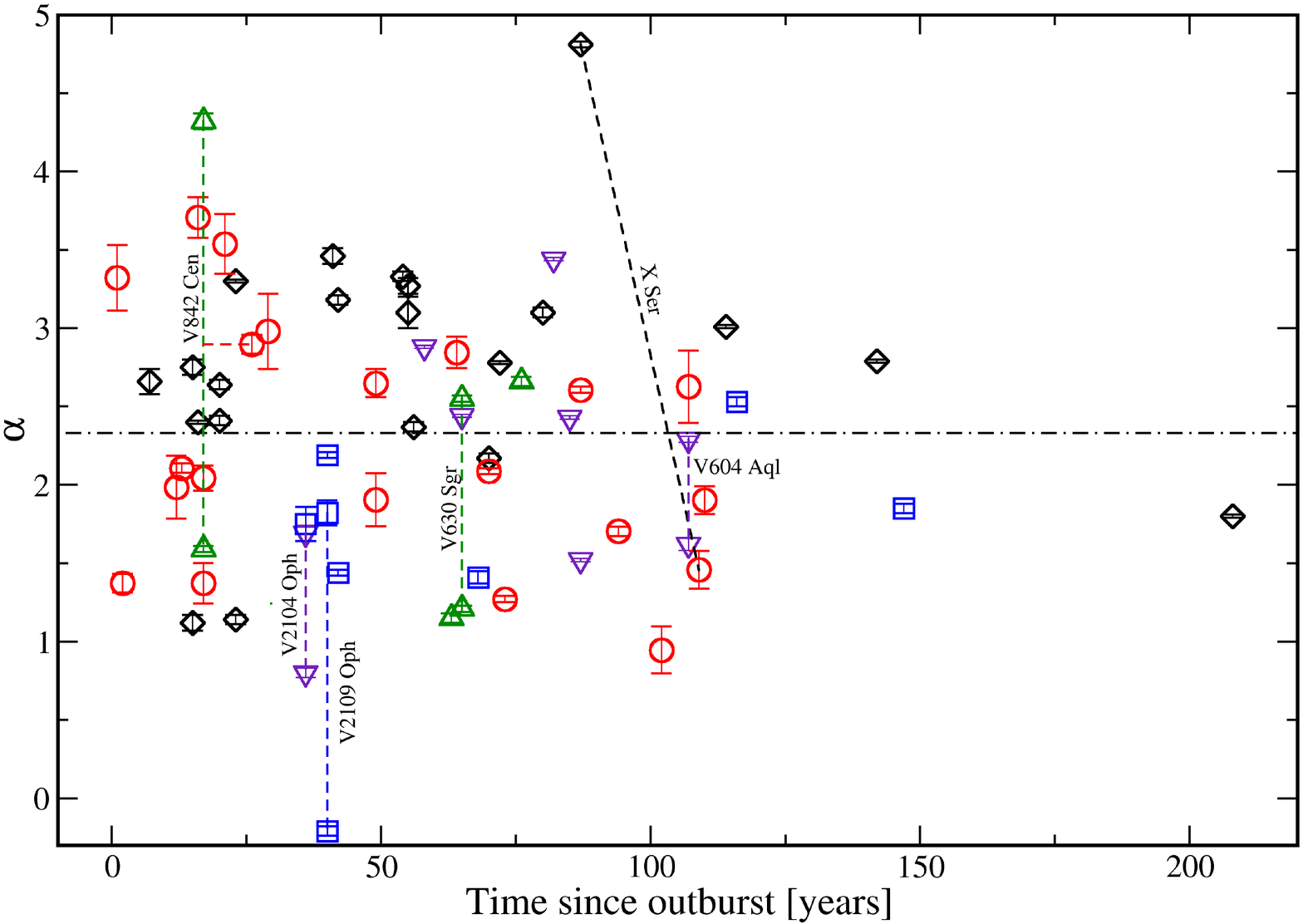}}
\caption{Power law exponents $\alpha$ as a function of the time since outburst. The symbols represent as follows: circles -- this paper; diamonds -- \citet{1996MNRAS.281..192R}; squares -- \citet{2012MNRAS.423.2476T}; triangles up -- \citet{2005A&A...432..199S}; tiangles down -- \citet{2014arXiv1405.3635T}. The double exponents for \object{V630\,Sgr}, \object{V842\,Cen}, \object{V2109\,Oph}, \object{V604\,Aql} and \object{V2104\,Oph} \citep{2005A&A...432..199S,2012MNRAS.423.2476T,2014arXiv1405.3635T} are connected with dashed lines. Our exponent for \object{V842\,Cen}  is also connected by a dashed line to the line between its two exponents from \citet{2005A&A...432..199S}. The $\alpha$ values for the only common object in ours and \citet{1996MNRAS.281..192R} samples,  \object{X\,Ser}, are connected with a dashed line as well. The dot-dashed line represents the value $\alpha=2.33$ for a steady-state accretion disk.}
\label{power}
\end{figure}

In Fig.~\ref{power} our power-law exponents, together with these from \citet{1996MNRAS.281..192R}\footnote{There are some discrepancies between the values of $\alpha$ in Table~4 and Fig.~7 in \citet{1996MNRAS.281..192R}. For our Fig.~\ref{power} we used the values from their Table~4.}, \citet{2005A&A...432..199S}, and \citet{2012MNRAS.423.2476T,2014arXiv1405.3635T}, are presented as a function of the time since outburst.  From the mean values and Fig.~\ref{power}, it is obvious that most of the  \citet{1996MNRAS.281..192R} power-law exponents are above the steady-state disk exponent, while the majority of the \citet{2005A&A...432..199S} and \citet{2012MNRAS.423.2476T,2014arXiv1405.3635T} exponents are below this value. Our exponents are more or less symmetrically distributed around the steady disk exponent. Nine years after \citet{2005A&A...432..199S} we estimated a power law exponent for \object{V842\,Cen} $\alpha=2.90$ which is practically the mean of their two values 4.32 and 1.59. Somewhat surprising is 
the large difference in the exponents of \object{X\,Ser} 4.81 and 1.46 estimated by \citet{1996MNRAS.281..192R} and by us about 87 and 109 years, respectively, after its outburst. From one side, this difference could, at least partly, be caused by, as  \citet{1996MNRAS.281..192R} pointed out, may be too high a value of $E_\mathrm{B-V}$ they used to deredden the spectrum. On the other hand, it is possible that \object{X\,Ser} shows an  intrinsic variation of the power law  exponent.  A remarkable change from $\alpha=0.5$ in their 1997 spectrum to $\alpha=1.1$ in the spectrum obtained in 1999 was reported by \citet{2000MNRAS.312..629T}. 

Excluding the double exponents from \citet{2005A&A...432..199S} and \citet{2012MNRAS.423.2476T,2014arXiv1405.3635T} as well as the \object{X\,Ser} exponent from \citet{1996MNRAS.281..192R} we estimate an average $\alpha=2.37\pm0.74$, a value very close to that of a steady-state accretion disk.

\section{Individual objects}
\label{obj}

\subsection{\object{T\,Pyx}}
\label{tpyx}

\object{T\,Pyx} is one of the best studied recurrent novae ever. H.~Leavitt discovered its 1902 outburst on Harvard archival plates in 1913 \citep{1913HarCi.179....1L}. During a time interval slightly longer than one century, six eruptions of \object{T\,Pyx} were observed, in 1890, 1902, 1920, 1944, 1967, and 2011. The last outburst was excellently covered with systematic observations in the whole electromagnetic wavelength region from X-ray to radio. More details on the previous and the last outbursts of \object{T\,Pyx} can be found in \citet{2010ApJS..187..275S}, \citet{2013ApJ...773...55S}, \citet{2011A&A...533L...8S}, \citet{2013A&A...549A.140S}, \citet{2014A&A...562A..28D}, \citet{2014AJ....147..107S}, and references therein. An extended shell, different from the other nova shells and composed of many small knots, was discovered by \citet{1979Msngr..17....1D}. A recent study by \citet{2010ApJ...708..381S} sheds new light on the complex shell of \object{T\,Pyx}.

Our RSS spectrum of \object{T\,Pyx} was obtained about 342 days after the outburst of the nova in April 2011. The star was still decreasing from the maximum, and at the moment of our observations its V brightness was between 13 and 14 magnitudes. As seen  in Fig.~\ref{spectra}, numerous emission lines, some very intensive, are superimposed on a strong blue continuum. The strongest features in the spectrum are the nebular lines of [\ion{O}{III}] 4363\,\AA, 4959\,\AA, and 5007\,\AA. Another relatively strong nebular emission is [\ion{N}{II}] 5755\,\AA, and a blend of, most probably, [\ion{Fe}{VII}] 6084\,\AA\ and [\ion{Ca}{V}] 6087\,\AA\ is apparent as well.
In the blue part of the spectrum, the lines of [\ion{Ne}{III}] 3868\,\AA\ and 3968\,\AA are clearly visible. The Balmer emission lines are relatively strong and easily seen up to H$_\mathrm{8}$. In the strongest nebular and Balmer emission lines a four component profile structure is obvious. The radial velocities of these components are $-1284\pm82$\,km\,s$^{-1}$, $-391\pm29$\,km\,s$^{-1}$, $329\pm18$\,km\,s$^{-1}$, and $1128\pm107$\,km\,s$^{-1}$. These profiles, particularly the [\ion{O}{III}] ones, are very similar to the observed by \citet{2013A&A...549A.140S} in their NOT spectrum of \object{T\,Pyx} obtained on  April 8, 2012. Only 4686\,\AA\ \ion{He}{II} emission presents in the spectrum with an intensity slightly lower than H$\beta$ and almost equal to that of the closest blend of \ion{N}{III} and \ion{C}{III} emission lines at $\sim$4840\,\AA\ (hereafter Bowen blend). From the \ion{He}{I} lines, only a very faint emission 5876\,\AA\ is visible (Fig.~\ref{map_sp}). It is interesting to note that \citet{
2013A&A...549A.140S} report that in their spectrum of \object{T\,Pyx}, obtained about two weeks after our, \ion{He}{I} was not present. A remarkable decrease in the Bowen blend intensity in the April 2012 spectrum of \citet{2013A&A...549A.140S} in comparison to ours is evident, too.

\subsection{\object{U\,Sco}}
\label{usco}

The first outburst of \object{U\,Sco} was discovered by \citet{1865AN.....64..170P} in 1863. The recurrent nature of the nova was revealed by \citet{1940BHarO.912...10T} during a study of archival Harvard patrol plates. \object{U\,Sco} is the fastest known nova with $t_{\mathrm{3}}=2\fd6$ and at the same time as the recurrent nova with the maximum number of ten observed outbursts. A detailed overview of the historical and the last 2010 outbursts can be found in \citet{2010ApJS..187..275S, 2010arXiv1009.3197S} and references therein. This recurrent nova is a double-lined eclipsing binary with $P_\mathrm{orb}=1\fd23$ \citep{1995ApJ...447L..45S}, orbital inclination $i=82\fdg7$, containing a very massive white dwarf  $M_\mathrm{WD}=1.55$\,M$_\sun$ \citep{2001MNRAS.327.1323T}.

Two years after the last outburst, \object{U\,Sco} practically returned to its inter-outburst state. In Fig.~\ref{spectra} it can be seen that a relatively strong, not very steep hot continuum is dominating the spectrum. The strongest emission line is \ion{He}{II} 4686\,\AA. The only two additional, remarkably weaker emission features that can be identified in the spectrum are H$\alpha$ and \ion{He}{II} 5412\,\AA.

\subsection{\object{V445\,Pup}}
\label{v445pup}

\object{V445\,Pup} is an unusual classical nova discovered at the very end of 2000 by K.~Kanatsu \citep{2000IAUC.7552....1K}. \citet{2003A&A...409.1007A} argue that the outburst and the peak brightness occurred between September 26 and November 23, 2000. Based mainly on the hydrogen deficiency and the He/C enrichment in the optical and the IR spectra of \object{V445\,Pup}, \citet{2003A&A...409.1007A} supposed that it is a helium nova. Modeling its light curve, \citet{2003ApJ...598L.107K} and \citet{2008ApJ...684.1366K} present arguments for it being a helium nova on a very massive white dwarf. \citet{2009ApJ...706..738W} discovered an expanding bipolar shell and estimated a distance to \object{V445\,Pup} of 8.2\,kpc. Studying the \object{V445\,Pup} progenitor on archive plates, \citet{2010PZ.....30....4G} conclude that it was a common-envelope binary with SED similar to that of an A0V type star. They suggest that the outburst was caused by a helium flash on the surface of a CO-type white dwarf leading to the 
lost of the common-envelope system.

In the spectrum of \object{V445\,Pup}, shown in Fig.~\ref{spectra}, a weak continuum is seen, remarkably increasing in the blue. The only line features visible in the spectrum are the nebular lines of [\ion{O}{III}] at 4959\,\AA\ and 5007\,\AA. A two-component structure is very visible in both emissions with the red component more than twice weaker (Table~\ref{em}). The measured radial velocities of the components are $-619\pm8$\,km\,s$^{-1}$ and $592\pm22$\,km\,s$^{-1}$, respectively. Our spectrum with the slit PA=0\degr covers, only the central 1\farcs5 of the expanding bipolar shell, whose axis lies along a PA$\sim$66\degr, discovered by \citet{2009ApJ...706..738W}.

\subsection{\object{V382\,Vel}}
\label{v382vel}

One of the brightest novae ever observed in the southern skies, \object{V382\,Vel} was discovered by P.~Williams and A.~Gilmore \citep{1999IAUC.7176....1L}. It is a fast nova ($t_{\mathrm{3}}\sim 9^\mathrm{d}$) that reached maximum brightness $V = 2\fm3$ on 23 May 1999 \citep{2002A&A...390..155D}. It was found that \object{V382\,Vel} is a ONeMg nova \citep{1999IAUC.7220....3W} very similar to V1974 Cygni \citep{2003AJ....125.1507S}. An orbital period of 3\fh5 was detected by \citet{2006AJ....131.2628B}. \citet{2013ApJ...762..105D} estimate the white dwarf mass to be about 1.2\,M$_\sun$. Despite the very high expansion velocity $\sim$3000--5000\,km\,s$^\mathrm{-1}$ and the relatively short distance of about 2\,kpc \citep{2002A&A...390..155D, 2003AJ....125.1507S}, observations of an extended shell around \object{V382\,Vel} were missing in the literature. In this paper we present the first detection of the \object{V382~Vel} expanding shell (see Sect.~\ref{shells}).

To our knowledge, the spectrum of \object{V382\,Vel} in Fig.~\ref{spectra} is the first quiescence spectrum presented after its outburst. It is characterized by a strong and steep blue continuum and by numerous superimposed emission lines. Among the strongest emission features are the Balmer lines, which are very visible up to H$_\mathrm{11}$, as well as many \ion{He}{I} lines, the Bowen blend, and the \ion{He}{II} 4686\,\AA\ line (see Table~\ref{em}). Also seen well is the line \ion{C}{IV} 5805\,\AA.  The only evidence of a significant contribution of emission from the expanding shell is the multicomponent profile of the H$\alpha$+[\ion{N}{II}] blend \citep[see also][]{2012MmSAI..83..610S}. 

\subsection{\object{CP\,Cru}}
\label{cpcru}

\object{CP\,Cru} is a nova probably with a missed maximum brightness, which was discovered by W.~Liller on August 26, 1996 \citep{1996IAUC.6463....1L}. It is a fast nova with $t_{\mathrm{2}} \sim 4^\mathrm{d}$ and an extended shell discovered in a relatively short time after the outburst \citep{2000AJ....120.2007D}. From the expansion parallax \citet{2000AJ....120.2007D} derived a 3.2\,kpc distance to \object{CP\,Cru}, corrected later to $2.6\pm0.5$\,kpc by \citet{2003AJ....126..993L}. \citet{2003MNRAS.340.1011W} found that \object{CP\,Cru} is an eclipsing binary with $P_{\mathrm{orb}} = 22\fh7$ and orbital inclination $\sim 70\degr$.

A hot, steeper in the blue continuum dominates in the quiescence spectrum of \object{CP\,Cru} (Fig.~\ref{spectra}). Because of the low $S/N$ of our spectrum, weak H$\alpha$ and \ion{He}{II} 4686\,\AA\ are the only emission features that could be identified with some certainty.

\subsection{\object{BY\,Cir}}
\label{bycir}

\object{BY\,Cir} is a slow nova with $t_{\mathrm{3}}\sim124^\mathrm{d}$ \citep{2010AJ....140...34S} discovered on January 27, 1995 \citep{1995IAUC.6130....1L}. The observations of \citet{2000AJ....120.2007D} did not resolve an extended shell around \object{BY\,Cir}, but the authors noted that the remnant is too young. \citet{2003MNRAS.340.1011W} found that \object{BY\,Cir} is an eclipsing binary with $P_\mathrm{orb}=6\fh76$, which belongs to the group of the long-period, deeply eclipsing nova remnants.

There is no spectrum of the \object{BY\,Cir} remnant published before the one we show in Fig.~\ref{spectra}. The hot continuum is relatively weak and not so steep in the blue. Only H$\alpha$ and H$\beta$ are visible in the spectrum. Very weak emissions of \ion{He}{II} 4686\,\AA, Bowen blend, and [\ion{O}{III}] 5007\,\AA\ are also present in the spectrum. The nebular [\ion{O}{III}] line and the multicomponent H$\alpha$ profile could be evidence that the emission from the expanding shell is still significant. It is noticeable that the H$\beta$ emission line also shows a clear triple structure, with component velocities on the order of $-950$\,km\,s$^\mathrm{-1}$, $-300$\,km\,s$^\mathrm{-1}$, and $750$\,km\,s$^\mathrm{-1}$.

\subsection{\object{V888\,Cen}}
\label{v888cen}

Another nova discovered by W.~Liller in February 1995 is \object{V888\,Cen} \citep{1995IAUC.6139....1L}. Analysing its light curve and spectra, \citet{2001MNRAS.324..553Y} found that \object{V888\,Cen} is a very fast nova with $t_{\mathrm{3}}\sim12^\mathrm{d}$ and estimated the nova distance to about 7.4\,kpc. \citet{2000AJ....120.2007D} failed to reveal an extended shell around this nova. 

The RSS spectrum of \object{V888\,Cen} presented in Fig.~\ref{spectra} reveals a not very steep hot continuum with only a relatively weak H$\alpha$ and possible, also weak, Bowen blend emissions superimposed. \citet{2005PASP..117..944S} obtained "the first spectrum of \object{V888\,Cen} in quiescence" nine years after the outburst. In their spectrum, faint emission lines of H$\alpha$, H$\beta$, and H$\gamma$, as well as \ion{He}{II} 4686\,\AA\ and the Bowen blend, were presented. The higher Balmer series lines were instead seen in absorption. Our spectrum differs remarkably from that of \citet{2005PASP..117..944S} because of its low quality. From our S/N we get an upper limit for the EW of not identified lines $EW<5$\,\AA.

\subsection{\object{V868\,Cen}}
\label{v868cen}

\object{V868\,Cen} was discovered by W.~Liller on April 2, 1991 \citep{1991IAUC.5230....1L}. Because of the unknown exact moment of maximum brightness, many different decline rates are published. \citet{2003JAD.....9....3W} derived an average $t_{\mathrm{3}}=70\pm15$ days and classified \object{V868\,Cen} as a medium-fast nova. \citet{2000AJ....120.2007D} did not find any extended shell around this nova.

Figure~\ref{spectra} presents the spectrum of the remnant of nova \object{V868\,Cen} obtained about 22 years after the outburst.  Except the hot continuum only H$\alpha$ is seen in emission. The multicomponent H$\alpha$ profile indicates a possible contribution by the H$\alpha$+[\ion{N}{II}] expanding shell emission. 

\subsection{\object{V842\,Cen}}
\label{v842cen}

The nova was discovered by McNaught on November 22, 1986 \citep{1986IAUC.4274....1M}. \citet{1987MNSSA..46...72W} classified \object{V842\,Cen} as a moderately fast nova, based on the estimated $t_{\mathrm{3}}\sim48^\mathrm{d}$. The nova extended shell was revealed by \citet{1998MNRAS.300..221G} and observed again three years later by \citet{2000AJ....120.2007D}. We detected the extending shell around \object{V842~Cen} in both the H$\alpha$+[\ion{N}{II}] narrow-band images and the long-slit spectra, and the details are presented in Sect.~\ref{shells}.  \citet{2005A&A...432..199S} show the spectrum of the remnant obtained in 2003. \citet{2009MNRAS.395.2177W} estimated the orbital period of \object{V842\,Cen} to 3\fh94 and the white dwarf rotation period to $\sim$57 seconds. They suggest that \object{V842\,Cen} is an intermediate polar with ``the fastest rotating white dwarf among the intermediate polars and the third fastest known in a cataclysmic variable''. A recent multiwavelength photometry and 
HST spectroscopy study \citep{2013ApJ...772..116S} has revealed some problems with the intermediate polar interpretation of \object{V842\,Cen}.

Our spectrum of \object{V842\,Cen} obtained about 26 years after the nova outburst can be compared to the spectrum obtained by \citet{2005A&A...432..199S} nine years earlier (Fig.~\ref{spectra} and Table~\ref{em}). An obvious difference is that the hot continuum in our spectrum is not very steep in the blue. Comparing the plotted spectra, the EW, and the FWHM, it looks like the Balmer and the \ion{He}{I} emission lines remain more or less the same. Some higher excitation emissions like \ion{C}{IV} 5805\,\AA\ and \ion{He}{II} 5412\,\AA\ are not visible in our spectrum. Other such emissions, for example, \ion{He}{II} 4686\,\AA\ and the Bowen blend are significantly fainter than nine years ago. The multicomponent structure of H$\alpha$ testifies that the [\ion{N}{II}] emission from the expanded shell continue to be significant. Also, the [\ion{O}{III}] 5007\,\AA\ nebular emission (not \ion{He}{I} as identified this feature \citet{2005A&A...432..199S}) is still seen clearly in the spectrum.

\subsection{\object{GQ\,Mus}}
\label{gqmus}

W.~Liller discovered this nova on January 18, 1983 \citep{1983IAUC.3764....1L}. \object{GQ\,Mus} was classified as a moderately fast nova with $t_{\mathrm{3}} \sim 45^\mathrm{d}$ \citep{1987SSRv...45....1D}. The most important observations after the outburst in a wide wavelength range are listed in \citet{1987SSRv...45....1D}. An orbital period of 85 minutes, confirmed by time-resolved photometry and spectroscopy, was found by \citet{1989ApJ...339L..41D, 1990RMxAA..21..369D, 1994ApJ...425..252D}. They also suggested that the outbursting component in \object{GQ\,Mus} is a magnetic white dwarf. The nova is a known super soft X-ray source \citep[and references therein]{2011ApJS..197...31S}. \citet{2008ApJ...687.1236H} estimate the mass of the white dwarf in the system to be $0.7\pm0.05$\,M$_\sun$.  A spectrum of the remnant, obtained in 1994, is shown by \citet{1995MNRAS.277..959D}. Attempts by \citet{1998MNRAS.300..221G} and \citet{2000AJ....120.2007D} failed to detect an extended shell around \object{GQ\,Mus}.

A comparison of the spectrum of \object{GQ\,Mus} in Fig.~\ref{spectra} to the one obtained in 1994 by \citet{1995MNRAS.277..959D} shows that, most probably, ours is the first really quiescence spectrum observed. Their spectrum is very rich in emission features, mainly Balmer and nebular lines, while in our spectrum a weak H$\alpha$ and maybe only a trace of the \ion{He}{II} 4686\,\AA\ emissions can be seen. The flat top of the H$\alpha$ profile suggests that the contribution of the [\ion{N}{II}] shell emission could still be significant. A hot continuum, which is steeper in the blue, is also evident.

Recently, \citet{2014BaltA..23....1N} have found that the total amplitude of the \object{GQ\,Mus} orbital modulation brightness decreased from $\sim0\fm9$ in 1992 to $\sim0\fm2$ in 2010. In the same time period, they found no evidence of period changes between 1989 and 2011. The optical spectrum obtained in 2001 and ours obtained in 2012 indicate continuing activity of \object{GQ\,Mus}.

\subsection{\object{Nova\,Car\,1972}}
\label{ncar1972}

\object{Nova\,Car\,1972} has been a questionable object even since its discovery. It was communicated as a possible nova even by its discoverers \citep{1978IBVS.1476....1M}. \citet{1987SSRv...45....1D} mentioned it as a poorly known nova. In a later attempt, \citet{2000AJ....120.2007D} were not able to find a H$\alpha$-bright candidate and expressed doubts about the reality of \object{Nova\,Car\,1972}.

It looks like there is a very weak blue continuum in the spectrum of \object{Nova\,Car\,1972} (Fig.~\ref{spectra}). But this spectrum is dereddened when using a relatively large E$_\mathrm{B-V}\sim0\fm48$, corresponding to the galactic coordinates of the object \citep[NASA's IPAC Infrared Science Archive, see Sect.~\ref{ebmv_sed} and Table~\ref{redd}]{2011ApJ...737..103S}. The originally extracted, not reddening-corrected spectrum (Fig.~\ref{map_sp}) indicates a rather very weak and red continuum.  There are no emission lines in the spectrum. We agree with the suggestion of \citet{2000AJ....120.2007D} about the reality of this object, and our conclusion is that \object{Nova\,Car\,1972} is \textsl{not} a nova. 

\subsection{\object{HS\,Pup}}
\label{hspup}

\object{HS\,Pup} reached a maximum brightness of $\sim\,8^\mathrm{m}$ in December 1963 \citep{1964IBVS...60....1H} but was discovered in February 1964 \citep{1964IBVS...59....1S}. \citet{1987SSRv...45....1D} classified it as a moderately fast nova ($t_{\mathrm{3}}\sim 65^\mathrm{d}$) with no spectroscopic observation available. The first quiescent spectra of the star were obtained by \citet{1991Msngr..64...32B,1992Msngr..69...42B}, but they published only an extremely short description. The last one, reported in the literature, was obtained by \citet{1995A&AS..114..575Z} several years later. An extended shell with a diameter <\,2\farcs5 was  found around \object{HS\,Pup} by \citet{1998MNRAS.300..221G} but they were not able to estimate the distance because of the unknown expansion velocity. \citet{2010MNRAS.403..398W} studied the rapid variations in the star brightness and suggest a tentative orbital period of 3\fh244.

Recently, \citet{2013MNRAS.436.2412T} published a better quality spectrum of \object{HS\,Pup} obtained in 2009. The Balmer emission lines and several \ion{He}{I} lines are visible. They estimate that the continuum corresponds to an early-to-mid K star. Based on an analysis of the radial velocities, they suggest an orbital period for \object{HS\,Pup} $P_{\mathrm{orb}}=6\fh41$.

A relatively weak continuum, increasing toward the blue, is obvious in our spectrum of \object{HS\,Pup} (Fig.~\ref{spectra}). The only line that can be definitely identified, is a comparatively strong H$\alpha$ emission. The spectrum obtained in 1994 and shown by \citet{1995A&AS..114..575Z} is very similar to ours with also only H$\alpha$ emission apparent. This does not exclude the presence of other lines in the \object{HS\,Pup} spectrum as the low S/N gives an upper limit for the equivalent width of unidentified lines $<11$\,\AA.

\subsection{\object{HZ\,Pup}}
\label{hzpup}

\object{HZ\,Pup} was discovered on Sonneberg archive plates by \citet{1964IBVS...45....1H} after its maximum brightness $\sim$7\fm7 occurred around January 20, 1963. Low-resolution spectra of the remnant were obtained by  \citet{1991Msngr..64...32B,1992Msngr..69...42B} and \citet{1995A&AS..114..575Z}. \citet{1997ASPC..121..679A} suggested that \object{HZ\,Pup} is an intermediate polar with a spin period of the magnetic white dwarf $\sim$13 minutes and $P_{\mathrm{orb}} \sim 5\fh11$. The observations of \citet{1998MNRAS.300..221G} did not show an extended shell around \object{HZ\,Pup} remnant.

A comparison of our RSS observations presented in Fig.~\ref{spectra} with the published by \citet{1991Msngr..64...32B,1992Msngr..69...42B} and \citet{1995A&AS..114..575Z} shows that the spectrum of \object{HZ\,Pup} remained more or less the same during the last 22 years. It is characterized by a strong blue continuum and numerous emission lines. All stronger lines of \ion{H}{I}, \ion{He}{I}, and \ion{He}{II} are visible in emission (Table~\ref{em}). Whereas the intensities of H$\beta$ and \ion{He}{II} 4686\,\AA\ are almost identical in ours and \citet{1995A&AS..114..575Z} spectra, in the spectrum of  \citet{1991Msngr..64...32B}, H$\beta$ is significantly stronger. The hump seen in the continuum in the region $\sim$\,4200--5200\,\AA\ is artificial, most probably caused by a very bright star in the slit at about 2\arcmin\ to the north of \object{HZ\,Pup}.  Our attempts to remove this hump during the data processing proved unsuccessful.

\subsection{\object{V365\,Car}}
\label{v365car}

This nova was first detected as a H$\alpha$ emission-line object by \citet{1967ApJS...14..125H}. Years later it was recognized by \citet{1975ApJ...200..694H} as a very slow nova with $t_{\mathrm{3}} > 530^\mathrm{d}$ \citep{1987SSRv...45....1D}. \citet{1975ApJ...200..694H} estimated the distance to \object{V365\,Car} to $3.5 \pm 1$\,kpc and found an upper limit for the ejection velocity of about 300\,km\,s$^\mathrm{-1}$. The remnant spectrum was only observed by \citet{1996A&AS..117..449Z}. Two attempts by \citet{1998MNRAS.300..221G} and \citet{2000AJ....120.2007D} to detect an extended shell around \object{V365\,Car} were unsuccessful. \citet{2002MNRAS.335...44W} found slow flickering in their time-resolved photometry of \object{V365\,Car}, but they were not able to detect any orbital period. \citet{2013MNRAS.436.2412T} show an average of their spectra obtained during 2011 and 2012 in the region $\sim6200-7100$\,\AA. They only mention the presence of relatively weak H$\alpha$ and \ion{He}{I} 6678\,\AA\ 
emissions. Based on the brightness and radial velocity variations, they suggest an orbital period for \object{V365\,Car} $P_{\mathrm{orb}}$ of $\sim 5\fh35$.

A blue continuum, similar to the observed by \citet{1996A&AS..117..449Z} in 1995 is apparent in our spectrum of \object{V365\,Car} shown in Fig.~\ref{spectra}. A very weak H$\alpha$ emission presents also in the spectrum but mentioned by \citet{1996A&AS..117..449Z} H$\beta$ and \ion{He}{II} 4686\,\AA\ emission lines are not seen. The reason is the low $S/N$ that gives an upper limit on the strength of not identified lines $EW<7$\,\AA.

\subsection{\object{CP\,Pup}}
\label{cppup}

\object{CP\,Pup} discovered by Dawson \citep[see][]{1942HarAC.637....2D} is one of the fastest ($t_{\mathrm{3}}\sim 8^\mathrm{d}$) and the brightest ($\sim0\fm5$) novae ever observed \citep{1987SSRv...45....1D}. Its outburst and post-outburst evolution are covered well by observations \citep[see, for example,][and references therein]{1987SSRv...45....1D,1989MNRAS.240...41O,1991PASP..103..964D,1993ApJ...412..278W,1998PASP..110.1026P,2012A&A...539A..94B}. However, some parameters of the nova, such as spectral and photometric orbital periods, white dwarf mass, and the existence of a magnetic field, are still discussed in the literature \citep[for more details see][]{2012A&A...539A..94B}. The extended shell around \object{CP\,Pup} was discovered by Zwicky about 13 years after the outburst \citep[see][]{1956AJ.....61..338B} and then was observed and studied by many authors \citep{1980MitAG..50...70S,1982ApJ...261..170W,1983ApJ...268..689C,1987Msngr..50....8D,1998MNRAS.300..221G}. The last observations of the 
expanding shell, together with a critical discussion of the results obtained on the basis of the previous observations, are presented by \citet{2000AJ....120.2007D}. They estimate a distance to \object{CP\,Pup} of 1140 pc and suggest that there is fast- and slow-moving material in the shell.

The spectrum of \object{CP\,Pup} shown in Fig.~\ref{spectra} does not differ substantially from the spectra of this nova, which is presented and described by other authors \citep{1987MNRAS.229..653D, 1989MNRAS.240...41O, 1991Msngr..64...32B, 1992Msngr..69...42B, 1993ApJ...412..278W, 2012A&A...539A..94B}. The continuum is strong and very steep in the blue. The strongest emissions in the spectrum belong to \ion{H}{I}, \ion{He}{I}, and \ion{He}{II}. The Bowen blend is only slightly weaker than \ion{He}{II} 4686\,\AA, and the \ion{C}{IV} 5805\,\AA\ emission line is seen as well. The Balmer lines from H$\alpha$ to H$_\mathrm{8}$ are the most intensive emissions in the spectrum with FWHM of about $1347\pm53$\,km\,s$^{-1}$ which is remarkably greater in comparison to the $1061\pm63$\,km\,s$^{-1}$ FWHM of \ion{He}{II} (Table~\ref{em}). 

\subsection{\object{BT\,Mon}}
\label{btmon}

\object{BT\,Mon} was discovered by Wachmann \citep{1987SSRv...45....1D} on December 17, 1939. The exact moment of the maximum and the nova's maximum brightness are unknown. Many different values for $t_{\mathrm{3}}$ of the nova are found in the literature, but there is no consensus about its speed class (see, for instance, the discussion in \citet{1998MNRAS.300..221G}). \object{BT\,Mon} is an eclipsing binary with $P_{\mathrm{orb}} \sim 8\fh01$, eclipse amplitude $\sim 2\fm7$, and orbital inclination $i \sim 82\degr$ \citep{1982ApJ...254..646R, 1983ApJ...268..710S, 1998MNRAS.296..465S}. The exact nature of the system is not known, and the following are still under discussion: high-velocity gas flows, existence of accretion disk, white dwarf magnetic field, etc. \citep[and references therein]{1984Ap&SS..99...95S, 1994ApJS...93..519K, 1996ApJ...456..777W, 1998MNRAS.296..465S}. \citet{1983MNRAS.205P..33M} detected a nebulae around \object{BT\,Mon} using H$\alpha$/[\ion{N}{ii}] spectral observations and 
estimated a distance of $\sim\ 1800$\,pc. Optical direct imaging observations presented by \citet{1998MNRAS.300..221G} were in good agreement with the results of \citet{1983MNRAS.205P..33M}. It is interesting that \citet{2000MNRAS.314..175G} were not able to detect the nebula in \textsl{HST} images of \object{BT\,Mon}.

In the RSS spectrum of \object{BT\,Mon}, obtained at orbital phase $0.605\pm0.002$ \citep[ephemeris from][]{1998MNRAS.296..465S}, a relatively strong but not so steep blue continuum is apparent. Our spectrum is very similar to what was observed by \citet{1983ApJS...53..523W} in 1980-1981. The Balmer lines are very intensive and easily seen in emission up to H$_\mathrm{10}$. The \ion{He}{II} emission at 4686\,\AA\ is stronger than H$\beta$, and the line at 5412\,\AA\ is also visible. The emission lines of \ion{He}{I} like 5876\,\AA, 4922\,\AA, 4471\,\AA\, and 4026\,\AA\ are obvious as well (Fig.~\ref{spectra}). The FWHM of \ion{He}{II} 4686\,\AA\ $852\pm11$\,km\,s$^{-1}$ is much smaller than the mean FWHM of  H$\beta$ and H$\gamma$ $1137\pm21$\,km\,s$^{-1}$. The FWHM of H$\beta$, H$\gamma$, and \ion{He}{II} 4686\,\AA\ and the EW of H$\beta$ and H$\gamma$ are in good agreement with those measured by \citet{1996ApJ...456..777W} at similar orbital phases, but the one measured by us EW for \ion{He}{II} 4686\,\AA\ 
is 
about twice smaller than their value (Table~\ref{em}). 

\subsection{\object{RR\,Pic}}
\label{rrpic}

\object{RR\,Pic} is among the brightest classical novae with a pre- and post-outburst magnitude $V\sim12^\mathrm{m}$. It exploded as a typical slow nova in 1925 \citep[time to decline by 3 mag $t_\mathrm{3}=150^\mathrm{d}$,][]{1987SSRv...45....1D} reaching  $V\sim1^\mathrm{m}$ at its maximum \citep{warner08}. Because of its brightness, \object{RR\,Pic} is one of the most studied novae \citep[see][and references therein]{jones31, 1986MNRAS.219..751W, 2003MNRAS.342..145S, 2006PASP..118...84R, 2008MNRAS.389.1345S}. An extended shell with a  complex shape around \object{RR\,Pic} has been known and has been well documented for long time \citep{1979ApJ...228..482W, 1979Msngr..17....1D, 1998MNRAS.300..221G, 2000AJ....120.2007D}. Some authors \citep{1982A&A...109..171H, 1986MNRAS.219..751W} have suggested that the nova is an eclipsing binary with $P=3\fh481$ \citep{1975A&A....41...15V} whose orbital plane is inclined to the line of sight by about $\sim70\degr$. Others did not confirm the presence of an eclipse \
citep[see the discussion in][]{2008MNRAS.389.1345S}.

Our spectrum of \object{RR\,Pic} reveals a strong blue continuum. The strongest features are the Balmer series and \ion{He}{II} emission lines (Fig.~\ref{spectra}). The only additional, strong emission is the Bowen blend. The spectrum is very similar to the one observed by \citet{2006PASP..118...84R} during 2001-2003. Using the corrected ephemeris given in \citet{2005ASPC..335..333S}, we find that our spectrum was obtained at orbital phase $0.25 \pm 0.03$. The EW of $\sim 21$\,\AA\ for H$\alpha$ measured by us (Table~\ref{em}) is close to the values measured by \citet{2003MNRAS.342..145S} at similar phases. The FWHM of the Balmer and \ion{He}{II} emissions, which we estimated using H$\beta$, H$\gamma$, \ion{He}{II} 4686\,\AA, and 5412\,\AA, are practically the same, $1118\pm58$\,km\,s$^{-1}$ and $1061\pm116$\,km\,s$^{-1}$, respectively.

\subsection{\object{GI\,Mon}}
\label{gimon}

According to \citet{1987SSRv...45....1D}, \object{GI\,Mon} is a fast nova whose maximum is poorly covered by observations. It was discovered by M.~Wolf on February 4, 1918. \citet{2004MNRAS.351.1015W} suggest an orbital period of $P_{\mathrm{orb}} \sim 4\fh3$, but it was not confirmed by the observations of \citet{2005A&A...431..289R}. \citet{1998MNRAS.300..221G} failed to resolve a shell around \object{GI\,Mon}.

A steep blue continuum dominates our spectrum of \object{GI\,Mon}. Additionally, only weak emissions of H$\alpha$, H$\beta$, \ion{He}{II} 4886\,\AA, and the Bowen blend can be seen. The RSS spectrum of \object{GI\,Mon} is very similar to the only published quiescent spectrum obtained by \citet{2000ApJS..128..387L} in 1998. Even the 4.5\,\AA\ H$\alpha$ EW measured by them is very close to our value of 4.0\,\AA\ (Fig.~\ref{spectra} and Table~\ref{em}). Keeping the similarity of ours and \citet{2000ApJS..128..387L} spectra in mind, it is interesting to mention that \citet{1991Msngr..64...32B} reported weak Balmer and \ion{He}{II} emissions in the spectrum of \object{GI\,Mon}. In their next paper \citep{1992Msngr..69...42B}, however, they include this nova in the group of objects showing strong \ion{He}{II} emission lines.

\subsection{\object{OY\,Ara}}
\label{oyara}

\object{OY\,Ara} was discovered by W.~Fleming on Harvard plates during the outburst in 1910 \citep{1910BHarO.427....1P, 1912AnHar..56..165F}. It is a moderately fast nova with $t_{\mathrm{3}}\sim80^\mathrm{d}$ and a peculiar secondary maximum \citep{1987SSRv...45....1D}. \citet{1997ApJ...483..899Z} found that \object{OY\,Ara} is an eclipsing system
with an orbital period of $\sim$0.16 days and masses of the components $M_\mathrm{1}=0.82$\,M$_\sun$\ and $M_\mathrm{2}=0.34$\,M$_\sun$. They suggested a very high rate of mass transfer for this nova.

Comparing a single spectrum obtained in 1994 with the 38 obtained in 1995 and covering the full orbital period, \citet{1997ApJ...483..899Z} found that the spectrum of the \object{OY\,Ara} remnant varies markedly on a one-year timescale, as well as during the orbital cycle. In general, our spectrum of \object{OY\,Ara} (Fig.~\ref{spectra}) is very similar to the spectrum obtained in 1994 and the averaged 1995 spectra in both continuum and lines. The Balmer lines, well seen up to $H\delta$, are the strongest emission features. The emissions of \ion{He}{II}, \ion{He}{I}, and Bowen blend are obviously weaker in our spectrum. The EW measured by us (Table~\ref{em}) are systematically larger than the mean values shown by \citet{1997ApJ...483..899Z} in their Table~4. All these differences are most probably caused by the orbital motion. As  seen in Fig.~5 of \citet{1997ApJ...483..899Z}, our spectrum closely  resembles those obtained at phase 0. It is very difficult to estimate the exact orbital phase at which our 
spectrum of \object{OY\,Ara} was obtained using the \citet{1997ApJ...483..899Z} ephemeris because of the very large error accumulated in more than 38\,000 cycles since the moment T$_\mathrm{0}$.

\subsection{\object{CN\,Vel}}
\label{cnvel}

\object{CN\,Vel} is a poorly studied, very slow nova with $t_{\mathrm{3}} > 800^\mathrm{d}$ \citep{1987SSRv...45....1D}. It was discovered by H.~Leavitt on a photographic plate obtained on December 5, 1905. Perhaps the most complete light curve for the years before and after the outburst was published by \citet{1933AnHar..84..189W}. The only spectrum obtained about one and a half years after its discovery was described by \citet{1916AnHar..76...19C}. Two spectral observations of the remnant were reported later by \citet{1991Msngr..64...32B} and \citet{1996A&AS..117..449Z}.

The spectra of \object{CN\,Vel} obtained in 1991 and 1995 \citep{1991Msngr..64...32B, 1996A&AS..117..449Z} are dominated by a strong blue continuum. In addition only very weak emissions of H$\alpha$ and \ion{He}{II} 4686\,\AA\ are visible. \citet{1991Msngr..64...32B} reported on an intriguing "blue flare" observed in February 1991. A recent low-resolution spectrum obtained in  May 2009 \citep{2013MNRAS.436.2412T} demonstrates a strong blue continuum and not very strong emission features. Comparatively intensive are H$\alpha$ and the Bowen blend. The other visible Balmer and \ion{He}{I} emission lines are weak. The H$\alpha$ radial velocities periodogram analysis carried out by \citet{2013MNRAS.436.2412T} suggests an orbital period $P_{\mathrm{orb}} \sim 5\fh29$ for \object{CN\,Vel}.

The blue continuum is seen well in our spectrum of \object{CN\,Vel} (Fig.~\ref{spectra}). There are no obvious line features because of the very low $S/N$ that determines an upper limit for the equivalent width of unidentified lines $< 13$\,\AA.

\subsection{\object{X\,Ser}}
\label{xser}

\object{X\,Ser} was discovered by H.~Leavitt on Harvard plates, several years after its outburst in 1903 \citep{1908HarCi.142....1L}. 
This is a very slow nova, which remained near its maximum light between May 5 and September 21, 1903, whose $t_{\mathrm{3}}$ decline time was about 555 days \citep{1990LNP...369..165D}. \citet{2000MNRAS.312..629T} found a long ($P_\mathrm{orb}=1\fd48$) spectroscopic orbital period for \object{X\,Ser}. Brightness variations with different amplitudes and characteristic times are typical of this old nova during quiescence \citep{1990LNP...369..165D, 1990PASP..102..758H, 1998AJ....115.2527H}. Spectral observations during the \object{X\,Ser} outburst are missing. The numerous published spectra of the remnant \citep{1983ApJS...53..523W, 1986ApJ...311..163S, 1990LNP...369..165D, 1992ApJS...78..537S, 1996MNRAS.281..192R, 2000MNRAS.312..629T} are evidence of remarkable changes in the continuum and the spectral lines.

At the time of our observations, only the Balmer emission lines from H$\alpha$ to H$\delta$ are undoubtedly presented in the spectrum. They are superimposed on a not very strong or very steep hot continuum (Fig.~\ref{spectra}). Very weak \ion{He}{I} emission lines (e.g., 5876\,\AA) can hardly be seen in the spectrum. There are not even traces of features like \ion{He}{II} 4686\,\AA, sometimes observed as strong emissions in the past.

\subsection{\object{DY\,Pup}}
\label{dypup}

\object{DY\,Pup} was discovered by I.E.~Woods about twenty years after its outburst in January 1902 \citep[see][]{1921BHarO.760Q...1S}. In \citet{1987SSRv...45....1D} the object is classified as a slow nova with brightness in the maximum $\sim$7$^\mathrm{m}$ and $t_{\mathrm{3}}\sim 160^\mathrm{d}$. The first and so far the only spectral observations of the post-nova remnant were published by \citet{1996A&AS..117..449Z}. \citet{1998MNRAS.300..221G} revealed an extended shell around \object{DY\,Pup} with dimensions about 7\arcsec$\times$5\arcsec. An orbital period of 3\fh336 is mentioned by \citet{2003MNSSA..62...74W}.

A careful extraction of our spectrum of \object{DY\,Pup} presented in Fig.~\ref{spectra} shows a strong blue continuum and, maybe, only a very weak trace of the H$\alpha$ emission. (The upper limit on the strength of unidentified lines is $EW<5$\,\AA.) Eighteen years ago \citet{1996A&AS..117..449Z} observed a very faint H$\alpha$ emission superimposed on a strong blue continuum, very similar to the present one.

\begin{sidewaystable*}
\addtolength{\tabcolsep}{-2pt}
\caption{Equivalent widths and FWHM of the emission lines in the targets spectra.}
\label{em}
\centering
\begin{tabular}{lrrrrrrrrrrrr}
\hline\hline
Star      & \multicolumn{2}{c}{H$\alpha$}   & \multicolumn{2}{c}{H$\beta$}    & \multicolumn{2}{c}{H$\gamma$}   & \multicolumn{2}{c}{H$\delta$}   & \multicolumn{2}{c}{H$\varepsilon$}  & \multicolumn{2}{c}{H8}      \\
          &   \multicolumn{1}{c}{EW} & \multicolumn{1}{c}{FWHM}  & \multicolumn{1}{c}{EW} &  \multicolumn{1}{c}{FWHM}  & \multicolumn{1}{c}{EW} &  \multicolumn{1}{c}{FWHM}  & \multicolumn{1}{c}{EW} &  \multicolumn{1}{c}{FWHM}  & \multicolumn{1}{c}{EW} & \multicolumn{1}{c}{FWHM}  & \multicolumn{1}{c}{EW} & \multicolumn{1}{c}{FWHM} \\
          &   \multicolumn{1}{c}{\AA} &\multicolumn{1}{c}{km\,s$^{-1}$} & \multicolumn{1}{c}{\AA}&\multicolumn{1}{c}{km\,s$^{-1}$}& \multicolumn{1}{c}{\AA} &\multicolumn{1}{c}{km\,s$^{-1}$} & \multicolumn{1}{c}{\AA} &\multicolumn{1}{c}{km\,s$^{-1}$} & \multicolumn{1}{c}{\AA} &\multicolumn{1}{c}{km\,s$^{-1}$}&  \multicolumn{1}{c}{\AA}&\multicolumn{1}{c}{km\,s$^{-1}$}\\
\hline
\object{T\,Pyx}		&95.4	$\pm$ 6.0	&468	$\pm$ 31	&46.9	$\pm$ 1.6	&1346	$\pm$ 53	&37.1	$\pm$ 2.8	&1369	$\pm$ 131	&25.2	$\pm$ 1.3	&2286	$\pm$ 125	&   &   &   &      \\
		&123.8	$\pm$ 7.6	&670	$\pm$ 37	&   &   &	&  &   &   &   &   &   &     \\
		&36.6	$\pm$ 4.8	&713	$\pm$ 88	&   &   &   &   &   &   &   &   &   &     \\
\object{U\,Sco}		&7.4	$\pm$ 0.8	&752	$\pm$ 100	&   &   &   &   &   &   &   &   &   &      \\
\object{V382\,Vel}	&71.6	$\pm$ 0.6\tablefootmark{a}	&1949	$\pm$ 19\tablefootmark{a}	&16.6	$\pm$ 0.2	&1677	$\pm$ 14	&8.6	$\pm$ 0.1	&1604	$\pm$ 19	&8.4	$\pm$ 0.2	&1654	$\pm$ 30	&7.4	$\pm$ 0.1	&1636	$\pm$ 20	&3.4	$\pm$ 0.1	&1295	$\pm$ 33   \\
\object{BY\,Cir}		&105.2	$\pm$ 3.1\tablefootmark{a}	&1790	$\pm$ 71\tablefootmark{a}	&24.8	$\pm$ 2.3	&1816	$\pm$ 178	&   &   &   &   &   &  &  &   \\
\object{V888\,Cen}	&8.1	$\pm$ 0.9	&962	$\pm$ 127	&   &   &   &   &   &   &   &   &     \\
\object{V868\,Cen}	&123.7 $\pm$ 9.4\tablefootmark{a}	&1303	$\pm$ 137\tablefootmark{a}	&   &   &   &   &   &   &   &   &   &    \\
\object{V842\,Cen}	&102.2	$\pm$ 0.4\tablefootmark{a}	&1705	$\pm$ 7\tablefootmark{a}	&16.3	$\pm$ 0.4	&1160	$\pm$ 31	&8.1	$\pm$ 0.2	&1086	$\pm$ 45	&8.8	$\pm$ 0.3	&1390	$\pm$ 48	&7.3	$\pm$ 0.2	&1174	$\pm$ 54	&   &      \\
\object{HS\,Pup}		&58.0	$\pm$ 3	&1252	$\pm$ 78	&   &   &   &   &   &   &   &   &   &   \\
\object{HZ\,Pup}		&59.9	$\pm$ 1.2	&958	$\pm$ 25	&20.5	$\pm$ 0.6	&1074	$\pm$ 39	&14.1	$\pm$ 0.5	&1109	$\pm$ 54	&13.7	$\pm$ 0.6	&1133	$\pm$ 49	&12.8	$\pm$ 0.5	&1280	$\pm$ 63	&9.1	$\pm$ 0.3	&713	$\pm$ 34  \\
\object{CP\,Pup}		&18.7	$\pm$ 0.7	&1360	$\pm$ 46	&14.6	$\pm$ 0.1	&1297	$\pm$ 12	&12.3	$\pm$ 0.1	&1355	$\pm$ 11	&13.1	$\pm$ 0.1	&1492	$\pm$ 15	&9.4	$\pm$ 0.1	&1241	$\pm$ 9	&8.7	$\pm$ 0.1	&1336	$\pm$ 11   \\
\object{BT\,Mon	}	&44.2	$\pm$ 0.4	&1306	$\pm$ 15	&21.0	$\pm$ 0.2	&1111	$\pm$ 12	&13.4	$\pm$ 0.2	&1162	$\pm$ 17	&10.7	$\pm$ 0.2	&1144	$\pm$ 23	&8.2	$\pm$ 0.2	&1172	$\pm$ 25	&5.4	$\pm$ 0.1	&1001	$\pm$ 28   \\
\object{RR\,Pic}		&20.6	$\pm$ 0.5	&929	$\pm$ 30	&7.7	$\pm$ 0.2	&1154	$\pm$ 30	&3.4	$\pm$ 0.1	&1082	$\pm$ 50	&3.5	$\pm$ 0.1	&1263	$\pm$ 46	&3.2	$\pm$ 0.1	&1658	$\pm$ 72	&   &  \\
\object{GI\,Mon}		&4.0	$\pm$ 0.3	&618	$\pm$ 51	&1.1	$\pm$ 0.2	&619	$\pm$ 93	&   &   &   &   &   &   &   &    \\
\object{OY\,Ara}		&53.4	$\pm$ 1.3	&1166	$\pm$ 29	&45.1	$\pm$ 1.0	&1127	$\pm$ 33	&34.0	$\pm$ 3.6	&1001	$\pm$ 89	& &	&   &   &   &     \\
\object{X\,Ser}		&20.0	$\pm$ 1.4	&538	$\pm$ 43	&14.2	$\pm$ 1.1	&752	$\pm$ 66	&15.4	$\pm$ 1.7	&1891	$\pm$ 249	&14.0	$\pm$ 1.2	&1425	$\pm$ 131	&   &   &   &      \\
\hline
 & & & & & & & & & & & & \\
\hline\hline
Star      & \multicolumn{2}{c}{\ion{He}{i}\,5876\,\AA}   & \multicolumn{2}{c}{\ion{C}{iv}\,5805\,\AA}    & \multicolumn{2}{c}{[\ion{N}{ii}]\,5755\,\AA} & \multicolumn{2}{c}{\ion{He}{ii}\,5412\,\AA}   & \multicolumn{2}{c}{[\ion{O}{iii}]\,5007\,\AA}  & \multicolumn{2}{c}{[\ion{O}{iii}]\,4959\,\AA}\\
          & \multicolumn{1}{c}{EW} & \multicolumn{1}{c}{FWHM} & \multicolumn{1}{c}{EW} & \multicolumn{1}{c}{FWHM} & \multicolumn{1}{c}{EW} & \multicolumn{1}{c}{FWHM} & \multicolumn{1}{c}{EW} & \multicolumn{1}{c}{FWHM} & \multicolumn{1}{c}{EW} & \multicolumn{1}{c}{FWHM} & \multicolumn{1}{c}{EW} & \multicolumn{1}{c}{FWHM} \\
          & \multicolumn{1}{c}{\AA} &\multicolumn{1}{c}{km\,s$^{-1}$} &\multicolumn{1}{c}{\AA} &\multicolumn{1}{c}{km\,s$^{-1}$}& \multicolumn{1}{c}{\AA} & \multicolumn{1}{c}{km\,s$^{-1}$}& \multicolumn{1}{c}{\AA} & \multicolumn{1}{c}{km\,s$^{-1}$}& \multicolumn{1}{c}{\AA} &\multicolumn{1}{c}{km\,s$^{-1}$}&\multicolumn{1}{c}{\AA} &\multicolumn{1}{c}{km\,s$^{-1}$}\\
\hline
\object{T\,Pyx}		&	&		&	      	&	      	&18.9	$\pm$1.7&330	$\pm$30	&	      	&	      	&82.6	$\pm$1.4&636	$\pm$10	&&	\\
		& &		&	      	&	      	&26.7	$\pm$3.5&574	$\pm$69	&	      	&	      	&477.5	$\pm$1.3&657	$\pm$2	&202.4	$\pm$1.2&707	$\pm$5	\\
		& &	&	      	&	      	&8.9	$\pm$3.7&1189	$\pm$536&	      	&	      	&475.9	$\pm$1.4&720	$\pm$2	&202.6	$\pm$1.3&724	$\pm$5\\
		&	      	&	      		&	      	&	      	&	      	&	      	&	      	&	      	&168.9	$\pm$1.6&766	$\pm$7	& 74.9	$\pm$2.1 &753 $\pm$18	\\
\object{V445\,Pup}	&	      	&	      		&	      	&	      	&	      	&	      	&	      	&	      	& 84.6	$\pm$2.6&623 $\pm$ 12 &38.9	$\pm$ 2.3&786	$\pm$36	\\
		&	      	&	      		&	      	&	      	&	      	&	      	&	      	&	      	&41.2	$\pm$3.6&1528	$\pm$144&17.8	$\pm$2.9&1494	$\pm$175\\
\object{V382\,Vel}	&5.6	$\pm$0.2&1654	$\pm$80		&2.6	$\pm$0.2&1642	$\pm$172&	      	&	      	&	      	&	      	&	      	&	      	&	      	&	      	\\
\object{BY\,Cir}		&&&	      	&	      	&	      	&	      	&	      	&	      	&20.1	$\pm$2.4&2215	$\pm$398&	      	&	      	\\
\object{V842\,Cen}	& 4.1 $\pm$ 0.4 &669 $\pm$ 71 &	      	&	      	&	      	&	      	&	      	&	      	&3.3	$\pm$0.3&572	$\pm$64	&	      	&	      	\\
		&&&	      	&	      	&	      	&	      	&	      	&	      	&1.4	$\pm$0.3&576	$\pm$150&	      	&	      	\\
\object{HS\,Pup}		&58.0	$\pm$3.0	&1399	$\pm$87		&	      	&	      	&	      	&	      	&	      	&	      	&	      	&	      	&	      	&	      	\\
\object{HZ\,Pup}		&9.7	$\pm$0.9&871	$\pm$105		&	      	&	      	&	      	&	      	&	      	&	      	&	      	&	      	&	      	&	      	\\
\object{CP\,Pup}		&2.4	$\pm$0.3&1011	$\pm$112		&2.5	$\pm$0.1&1113	$\pm$85	&	      	&	      	&3.7	$\pm$0.2&1072	$\pm$59	&	      	&	      	&	      	&	      	\\
\object{BT\,Mon	}	&6.9	$\pm$0.2&1088	$\pm$40		&1.3	$\pm$0.2&774	$\pm$133&	      	&	      	&1.6	$\pm$0.2&840	$\pm$113&	      	&	      	&	      	&	      	\\
\object{RR\,Pic}		& &		&3.8	$\pm$0.5&2162	$\pm$313&	      	&	      	&3.0	$\pm$0.2&1164	$\pm$115&	      	&	      	&	      	&	      	\\
\object{OY\,Ara}		&16.9 $\pm$ 0.9&1224 $\pm$ 74&	      	&	      	&	      	&	      	&	      	&	      	&	      	&	      	&	      	&	      	\\
\hline
\end{tabular} 
\tablefoot{
\tablefoottext{a}{Blended with the [\ion{N}{ii}] emission lines.}
}
\end{sidewaystable*}
\addtocounter{table}{-1}

\begin{sidewaystable}
\caption{Continued}
\centering
\begin{tabular}{lrrrrrrrrrrrr}
\hline\hline
Star      & \multicolumn{2}{c}{\ion{He}{i}\,4922\,\AA}   & \multicolumn{2}{c}{\ion{He}{i}\,4713\,\AA}    & \multicolumn{2}{c}{\ion{He}{ii}\,4686\,\AA}   & \multicolumn{2}{c}{Bowen bl.}   & \multicolumn{2}{c}{\ion{He}{i}\,4471\,\AA}  & \multicolumn{2}{c}{[\ion{O}{iii}]\,4363\,\AA}      \\
          & \multicolumn{1}{c}{EW} & \multicolumn{1}{c}{FWHM} & \multicolumn{1}{c}{EW} & \multicolumn{1}{c}{FWHM} & \multicolumn{1}{c}{EW} & \multicolumn{1}{c}{FWHM} & \multicolumn{1}{c}{EW} & \multicolumn{1}{c}{FWHM} & \multicolumn{1}{c}{EW} & \multicolumn{1}{c}{FWHM} & \multicolumn{1}{c}{EW} & \multicolumn{1}{c}{FWHM} \\
          & \multicolumn{1}{c}{\AA} &\multicolumn{1}{c}{km\,s$^{-1}$}&\multicolumn{1}{c}{\AA} &\multicolumn{1}{c}{km\,s$^{-1}$}& \multicolumn{1}{c}{\AA} & \multicolumn{1}{c}{km\,s$^{-1}$}& \multicolumn{1}{c}{\AA} & \multicolumn{1}{c}{km\,s$^{-1}$}& \multicolumn{1}{c}{\AA} &\multicolumn{1}{c}{km\,s$^{-1}$}&\multicolumn{1}{c}{\AA} &\multicolumn{1}{c}{km\,s$^{-1}$}\\
\hline
\object{T\,Pyx}		&	      	&	      	&	      	&	      	&41.1	$\pm$3.3&1984	$\pm$148&73.4	$\pm$4.0	&3496	$\pm$21	&	      	&	      	&10.9	$\pm$1.5&813	$\pm$122\\
		&	      	&	      	&	      	&	      	&	      	&	      	&	      	&	      	&	      	&	      	&32.4	$\pm$2.0	&587	$\pm$34	\\
		&	      	&	      	&	      	&	      	&	      	&	      	&	      	&	      	&	      	&	      	&29.4	$\pm$1.7&530	$\pm$29	\\
\object{U\,Sco}		&	      	&	      	&	      	&	      	&8.9	$\pm$0.5&668	$\pm$45	&	      	&	      	&	      	&	      	&	      	&	      	\\
\object{V382\,Vel}	&2.2	$\pm$0.1&1656	$\pm$163&2.0	$\pm$0.3&1544	$\pm$215&10.9	$\pm$0.3&1442	$\pm$36	&9.2	$\pm$0.2&2332	$\pm$70	&3.6	$\pm$0.1&2073	$\pm$97	&	      	&	      	\\
\object{BY\,Cir}		&	      	&	      	&	      	&	      	&12	$\pm$4.8&2035	$\pm$747&16.4	$\pm$4.0	&2175	$\pm$606&	      	&	      	&	      	&	      	\\
\object{V842\,Cen}	&	      	&	      	&	      	&	      	&6.8	$\pm$0.6&2172	$\pm$216&6.7	$\pm$0.6&1951	$\pm$18	& & &	      	&	      	\\
\object{HZ\,Pup}	&	      	&	      	&	      	&	      	&16.7	$\pm$0.6&1138	$\pm$47	&	      	&	      	&	      	&	      	&	      	&	      	\\
\object{CP\,Pup}		&2.1	$\pm$0.1&1193	$\pm$94	& & &8.4	$\pm$0.2&1050	$\pm$23	&8.8	$\pm$0.1&1638	$\pm$33	&2.0	$\pm$0.1&858	$\pm$49	&	      	&	      	\\
\object{BT\,Mon}		&2.7	$\pm$0.2&1038	$\pm$85	&3.1	$\pm$0.4&1786	$\pm$233&17.3	$\pm$0.3&852	$\pm$11	&6.9	$\pm$0.2&1136	$\pm$33	&3.9	$\pm$0.2&1314	$\pm$74	&	      	&	      	\\
\object{RR\,Pic}		&	      	&	      	& & &12.6	$\pm$0.2&958	$\pm$16	&8.2	$\pm$0.2&2087	$\pm$73	&	      	&	      	&	      	&	      	\\
\object{GI\,Mon}		&	      	&	      	&	      	&	      	&1.4	$\pm$0.2&876	$\pm$109&	      	&	      	&	      	&	      	&	      	&	      	\\
\hline
\end{tabular}
\end{sidewaystable}

\section{Emission line intensities}
\label{em_lines}

\begin{figure}
\centering
\resizebox{\hsize}{!}{\includegraphics{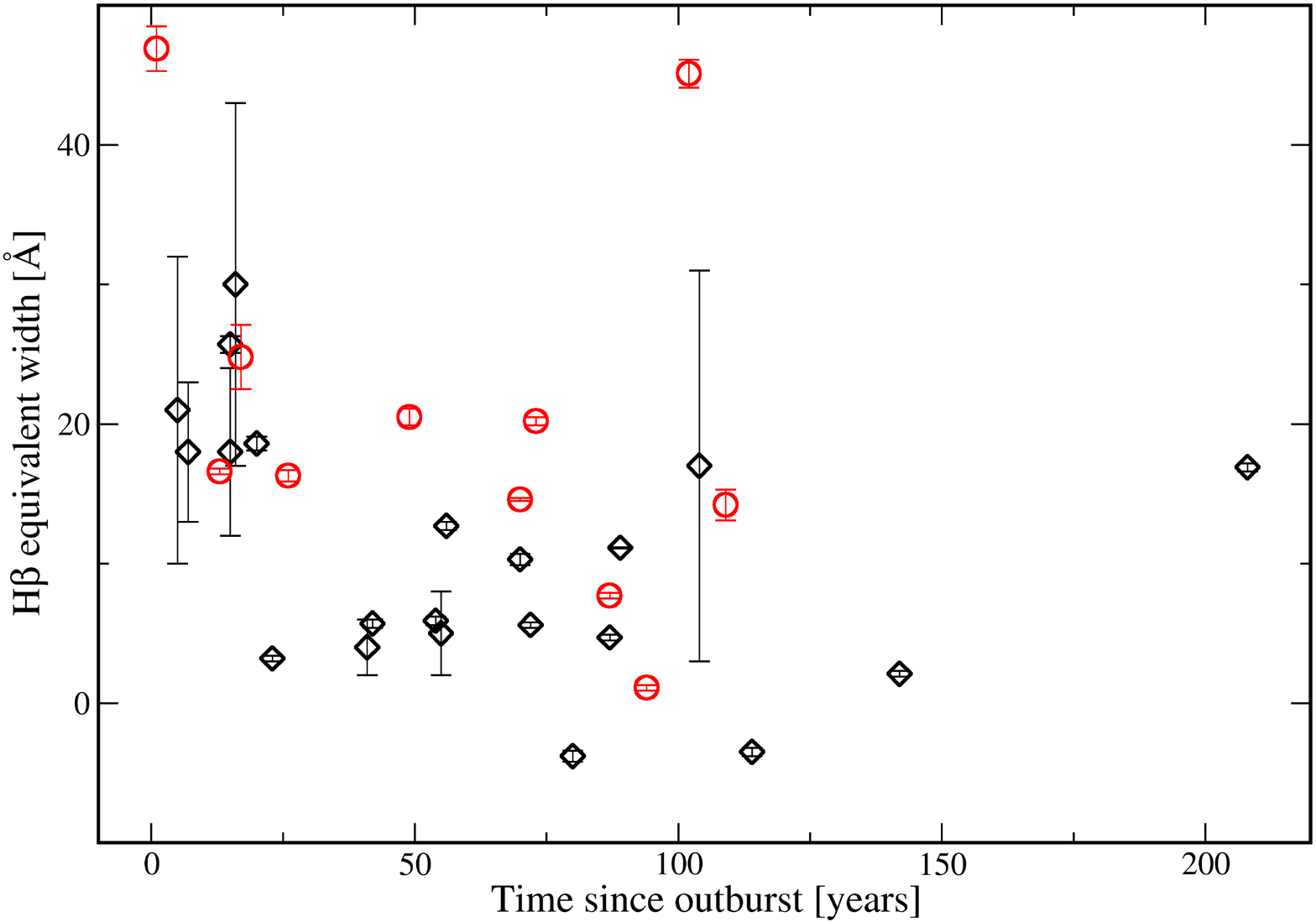}}
\caption{Equivalent width of H$\beta$ versus time since outburst. The symbols represents as follows: circles -- this paper; diamonds -- \citet{1996MNRAS.281..192R}}.
\label{hbeta}
\end{figure}

\begin{figure}
\centering
\resizebox{\hsize}{!}{\includegraphics{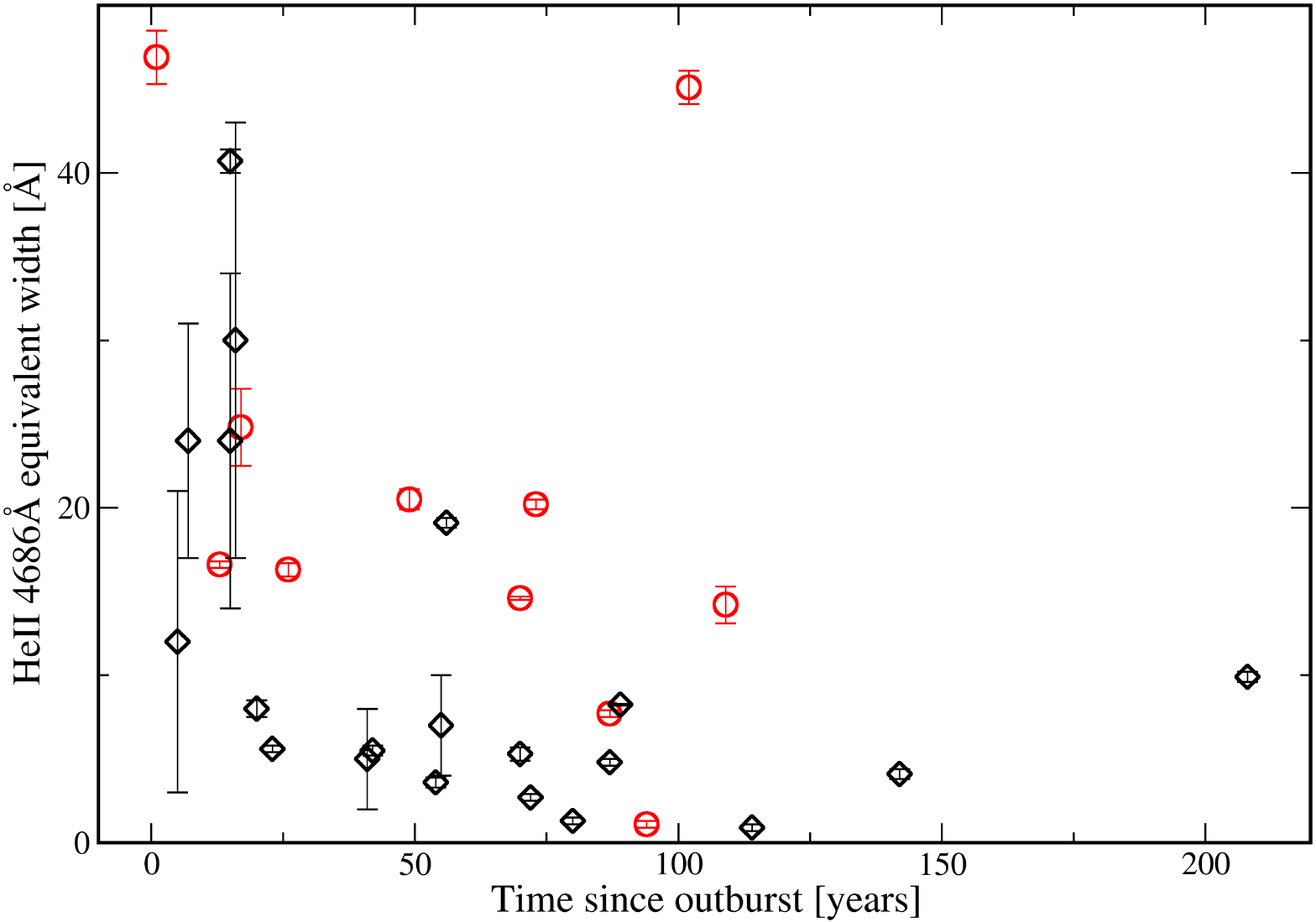}}
\caption{Equivalent width of \ion{He}{ii}\,4686\AA\ versus time since outburst. The symbols are as in Fig.~\ref{hbeta}}.
\label{heii}
\end{figure}

\begin{figure}
\centering
\resizebox{\hsize}{!}{\includegraphics{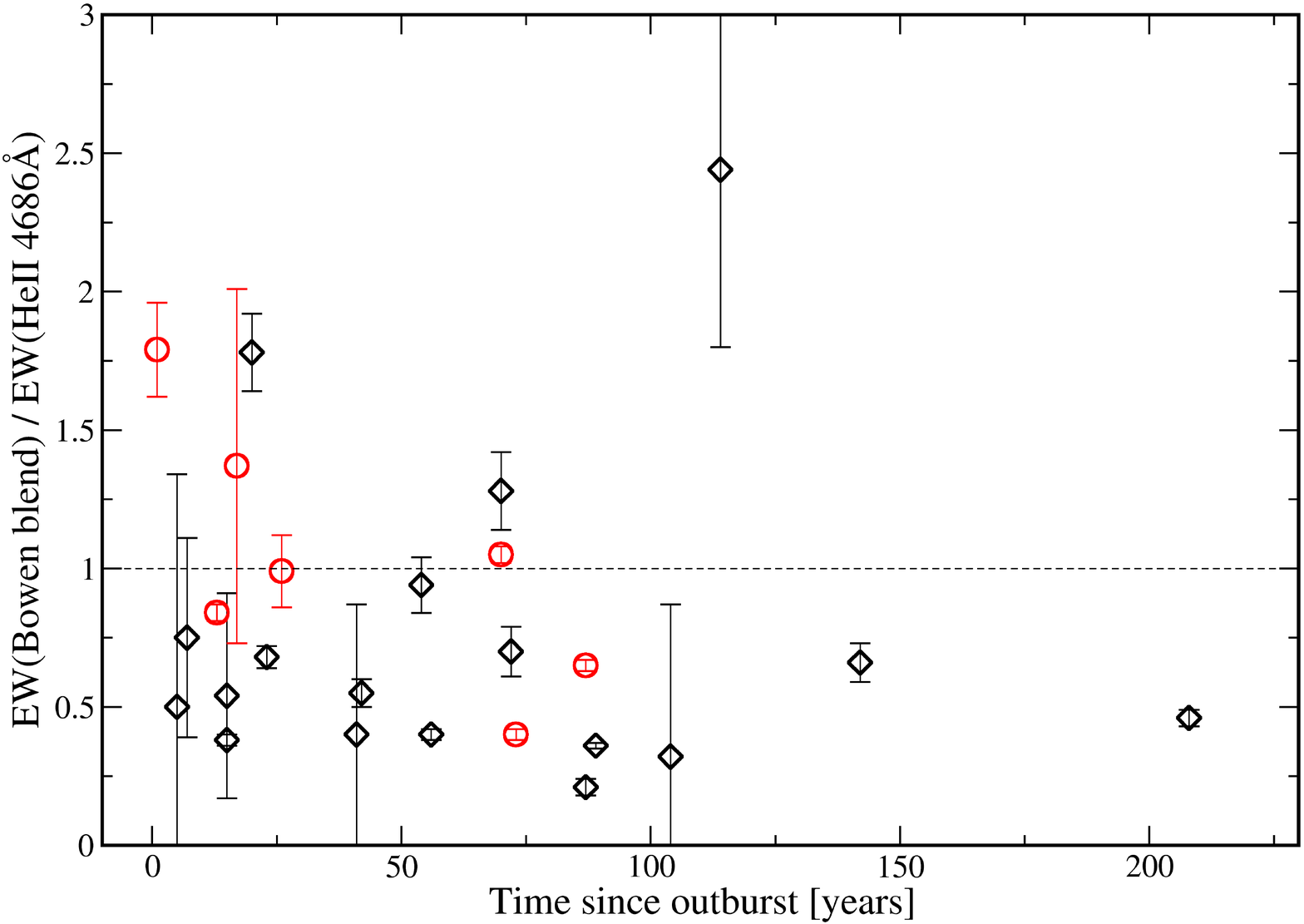}}
\caption{EW(Bowen blend)/EW(\ion{He}{ii}\,4686\AA) ratio versus time since outburst. The symbols are as in Fig.~\ref{hbeta}}.
\label{bow_heii}
\end{figure}

Because of the unsuitable placement of H$\alpha$ on the very edge of the CCD chip, the measured by us equivalent widths should only be considered as a rough estimate (see Sect.~\ref{obs}). Nevertheless, the change in the H$\alpha$ equivalent width with the time since the outburst is consistent with those described by \citet{1996MNRAS.281..192R}. The dependence of the H$\alpha$ equivalent width on the inclination of the orbit \citep{1987MNRAS.227...23W} shows that this EW should be about 70\,\AA\ even for systems with the greatest orbital inclination. Four of our objects (\object{V382\,Cen}, \object{V868\,Cen}, \object{V842\,Cen}, \object{BY\,Cir}) show larger EWs, but for all of them there are indications that the nebular spectrum has not disappeared completely and that H$\alpha$ is blended with the lines of [\ion{N}{ii}]. \object{T\,Pyx} demonstrates a very strong H$\alpha$, too, but our spectrum has only been obtained about one year after its last outburst in 2011.

In Figs.~\ref{hbeta} and \ref{heii} the changes in the H$\beta$ and \ion{He}{ii} 4686\,\AA\ EWs with the time since the outburst are compared to the published by \citet{1996MNRAS.281..192R}. The accordance in the EWs in the first 25 to 30 years is good, while our values are slightly higher in the next years. Two stars - \object{T\,Pyx} and \object{OY\,Ara} - show remarkable deviations. As mentioned above, \object{T\,Pyx} was observed shortly after the outburst. The \object{OY\,Ara} spectrum, on the other hand, was most probably obtained during eclipse, when the hot continuum is weaker (see Sec.~\ref{oyara}).

As \citet{1996MNRAS.281..192R} noted, the ratio of the EWs of the Bowen blend and \ion{He}{ii} 4686\,\AA\ do not correlate with the time since outburst (Fig.~\ref{bow_heii}), which can be considered as observational evidence that the irradiation does not change remarkably over a relatively long time interval.

\section{Nova shells}
\label{shells}

We searched for nova shells in both the H$\alpha$+[\ion{N}{II}] narrow-band images and the RSS long-slit spectra by the use of the appropriate \textsc{iraf} packages and tasks. In the narrow-band images of each of our objects, a PSF was generated using close, not blended field stars with brightness similar to the nova. After subtracting the PSF from the stars in the narrow-band image, we examined the resultant images looking for evidence of an extended nova shell. As a second approach, we examined the radial plot fit of the nova. Using three or four field stars for each nova, which were chosen like the stars used to generate the PSF, we determined their centers and next radial profiles were fit to the field stars and to the nova. We then compared the radial profiles of the nova and the field stars searching for extended shells. Obviously, the use of radial plots suggests that we assumed that the extended nova shells are spherically symmetric. 

For the long-slit spectra, we used a method similar to the one described by \citet{1983MNRAS.205P..33M}. To remove the contamination of the stellar continuum and the H$\alpha$ emission line, we determined the average spectral profile perpendicular to the dispersion in a featureless region of the continuum far enough away from the emission line.  The average spectral profile was normalized to the maximum of the observed spectrum in each column perpendicular to the dispersion, defined by the fitting of a parabola to the seven maximal points. The normalized profile was then extracted pixel by pixel from the observed long-slit spectrum. In the resulting images we were searching for evidence of extended nova shells. 

 Applying the described methods for all our targets, we were only able to detect expanding shells in both the long-slit spectra and the narrow-band direct images for two objects - \object{V842~Cen} and \object{V382~Vel}. 

\subsection{\object{V842~Cen}}
\label{v842cen_neb}

The extended shell of \object{V842~Cen} was detected for the first time by \citet{1998MNRAS.300..221G} in 1995 about eight years after its outburst. They reported a shell diameter $\sim 1\farcs6$. Three years later the observations of \citet{2000AJ....120.2007D} revealed a shell of size $5\farcs6 \times 6\farcs0$. 

\begin{figure}
\centering
\resizebox{\hsize}{!}{\includegraphics{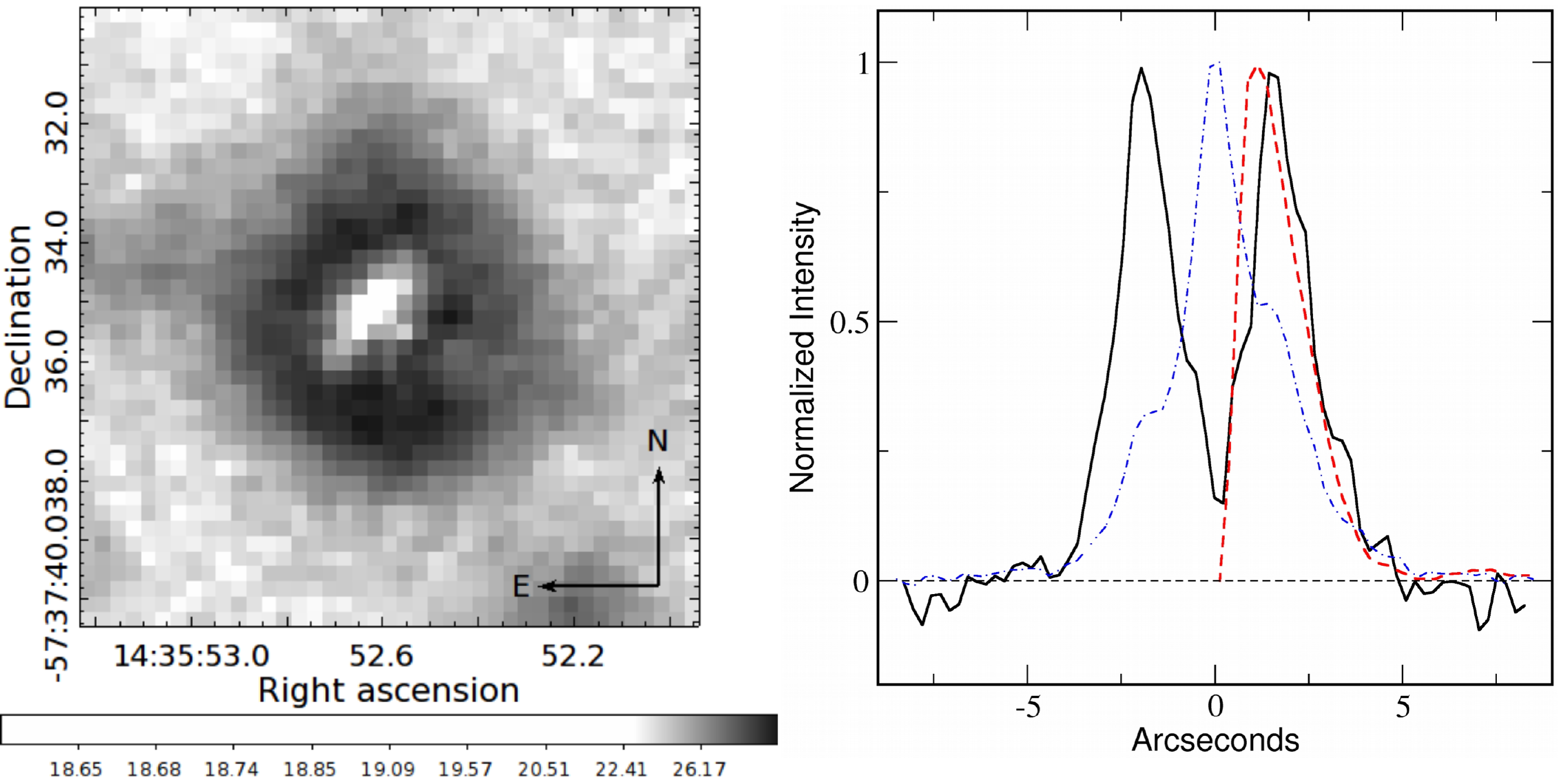}}
\caption{\textsl{Left}: H$\alpha$+[\ion{N}{II}] narrow-band image of the \object{V842~Cen} extended shell on a logarithmic gray scale. \textsl{Right}: comparison of the radial profile of \object{V842~Cen} in the H$\alpha$+[\ion{N}{II}] narrow-band image after subtracting the average standard star profile (\textsl{dashed line}) with cross sections averaged perpendicular to the dispersion ($\pm$5\AA\ around the H$\alpha$ emission line center) in the long-slit spectrum before (\textsl{dot-dashed line}, Fig.~\ref{sp_v842} left) and after (\textsl{continuous line}, Fig.~\ref{sp_v842} right) removing the stellar continuum and the H$\alpha$ emission line contamination. The profile and the cross sections were normalized to their maxima.}
\label{di_ort_v842}
\end{figure}

\begin{figure}
\centering
\resizebox{\hsize}{!}{\includegraphics{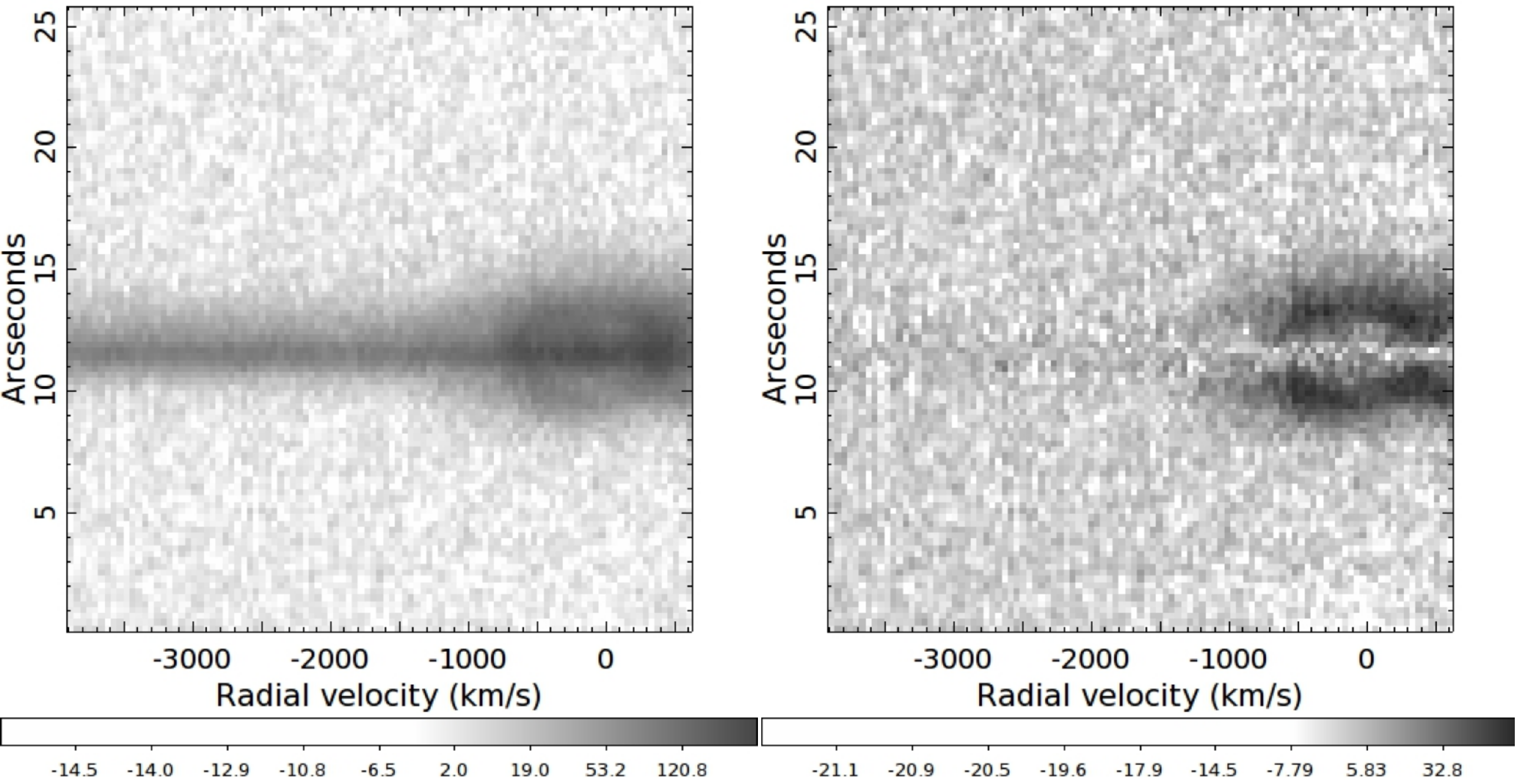}}
\caption{\textsl{Left}: a small part of the observed long-slit spectrum of \object{V842~Cen} in the region of H$\alpha$. \textsl{Right}: the same spectrum, after the contamination of the stellar continuum and the H$\alpha$ emission line has been removed. The slit position was NS with N at the top of the images. A logarithmic gray scale is used for both images.}
\label{sp_v842}
\end{figure}

The H$\alpha$+[\ion{N}{II}] narrow-band \object{V842~Cen} image, presented in Fig.~\ref{di_ort_v842}, shows an approximately spherically symmetric shell. The shell is visible in the long-slit spectrum before and after the correction for the contamination of the stellar continuum and the H$\alpha$ emission line as shown in Fig.~\ref{sp_v842}. In the image with the uncorrected  spectrum (Fig.~\ref{sp_v842}), as well as in its cross section shown in Fig.~\ref{di_ort_v842}, two maxima on both sides of the stellar spectrum are apparent. These maxima coincide with the peaks in the cross section of the corrected spectrum (Fig.~\ref{di_ort_v842}). The separation of these peaks is $3\farcs6 \pm 0\farcs3$. From the cross sections and the radial profile (Fig.~\ref{di_ort_v842}) for the outer diameter of the shell we obtain $10\farcs6 \pm 0\farcs3$.

Two expansion velocities were suggested for \object{V842~Cen} \citep[see the discussion in][]{2000AJ....120.2007D}. A velocity of about $525$\,km\,s$^{-1}$ for the dense regions of the shell and $\sim2000$\,km\,s$^{-1}$ for the low-density material. Assuming that the $3\farcs6$ shell ring is caused by the slowly expanding dense material and that the outer size $10\farcs6$ of the shell reflects the expansion of the low-density material, their velocities measured in our spectrum are $500 \pm 10$\,km\,s$^{-1}$ and $1500 \pm 90$\,km\,s$^{-1}$ respectively. These velocities give an average distance to \object{V842~Cen} of about $1.5 \pm 0.2$\,kpc. This value exceeds all previous estimations of the distance, independently of the method used. The reddening-based estimations of 0.9\,kpc \citep{1989MNRAS.241..311S} and 1.0\,kpc \citep{1994A&A...291..869A} and the derived by the expansion parallax 1.3\,kpc \citep{1998MNRAS.300..221G} and 0.42\,kpc \citep{2000AJ....120.2007D}. It should be noted, however, that the 
lower limit of our value for the \object{V842~Cen} distance agrees with the estimation of \citep{1998MNRAS.300..221G}. Our observations of the \object{V842~Cen} shell were obtained long after the previous ones, which permitted us to use them in our distance estimation, with the different shell components reflecting the expansion of the different density material for the first time. Moreover, the velocities measured in our spectrum seem to indicate a definite decrease in the expanding material velocities.

\subsection{\object{V382~Vel}}
\label{v382_vel_neb}

\begin{figure}
\centering
\resizebox{\hsize}{!}{\includegraphics{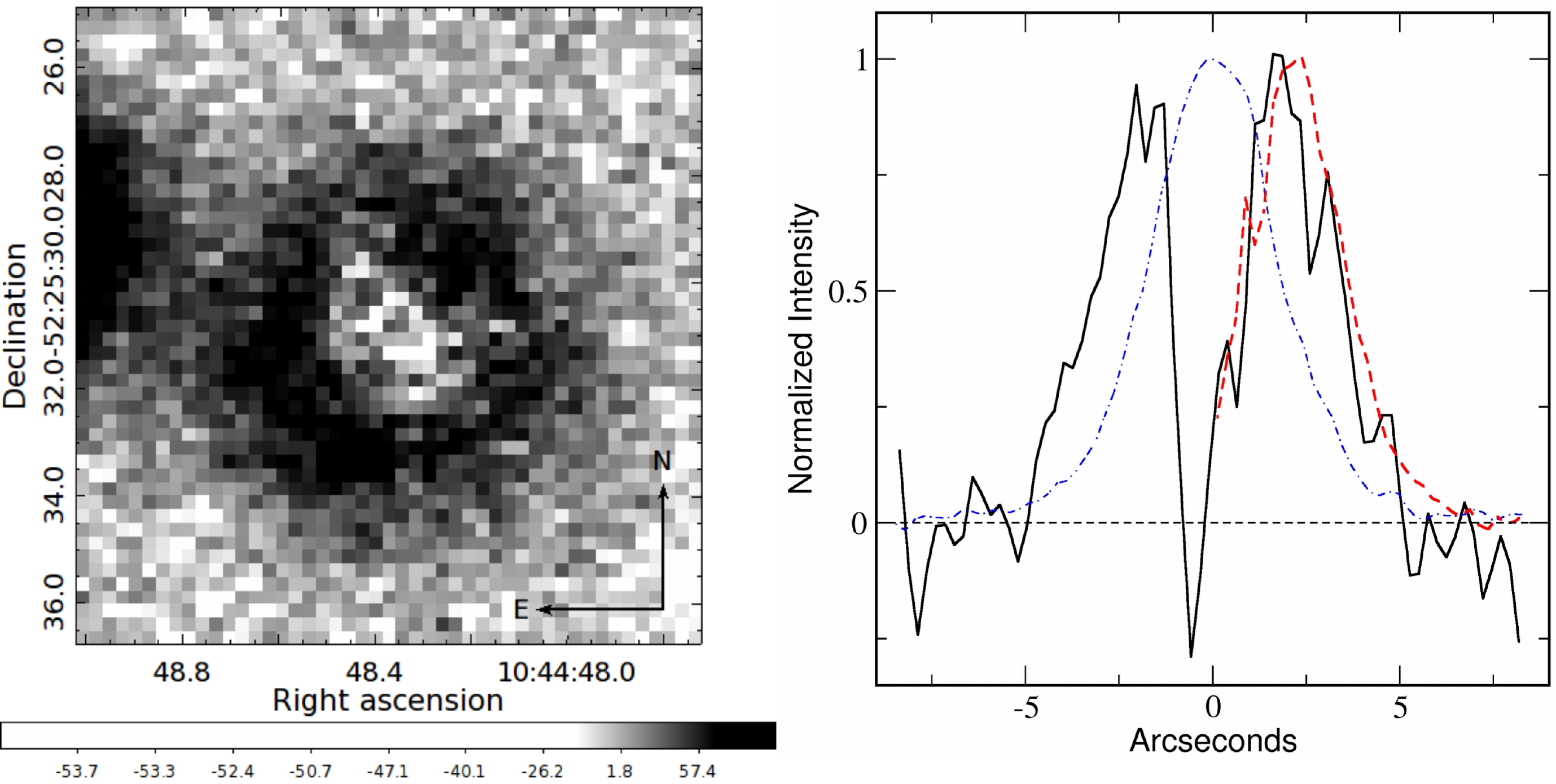}}
\caption{As in Fig.~\ref{di_ort_v842}, but for \object{V382~Vel}.}
\label{di_ort_v382}
\end{figure}

\begin{figure}
\centering
\resizebox{\hsize}{!}{\includegraphics{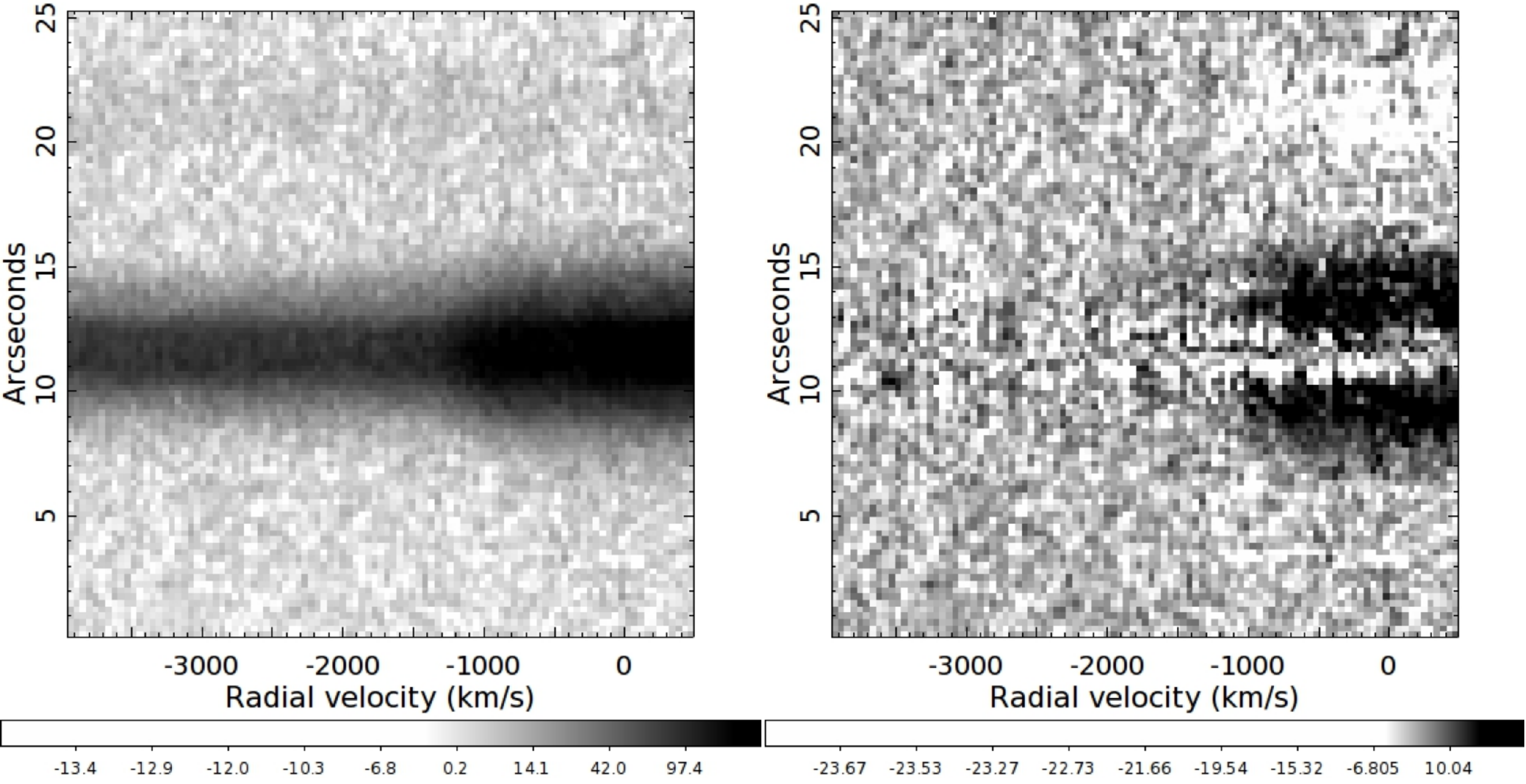}}
\caption{As in Fig.~\ref{sp_v842} but for \object{V382~Vel}.}
\label{sp_v382}
\end{figure}

Our observations are the first detection of the \object{V382~Vel} expanding shell simultaneously in H$\alpha$+[\ion{N}{II}] narrow-band imaging and long-slit spectrum (Figs.~\ref{di_ort_v382} and  \ref{sp_v382}). As seen in Fig.~\ref{di_ort_v382}, the \object{V382~Vel} expanding shell shows better spherical symmetry in comparison to the  \object{V842~Cen} (Fig.~\ref{di_ort_v842}). Based on the radial plot and the spectrum cross-sections shown in Fig.~\ref{di_ort_v382}, we obtain $12\farcs0 \pm 0\farcs5$ for the outer diameter of the shell. 

There are different expansion velocities published for the \object{V382~Vel} shell. \citet{2002A&A...390..155D} give a velocity $\ga 3500$\,km\,s$^{-1}$. \citet{2002AIPC..637..233A} find an average velocity of  1600\,km\,s$^{-1}$, decreasing with time. A maximum expansion velocity of more than 5000\,km\,s$^{-1}$ was estimated by \citet{2003AJ....125.1507S}. In our spectrum we measured a velocity of the expanding shell of about $1800 \pm 100$\,km\,s$^{- 1}$, which is different from all those published so far. This leads to an estimate of the distance to \object{V382~Vel}  of $0.8\pm 0.1$\,kpc. A distance to \object{V382~Vel} of about 1.7\,kpc was obtained by \citet{2002A&A...390..155D} using the maximum magnitude vs. rate-of-decline relationship method. Analyzing and modeling ultraviolet spectra, \citet{2003AJ....125.1507S} estimated that the distance to \object{V382~Vel} is about 2.5\,kpc.  A comparison of ours and previously published distances to \object{V382~Vel} shows large differences between 
the particular estimates. Obviously, a more accurate estimation of the distance to this post-nova requires additional observations.

\section{Conclusions}

We presented recent long-slit and narrow-band imaging observations of 22 southern post-novae. We compared our new spectra of the post-novae remnants with the previously published ones and discussed the observed changes. The changes with the time since outburst of the equivalent widths of the H$\beta$ and \ion{He}{ii} 4686\,\AA\ emissions, as well as the ratio Bowen blend/ \ion{He}{ii} 4686\,\AA, are consistent with those in the \citet{1996MNRAS.281..192R} survey. 

Combining our data and data published in the literature, we estimated an average value $\alpha=2.37\pm0.74$ for the exponent of the power law fitted to the post-novae  continua. Our observations confirm the suggestion of \citet{2000AJ....120.2007D} that \object{Nova Car 1972} is not a real nova and must be excluded from the old novae list. 

We observed the expanding nebula around \object{V842~Cen} again and showed that it consists of inner and outer parts with diameters $3\farcs6$ and $10\farcs6$, respectively, which are related to matter with different densities and current expansion velocities of about $500$\,km\,s$^{-1}$ and $1500$\,km\,s$^{-1}$. For the distance to \object{V842~Cen}, we obtained $1.5$\,kpc.

For the first time, we detected an expanding shell around \object{V382~Vel}. The outer diameter of the shell is around $12\arcsec$, which in combination with the measured expansion velocity $1800$\,km\,s$^{-1}$, gives a distance  $0.8$\,kpc.

\onlfig{\begin{figure*}
 \centering
 \includegraphics[width=0.3\textwidth]{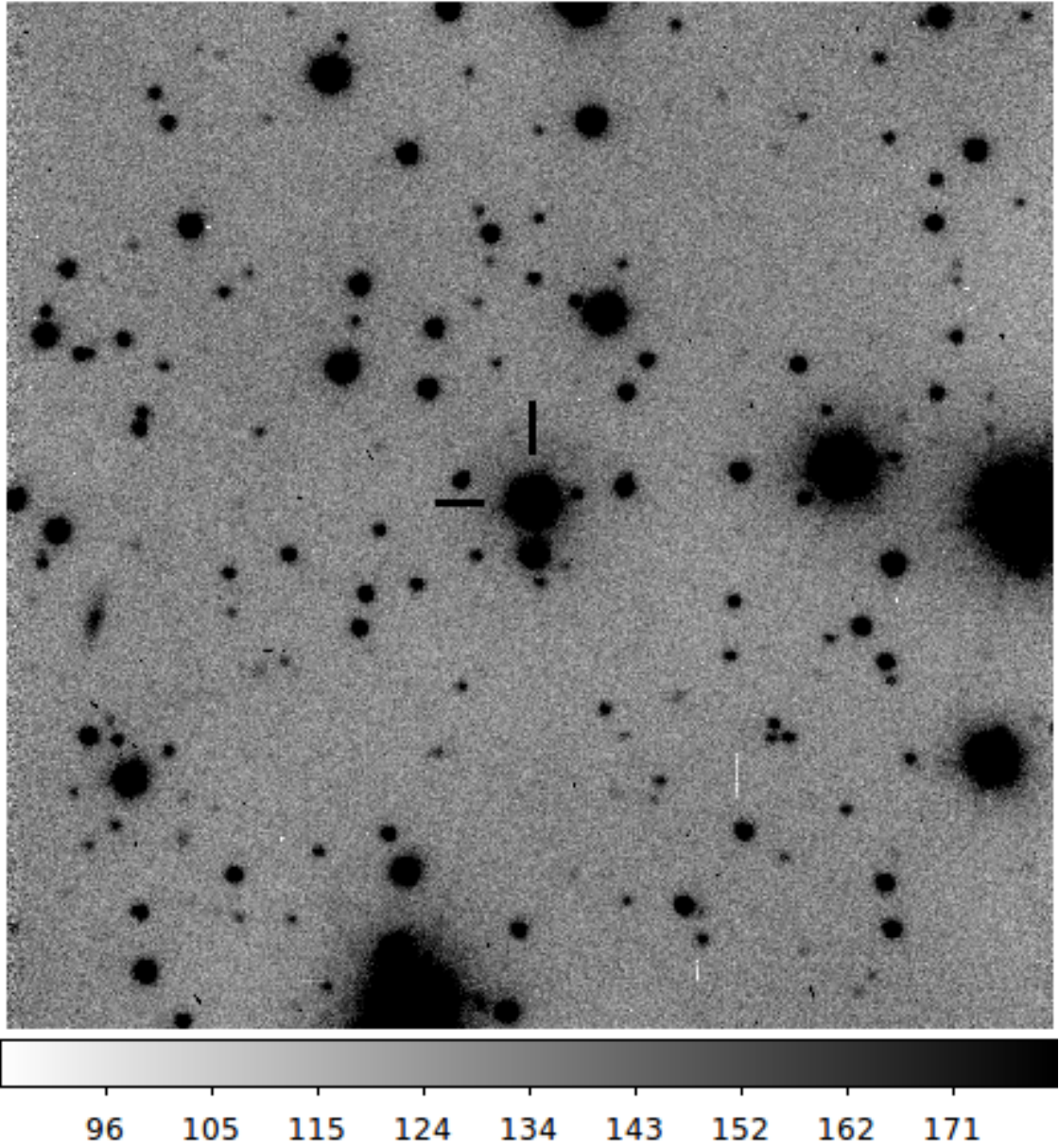}
  \hspace*{1mm}
 \includegraphics[width=0.65\textwidth]{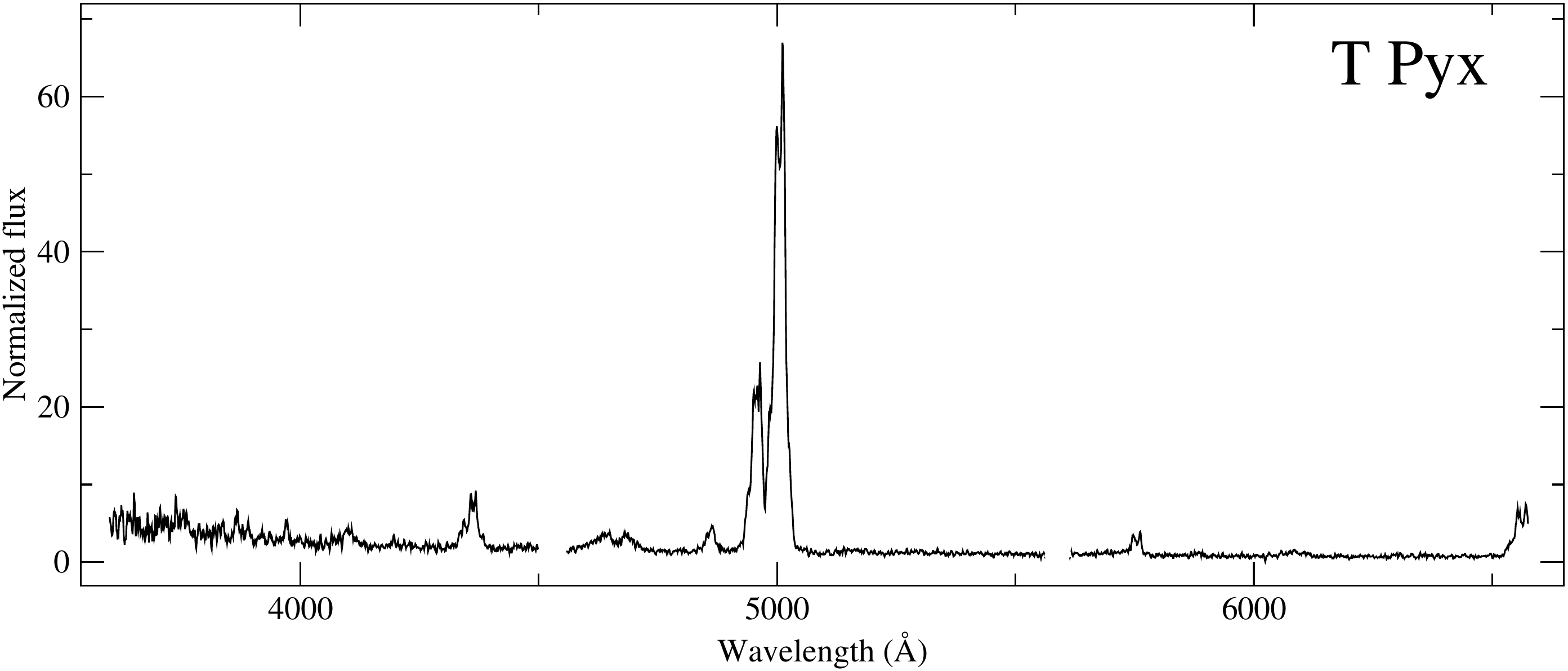} \\
\vspace{2mm}
\includegraphics[width=0.3\textwidth]{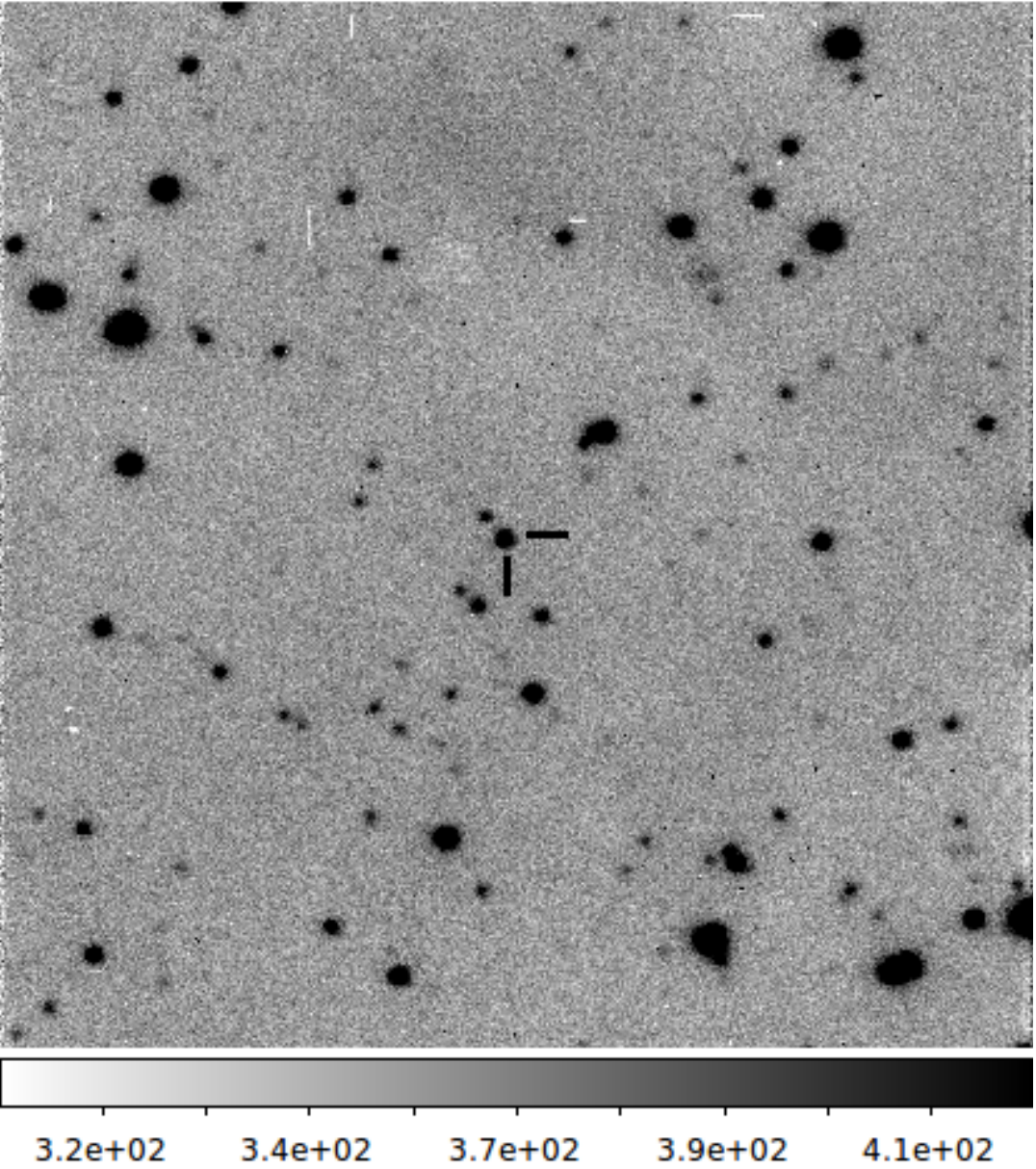}
  \hspace*{1mm}
 \includegraphics[width=0.65\textwidth]{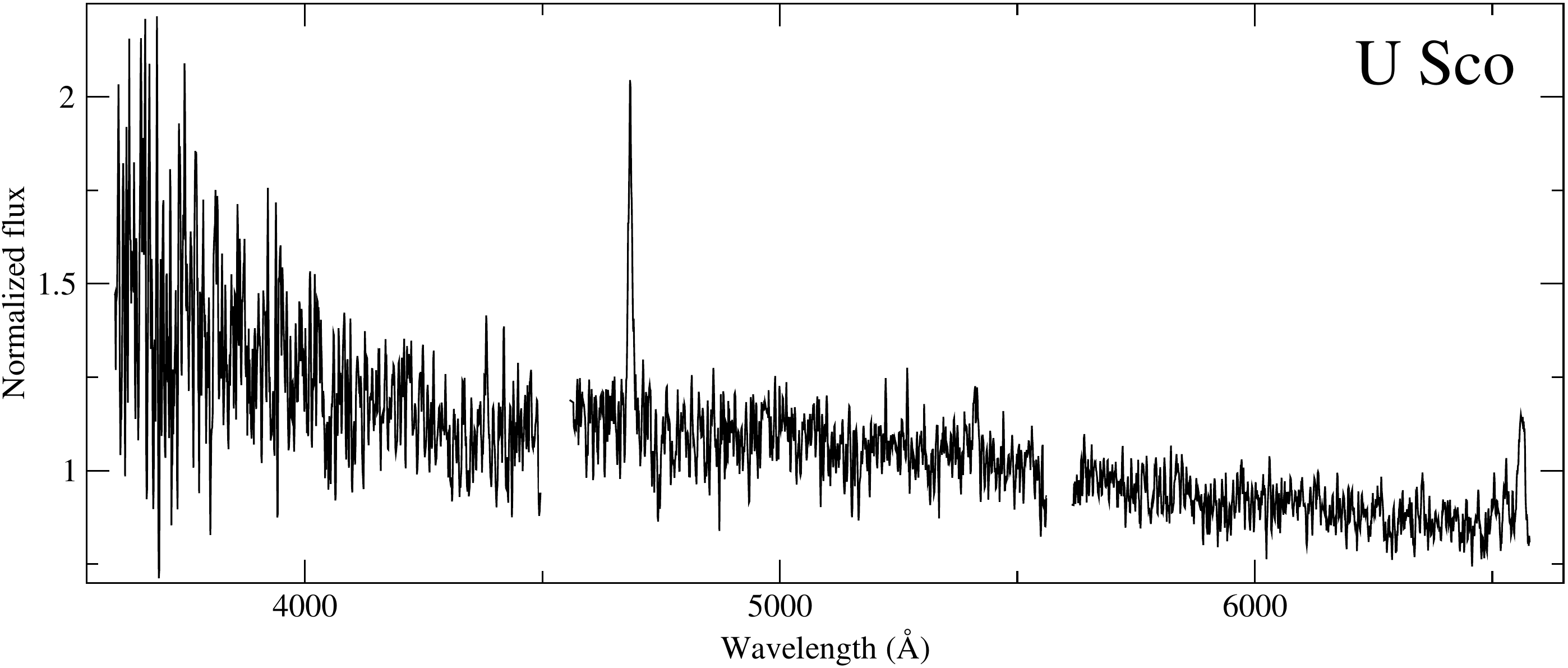} \\
\vspace{2mm}
\includegraphics[width=0.3\textwidth]{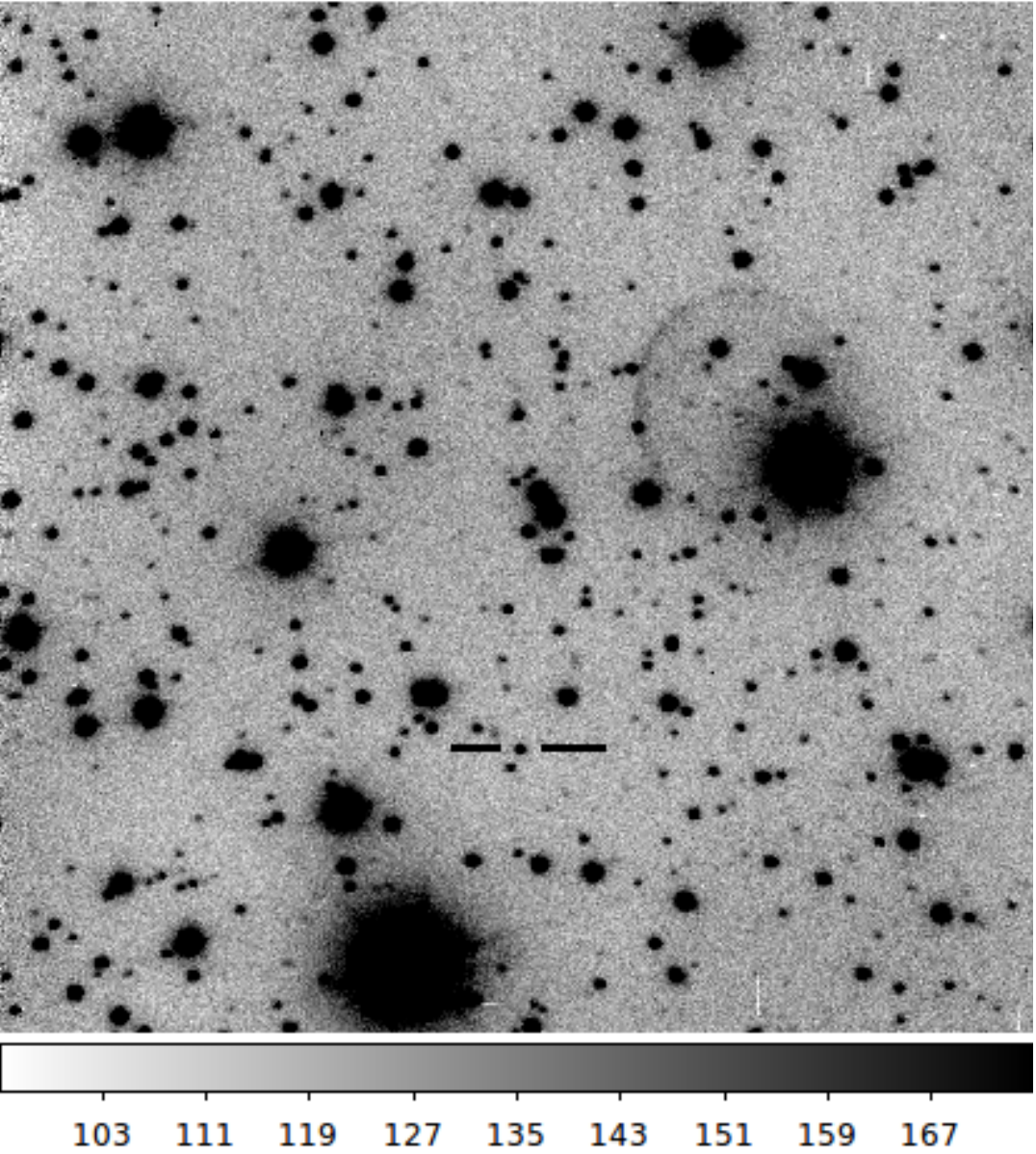}
  \hspace*{1mm}
 \includegraphics[width=0.65\textwidth]{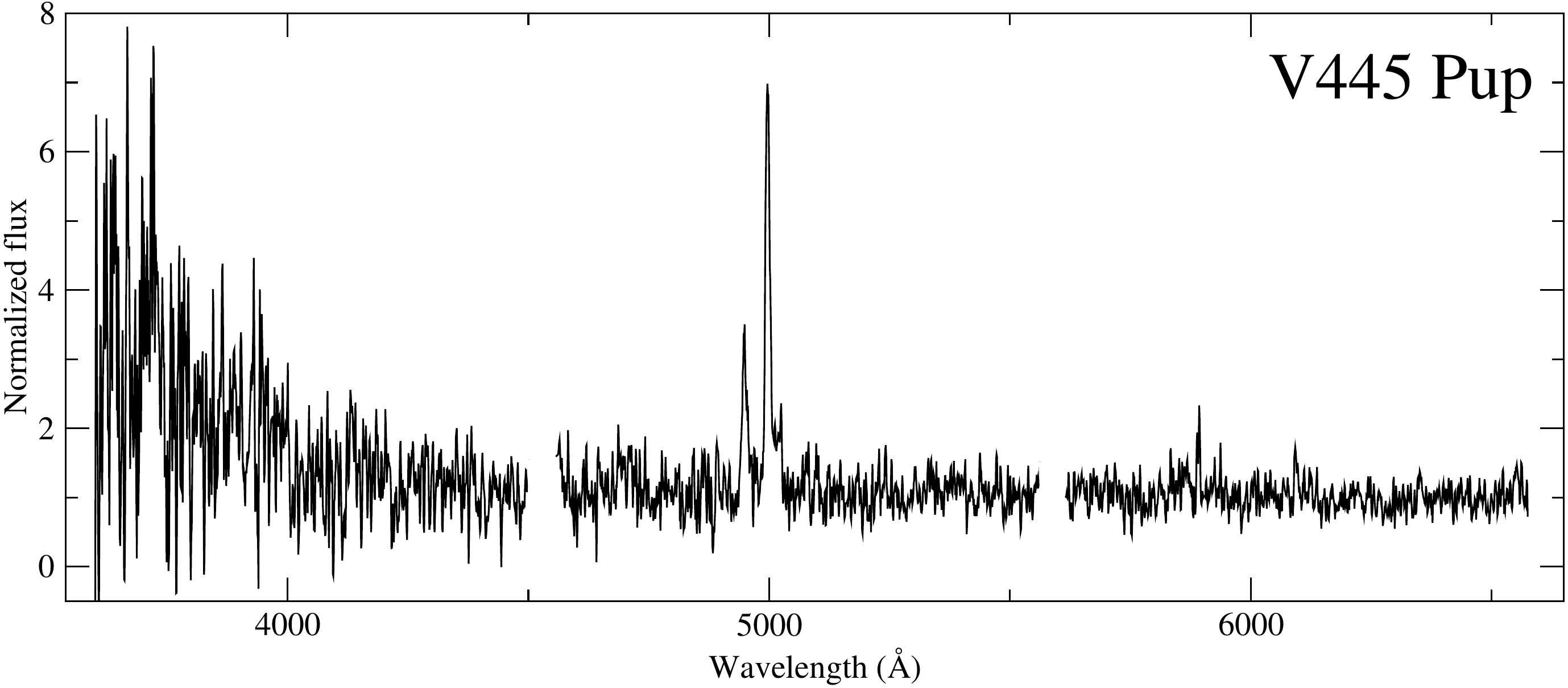} \\
\vspace{2mm}
 \includegraphics[width=0.3\textwidth]{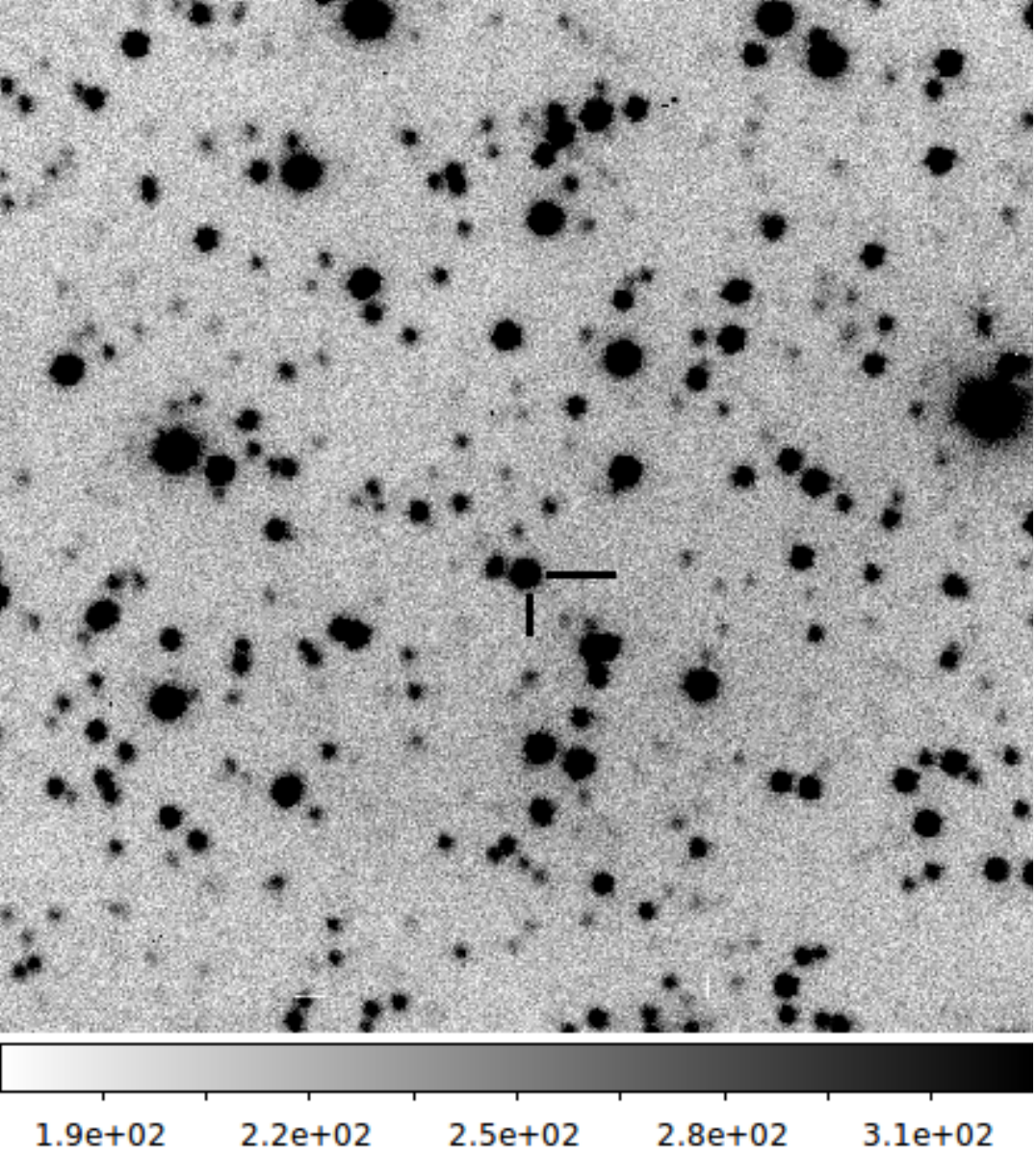}
  \hspace*{1mm}
 \includegraphics[width=0.65\textwidth]{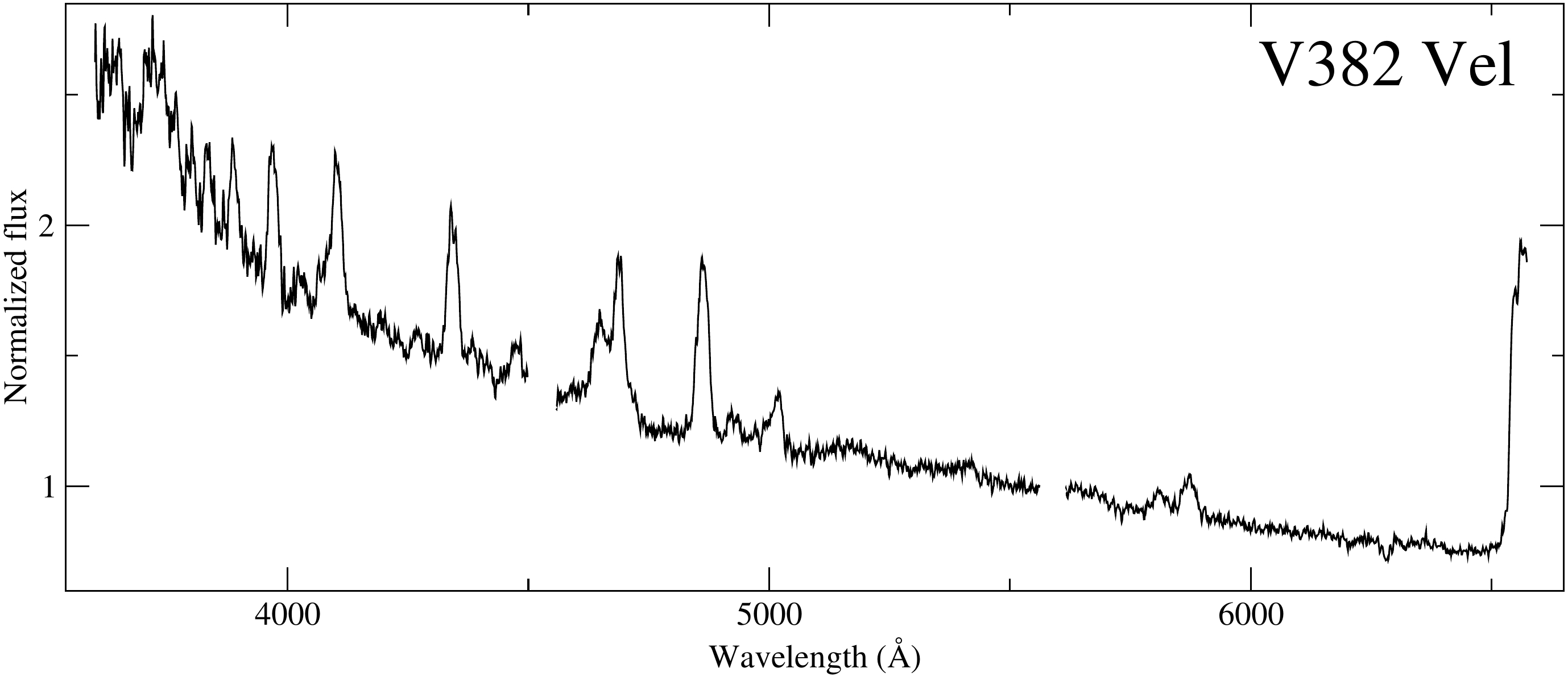} 
\caption{H$\alpha$\textbf{+[\ion{N}{II}]} narrow-band image \textbf{in linear gray scale} (left) and extracted, not dereddened spectrum (right) for each of our targets. The images are $4\arcmin \times 4\arcmin$ with north up and east to the left. The gaps in the spectra reflect the RSS interchip gaps.}
\label{map_sp}
\end{figure*}

\addtocounter{figure}{-1}
\begin{figure*}
 \centering
 \includegraphics[width=0.3\textwidth]{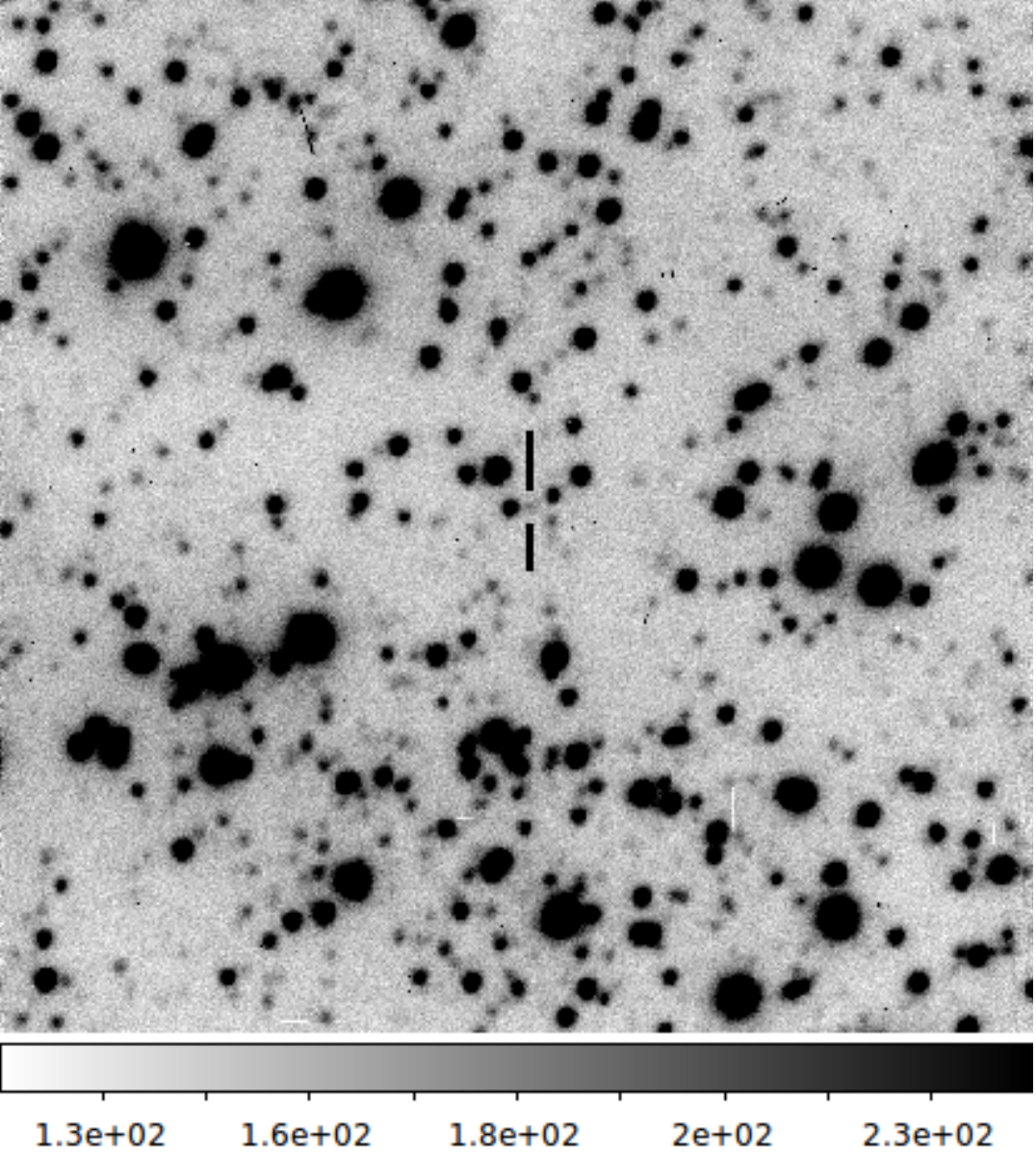}
  \hspace*{1mm}
 \includegraphics[width=0.65\textwidth]{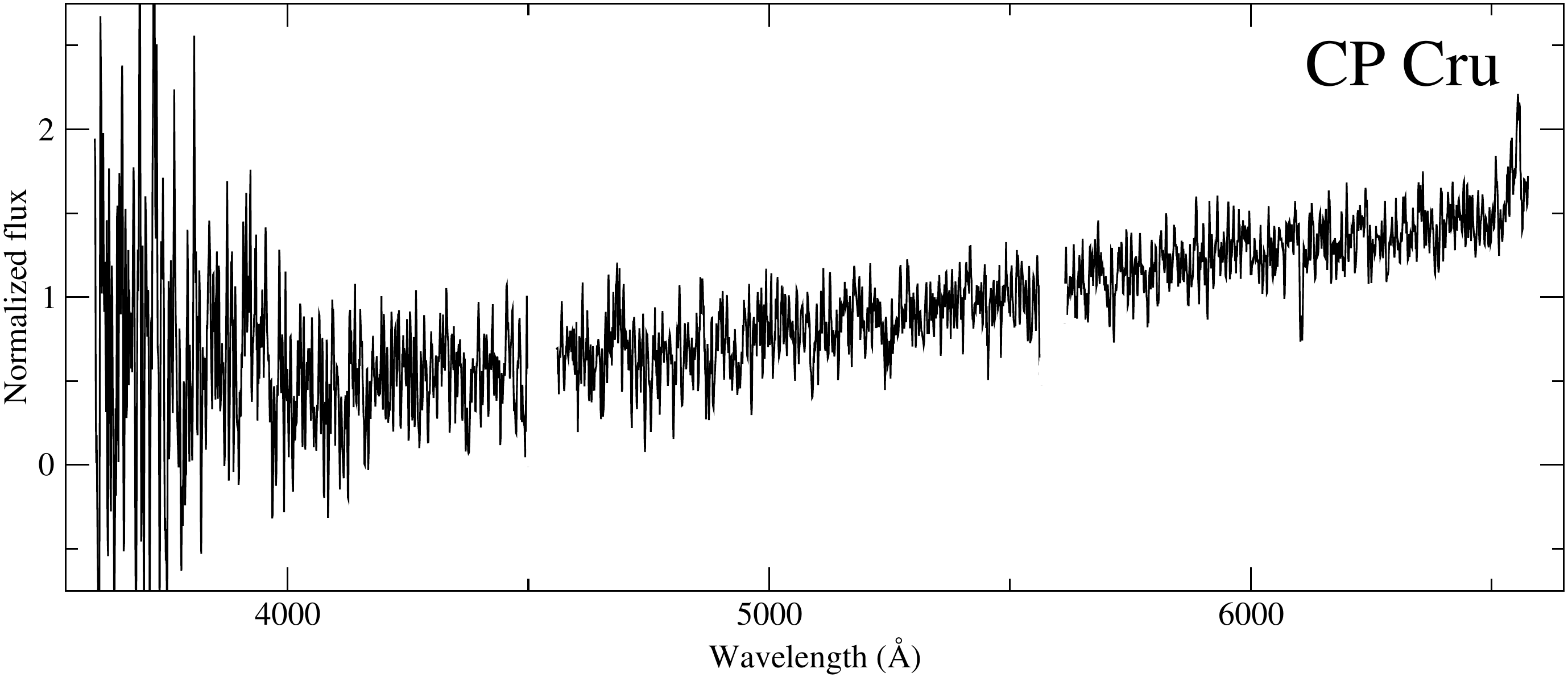} \\
\vspace{2mm}
\includegraphics[width=0.3\textwidth]{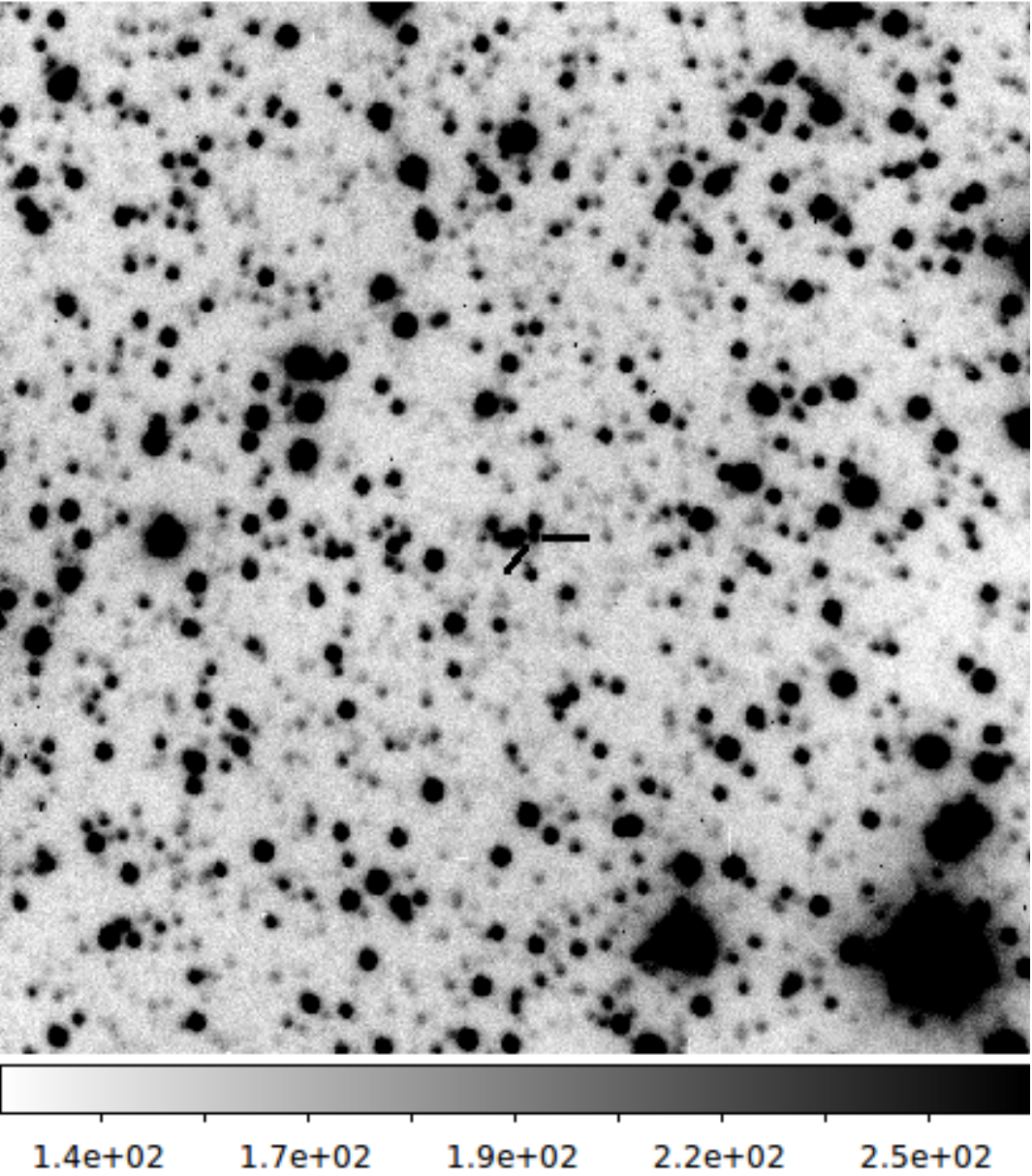}
  \hspace*{1mm}
\includegraphics[width=0.65\textwidth]{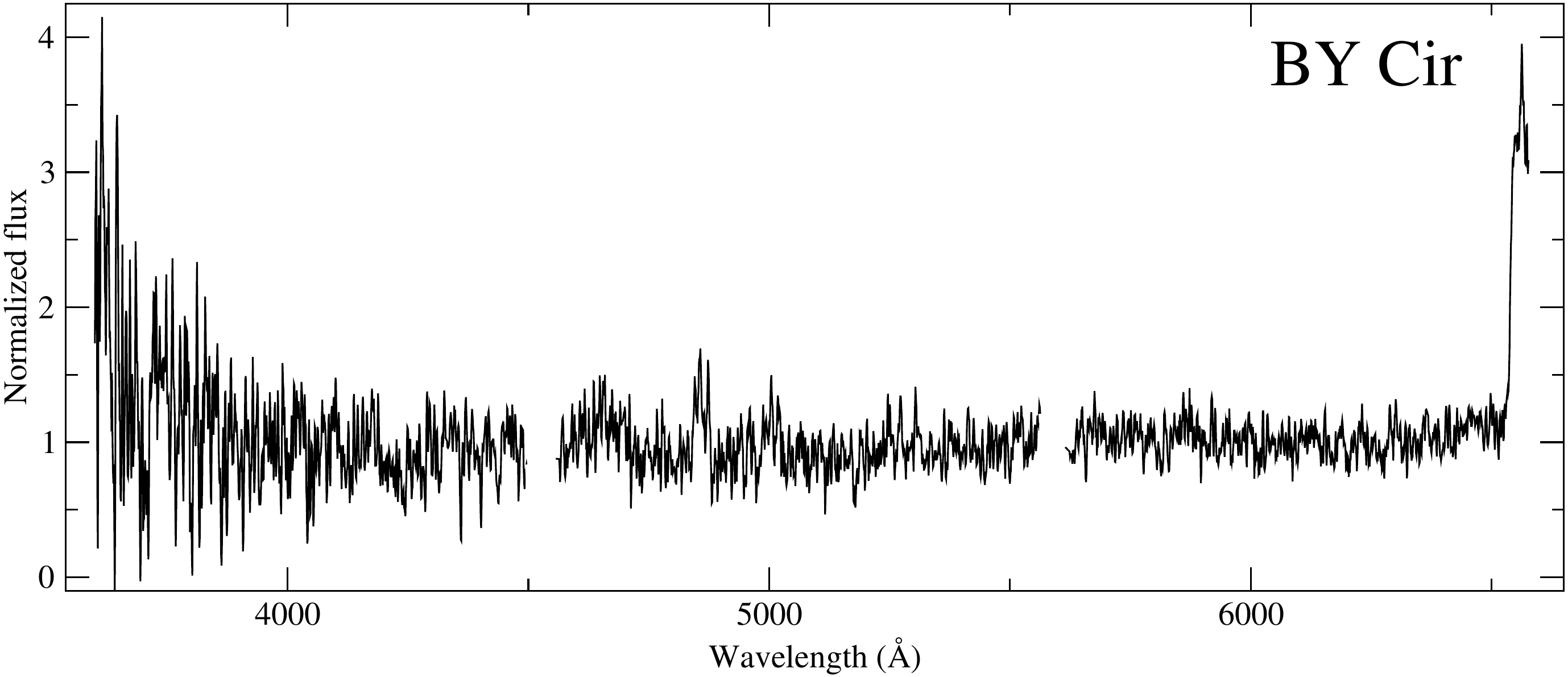} \\
\vspace{2mm}
\includegraphics[width=0.3\textwidth]{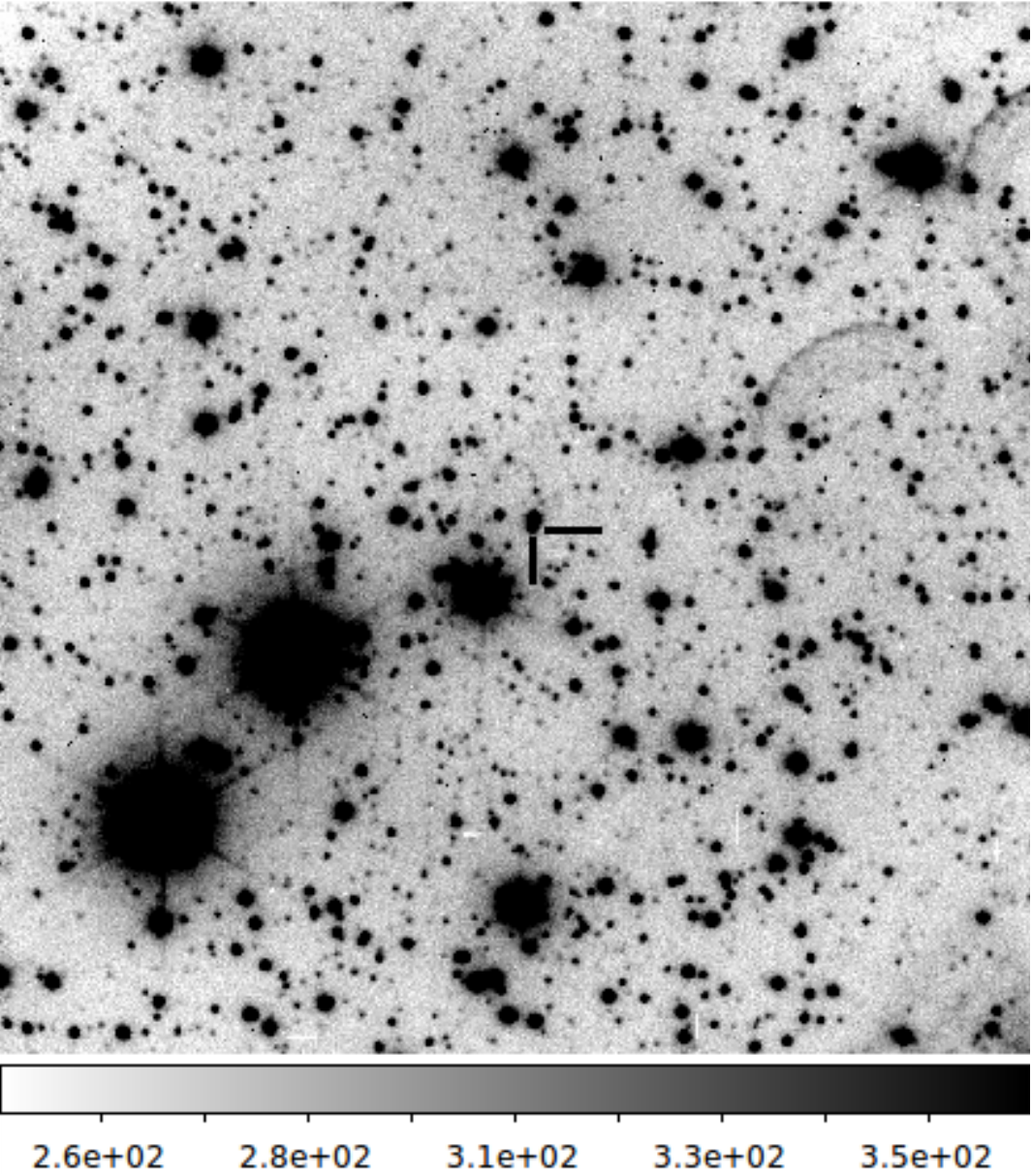}
  \hspace*{1mm}
 \includegraphics[width=0.65\textwidth]{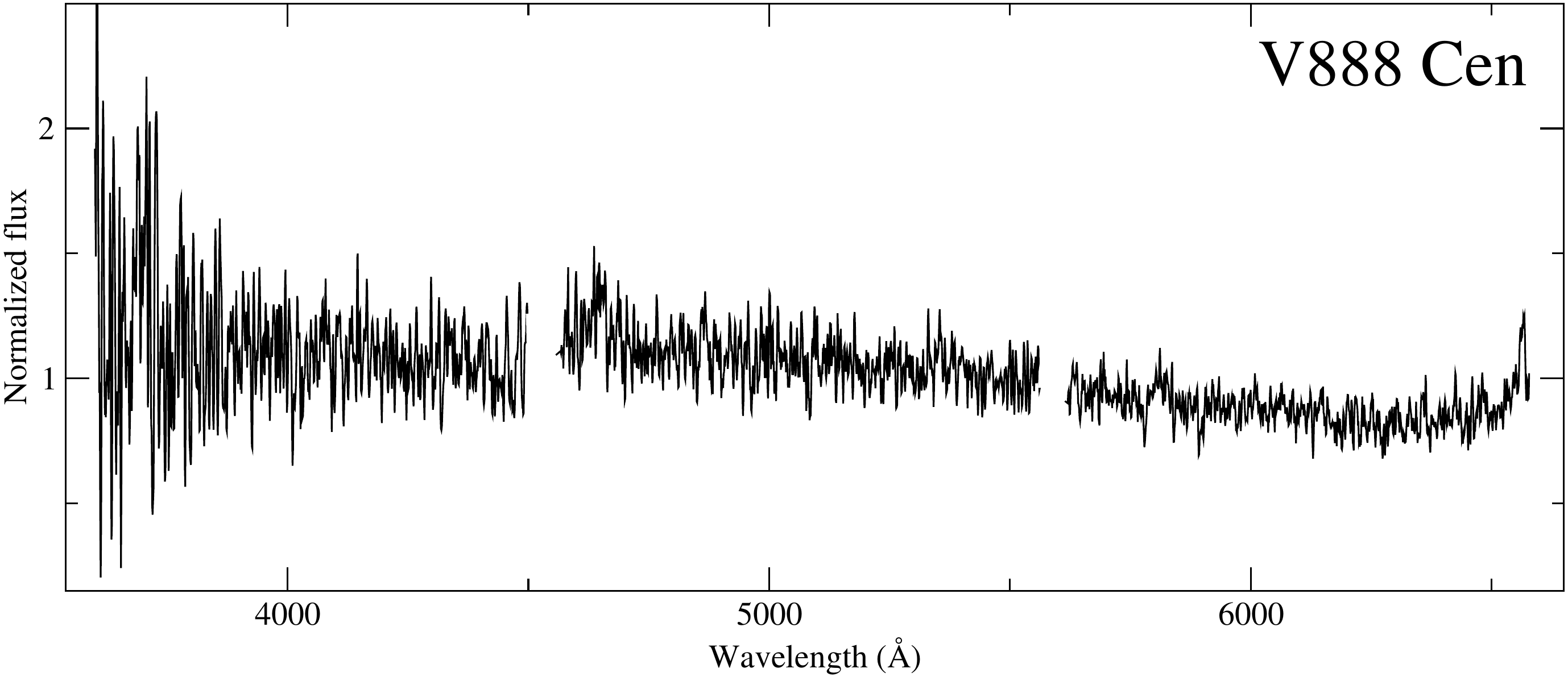} \\
\vspace{2mm}
\includegraphics[width=0.3\textwidth]{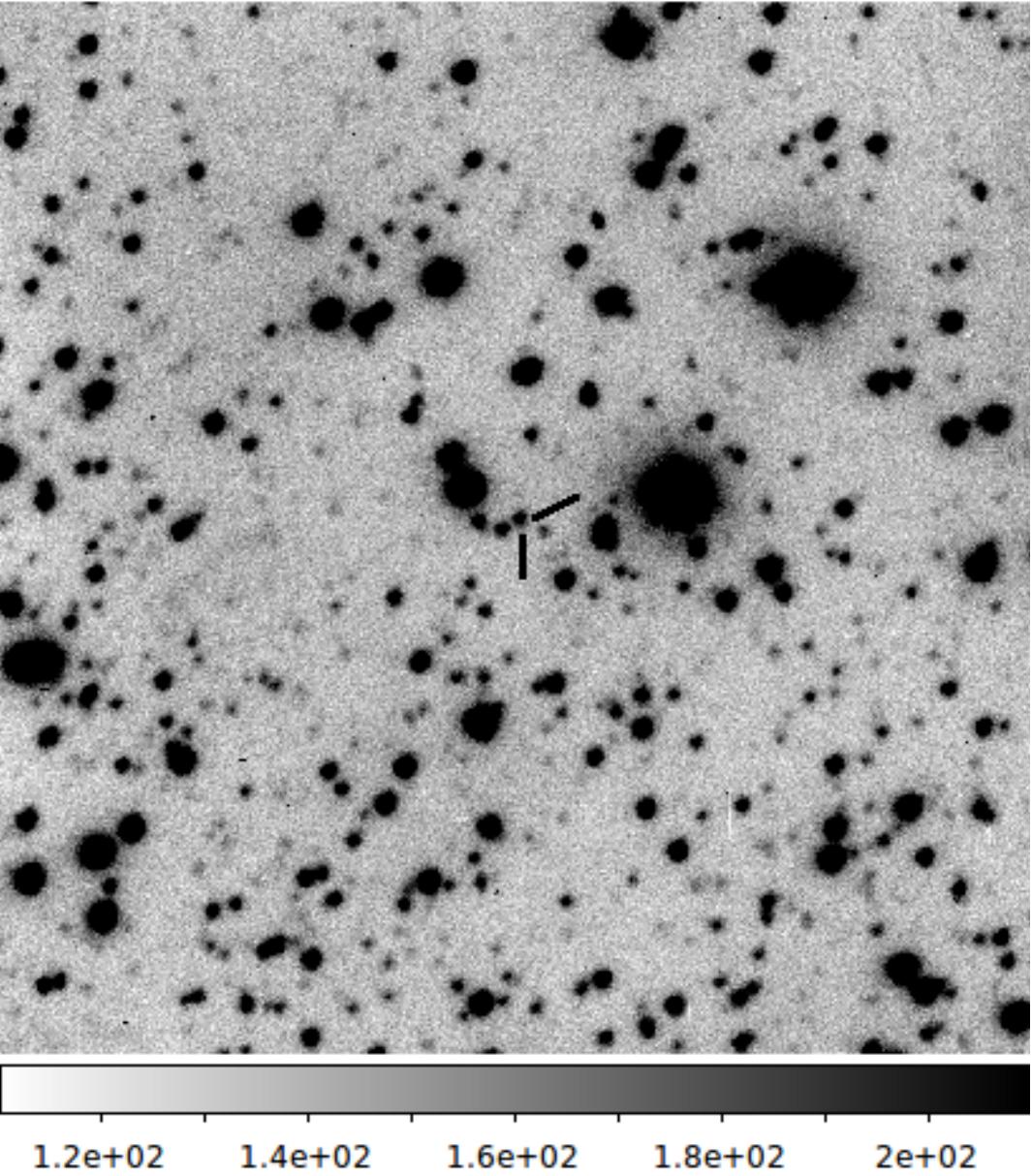}
  \hspace*{1mm}
\includegraphics[width=0.65\textwidth]{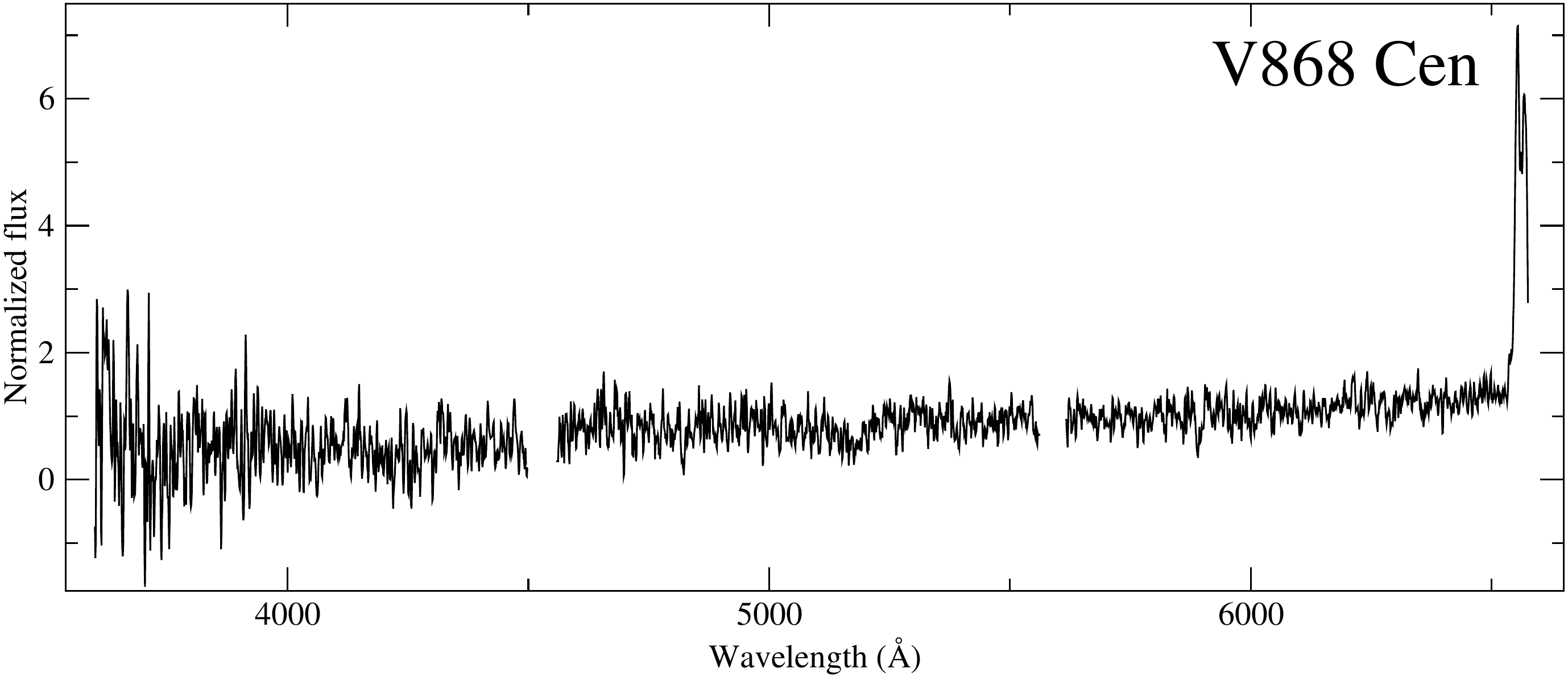} 
\caption{Continued}
\end{figure*}

\addtocounter{figure}{-1}
\begin{figure*}
 \centering
 \includegraphics[width=0.3\textwidth]{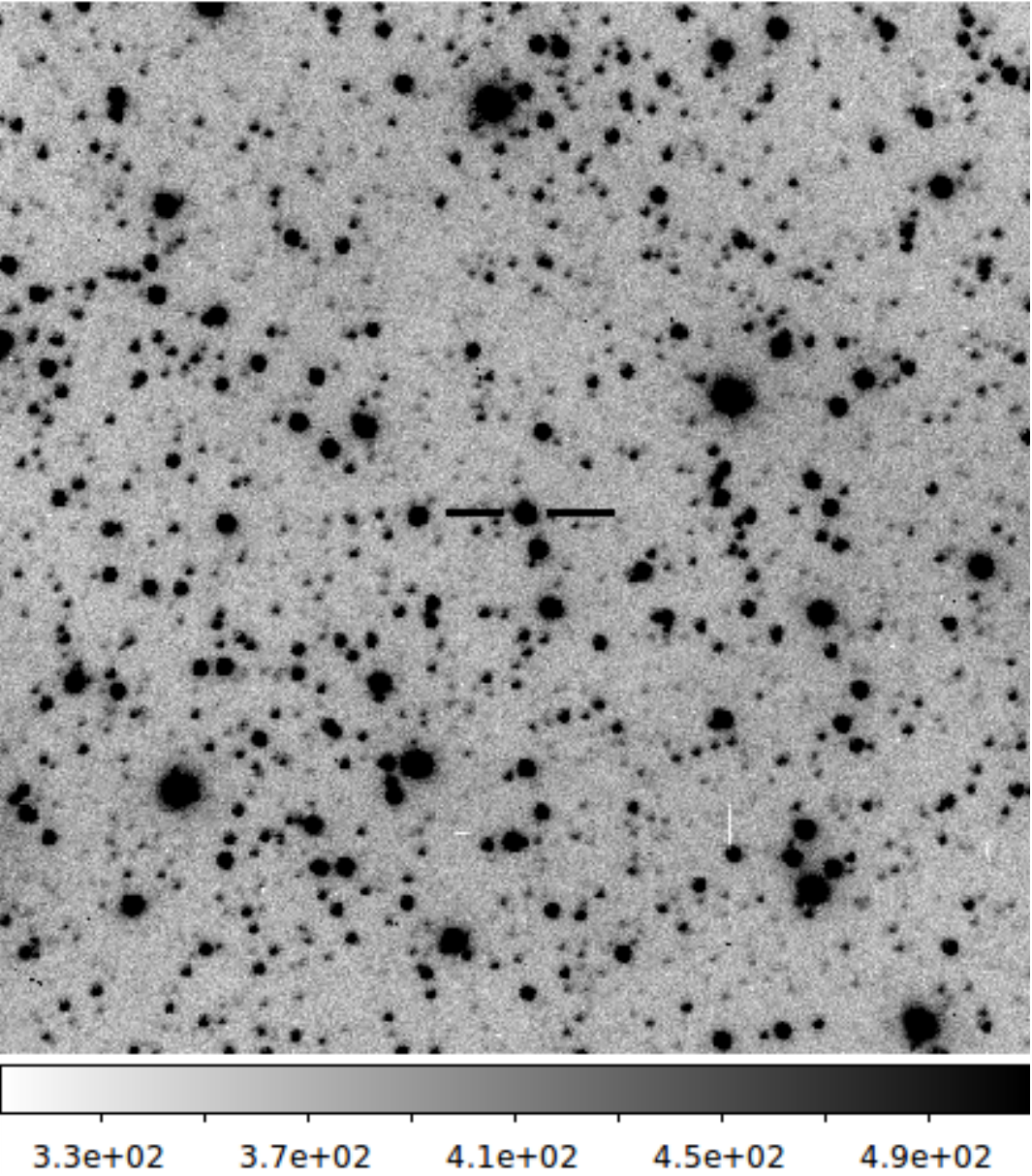}
  \hspace*{1mm}
 \includegraphics[width=0.65\textwidth]{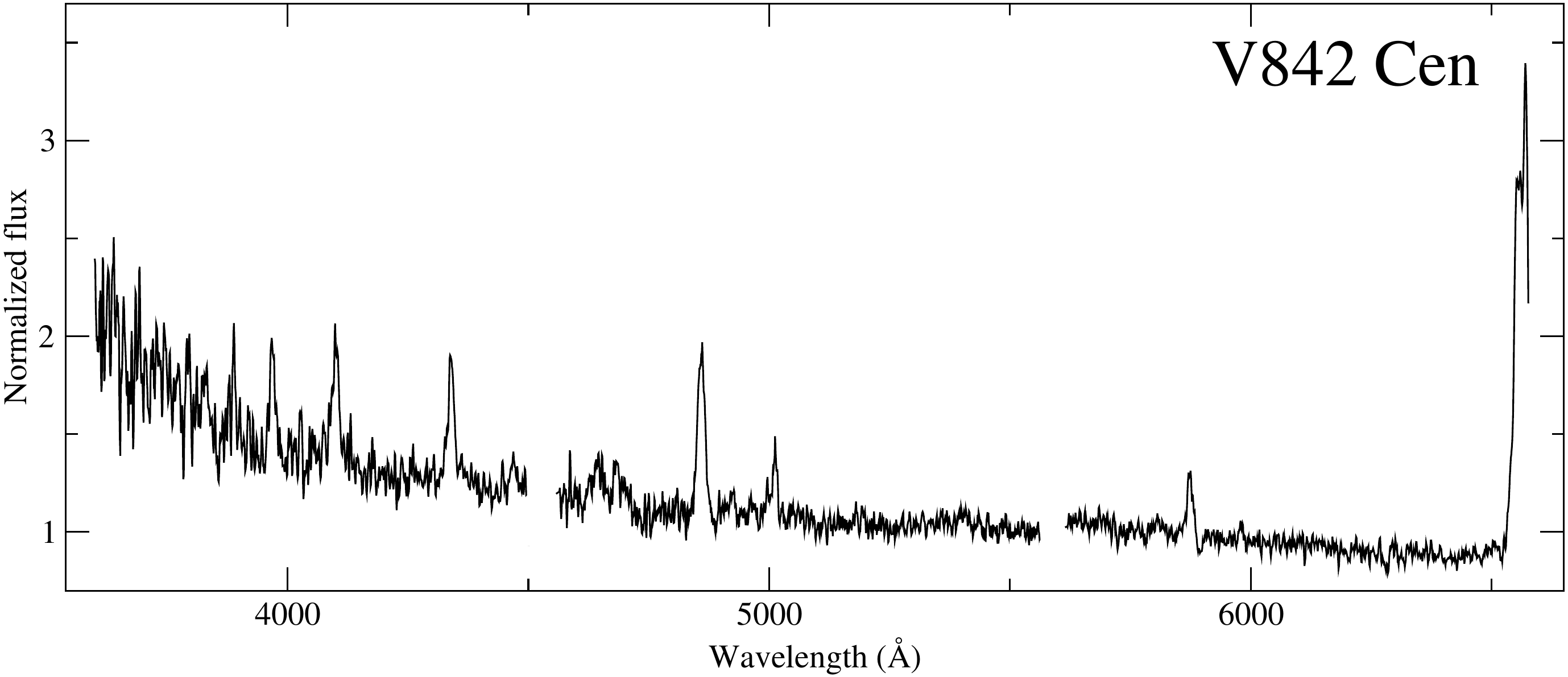} \\
\vspace{2mm}
\includegraphics[width=0.3\textwidth]{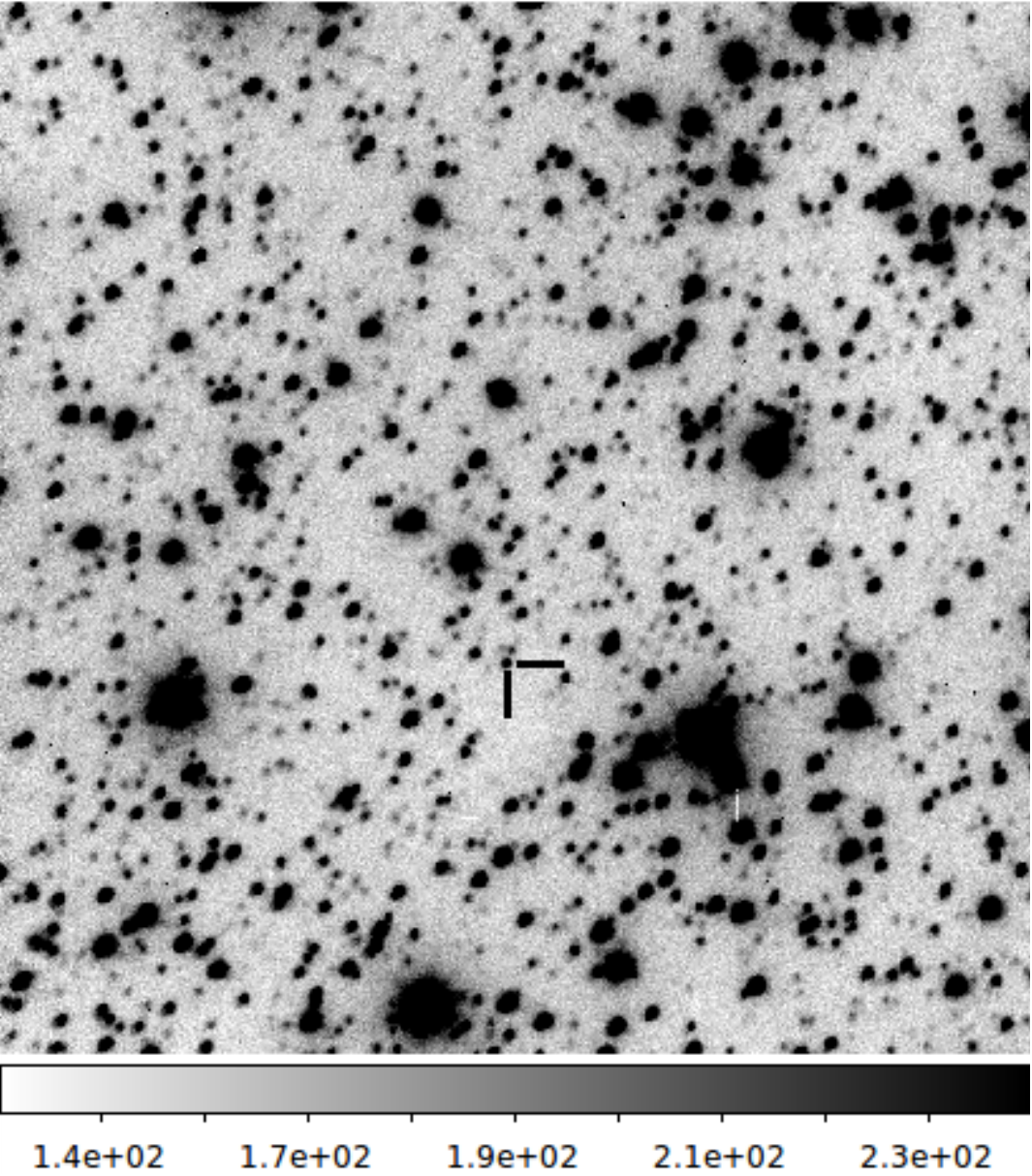}
  \hspace*{1mm}
\includegraphics[width=0.65\textwidth]{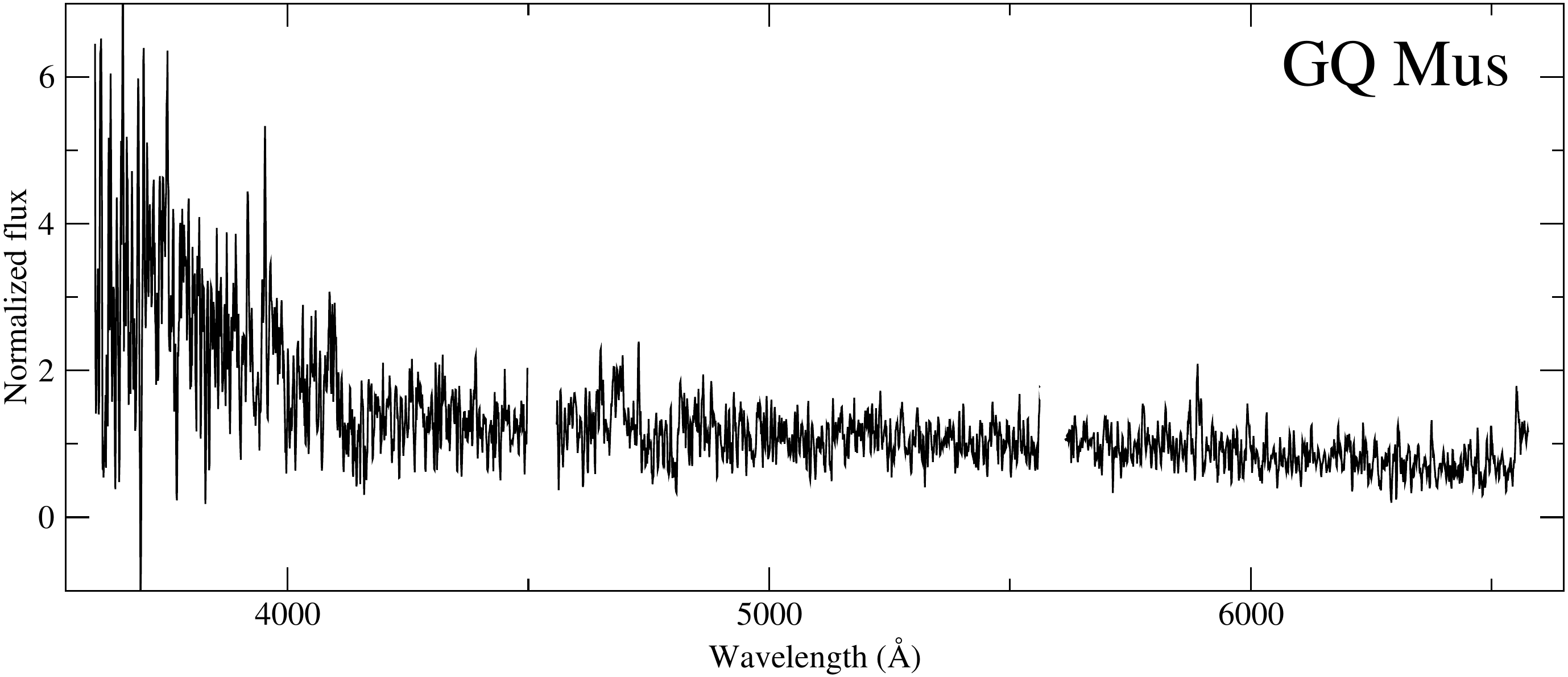} \\
\vspace{2mm}
\includegraphics[width=0.3\textwidth]{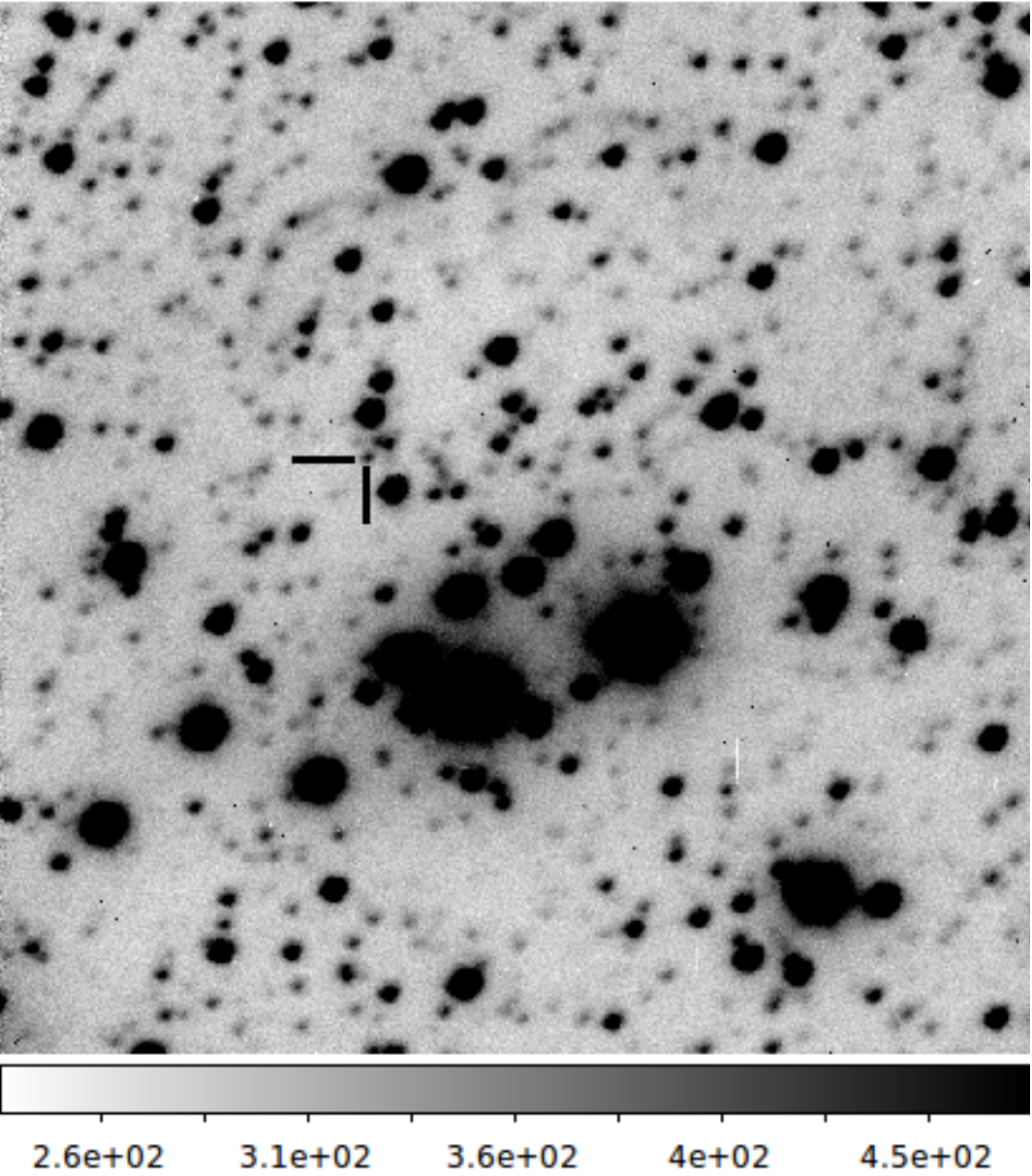}
  \hspace*{1mm}
 \includegraphics[width=0.65\textwidth]{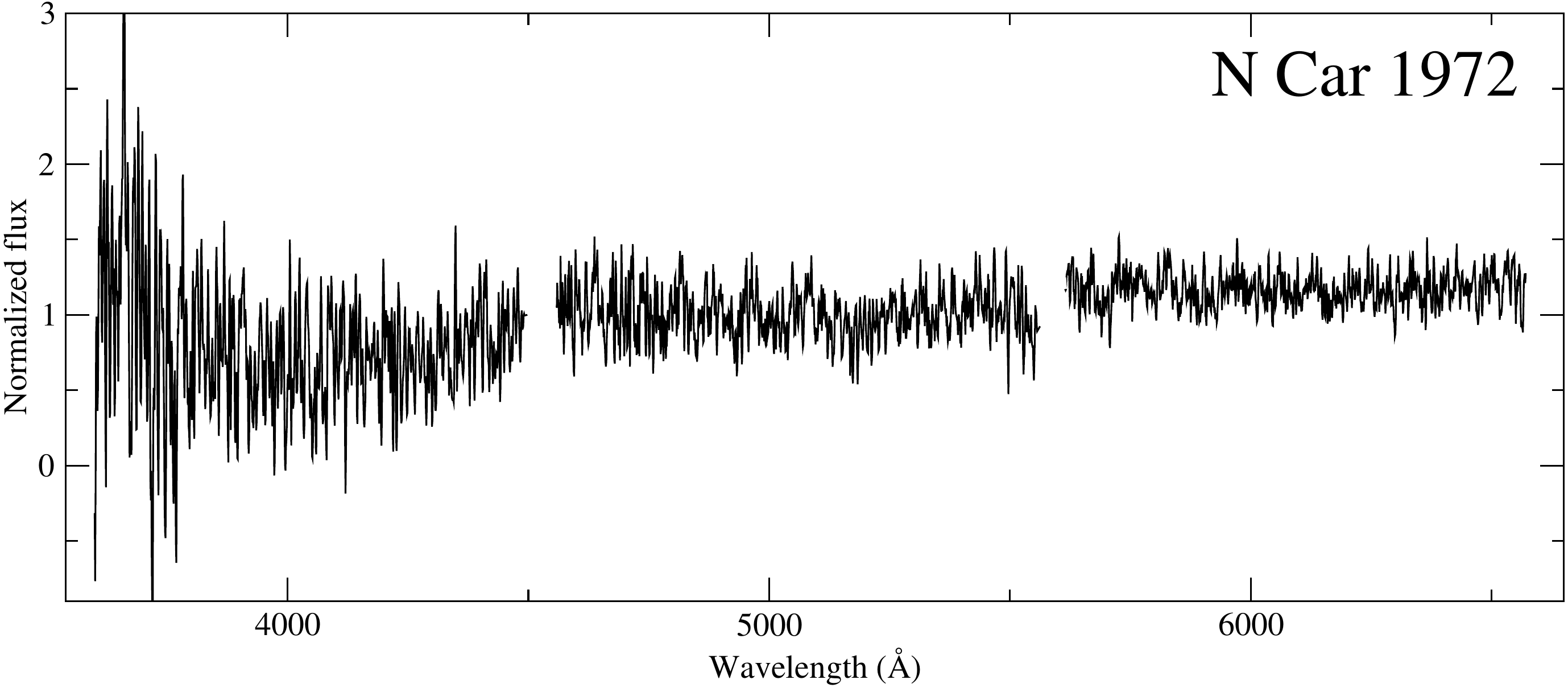} \\
\vspace{2mm}
\includegraphics[width=0.3\textwidth]{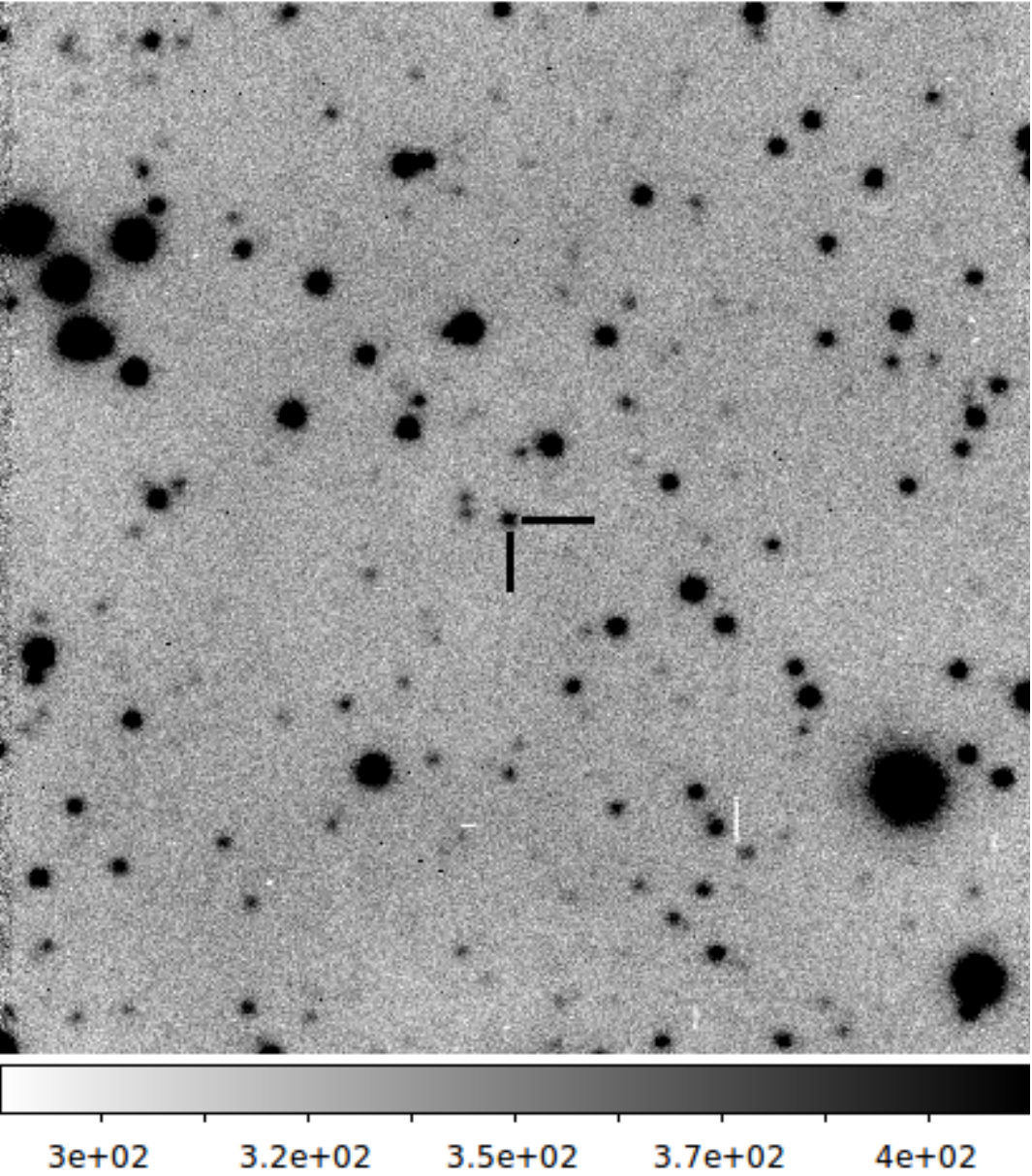}
  \hspace*{1mm}
\includegraphics[width=0.65\textwidth]{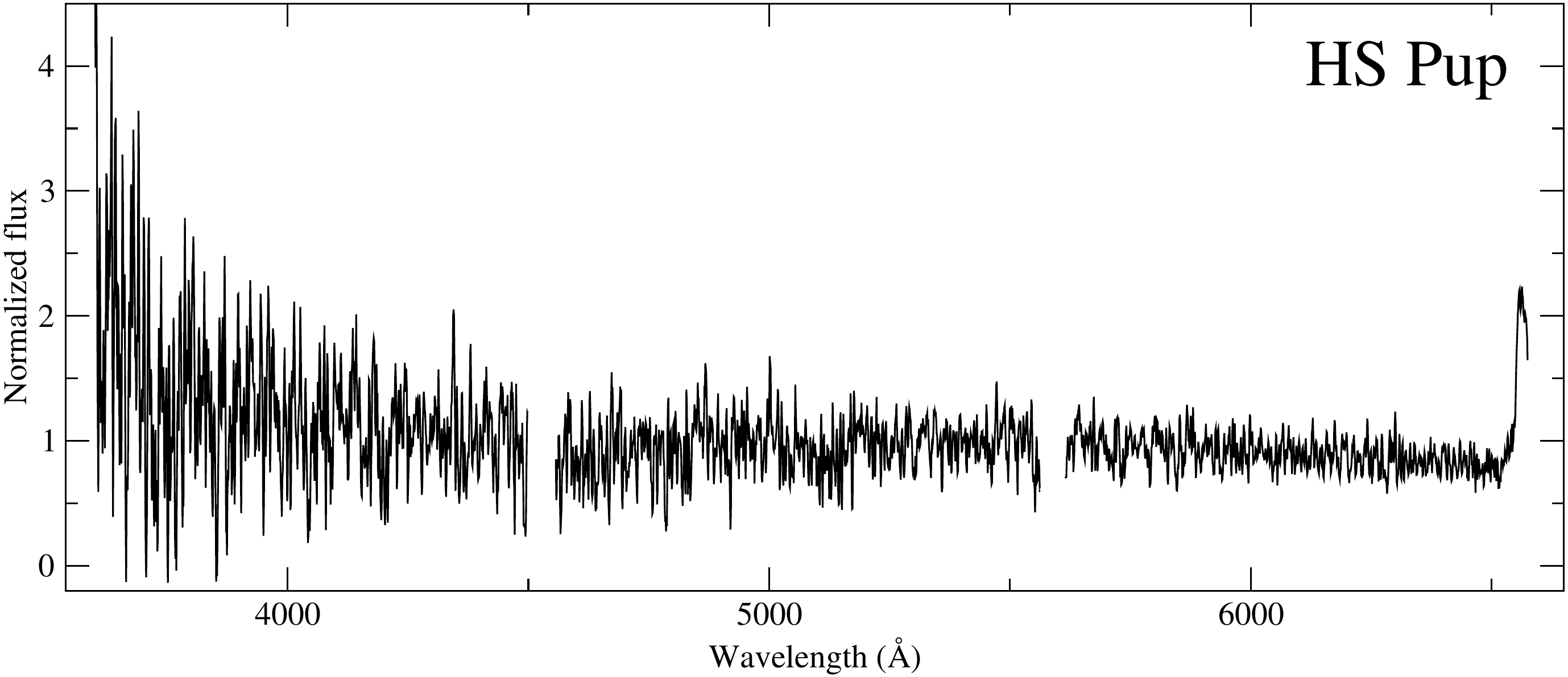} 
\caption{Continued}
\end{figure*}

\addtocounter{figure}{-1}
\begin{figure*}
 \centering
 \includegraphics[width=0.3\textwidth]{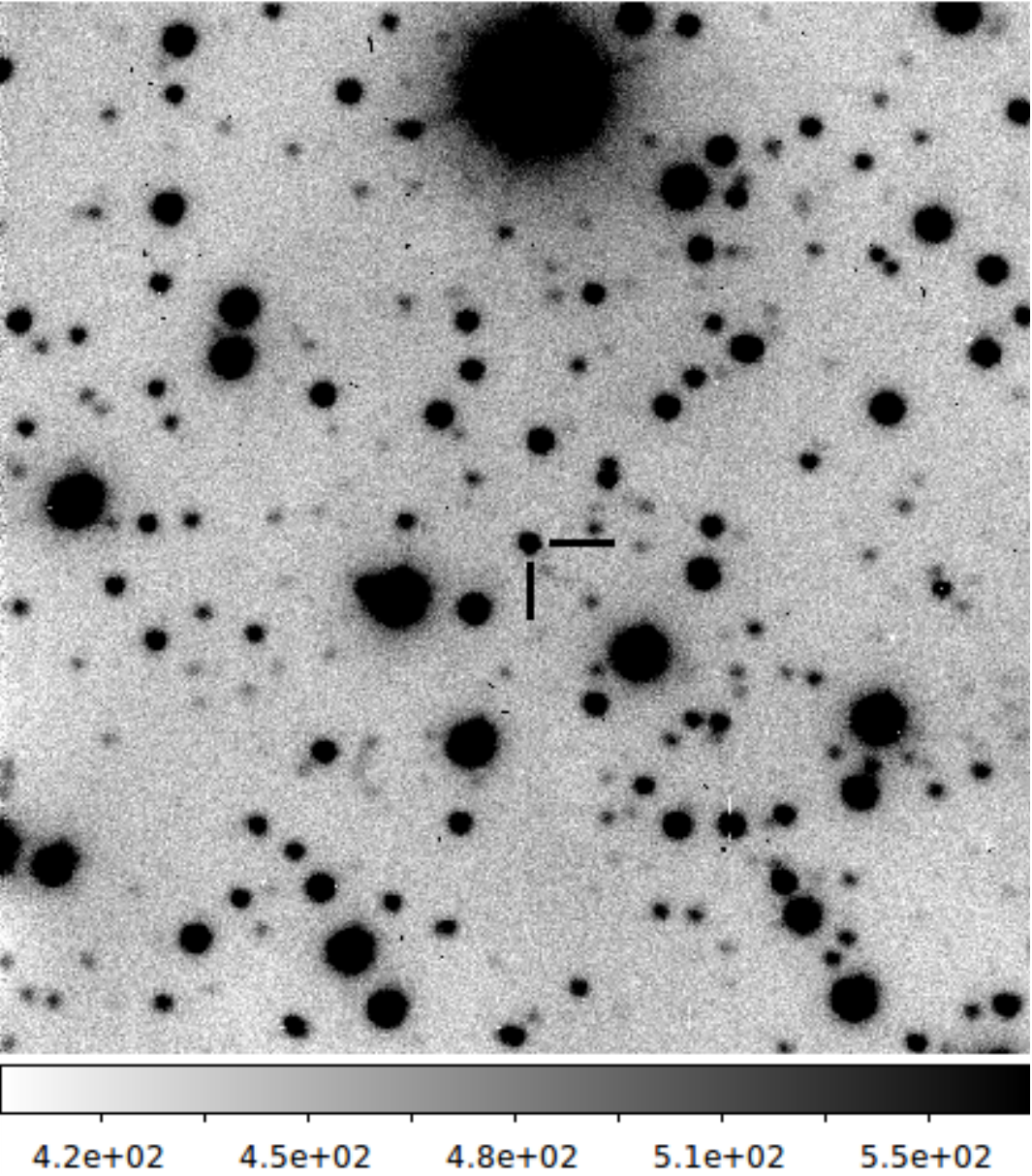}
  \hspace*{1mm}
 \includegraphics[width=0.65\textwidth]{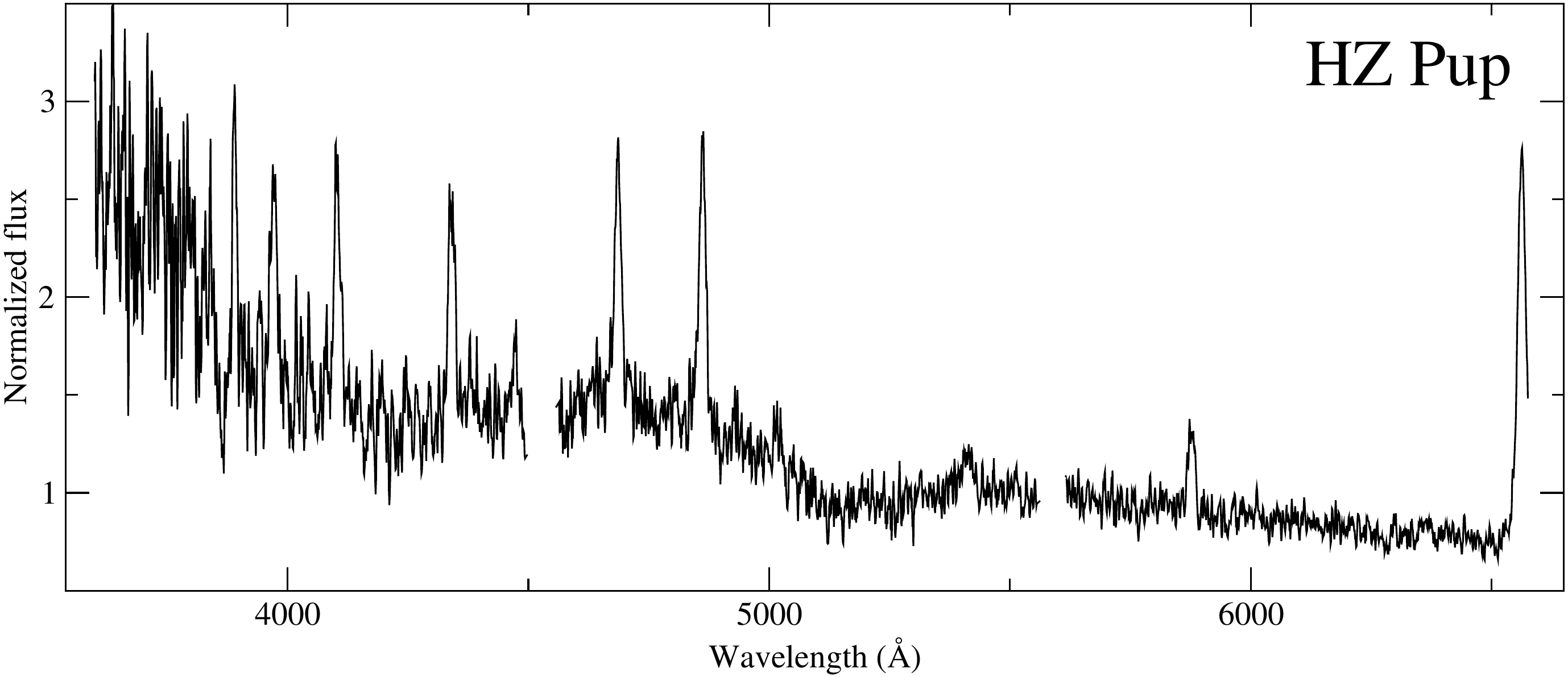} \\
\vspace{2mm}
\includegraphics[width=0.3\textwidth]{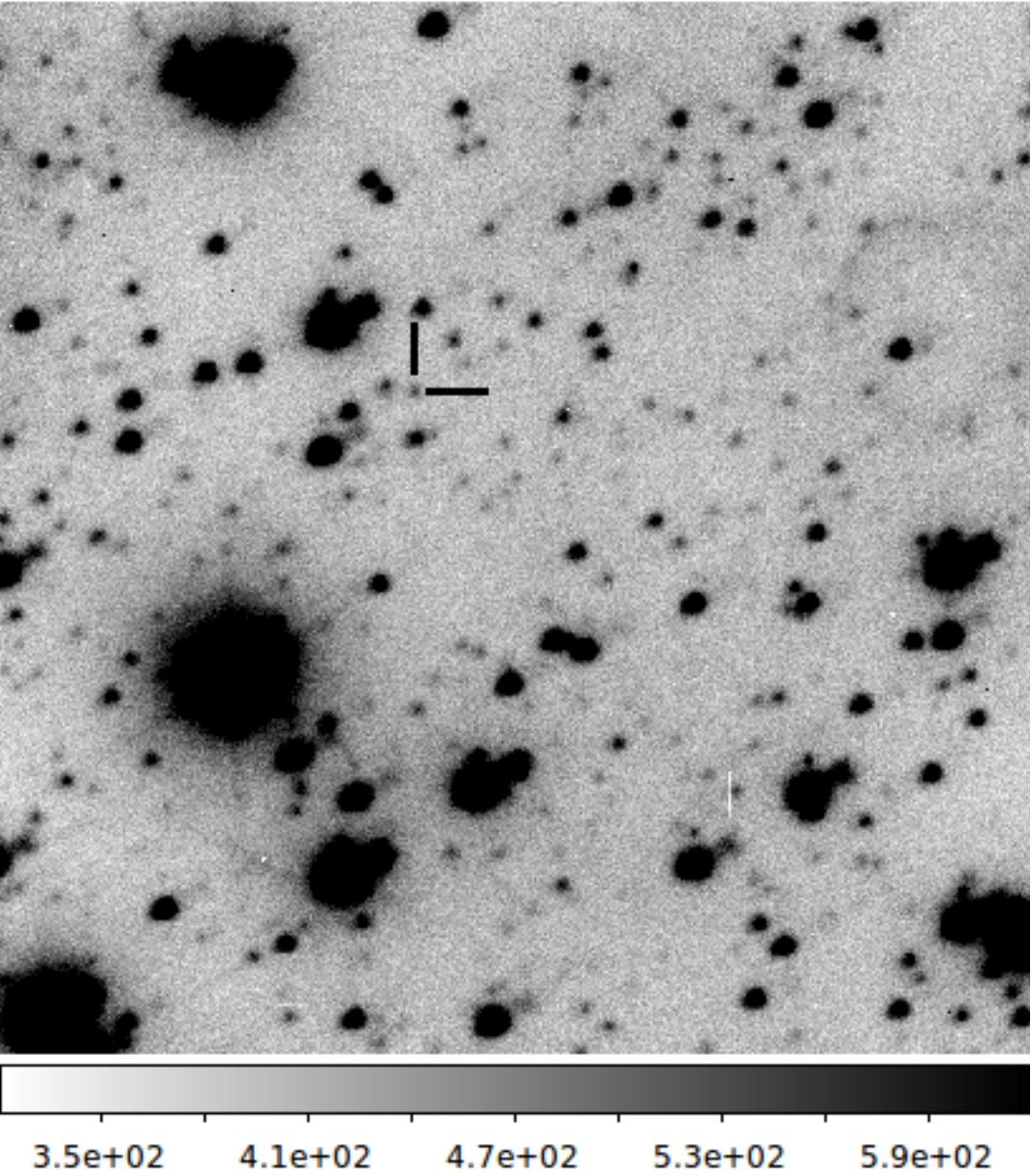}
  \hspace*{1mm}
\includegraphics[width=0.65\textwidth]{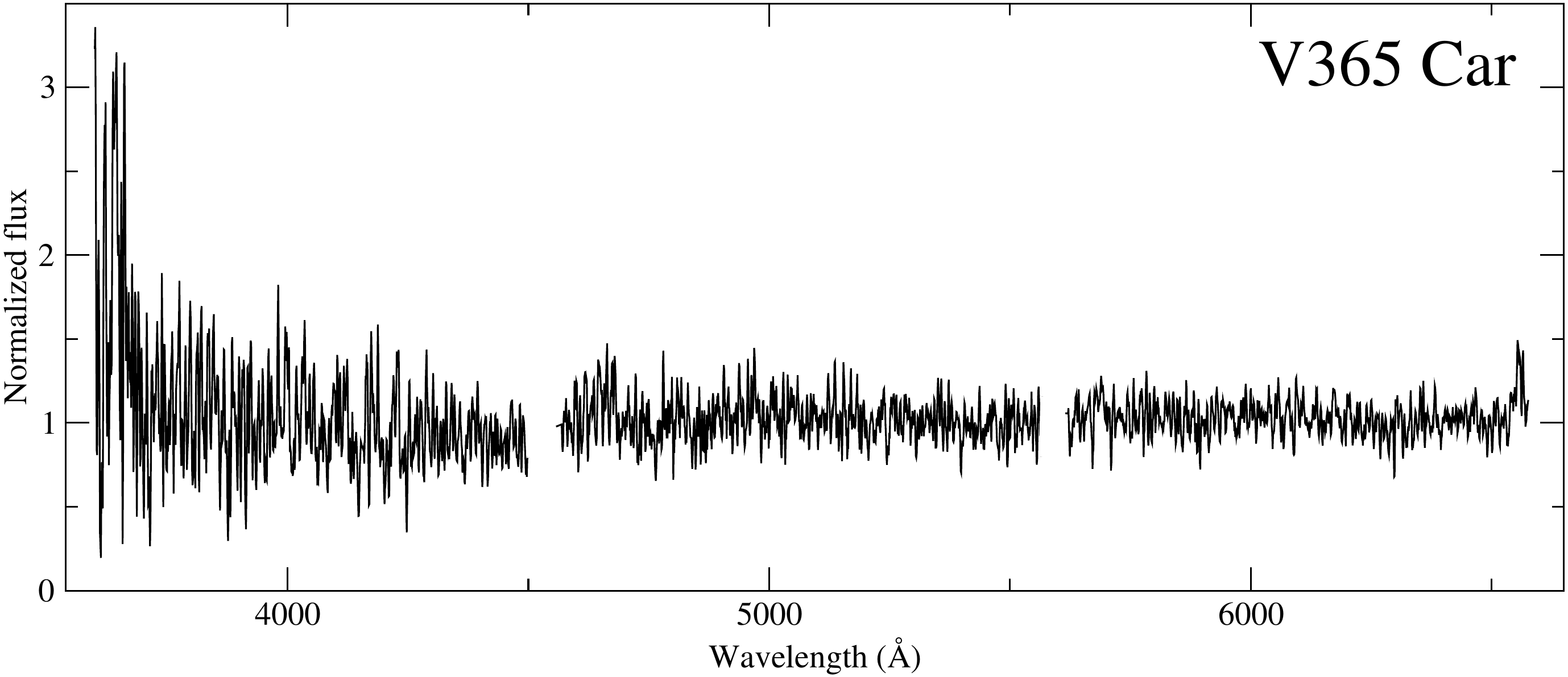} \\
\vspace{2mm}
\includegraphics[width=0.3\textwidth]{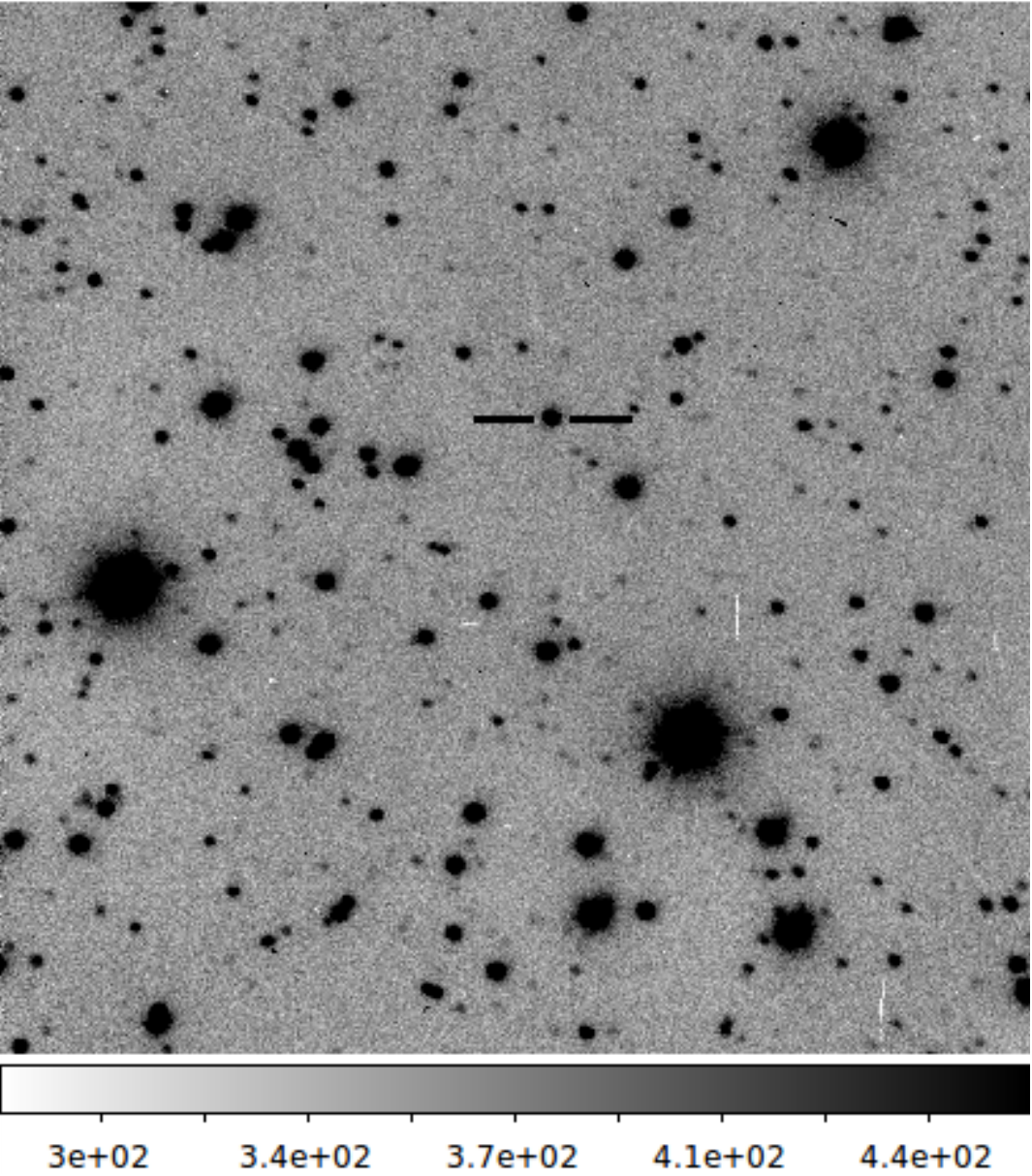}
  \hspace*{1mm}
 \includegraphics[width=0.65\textwidth]{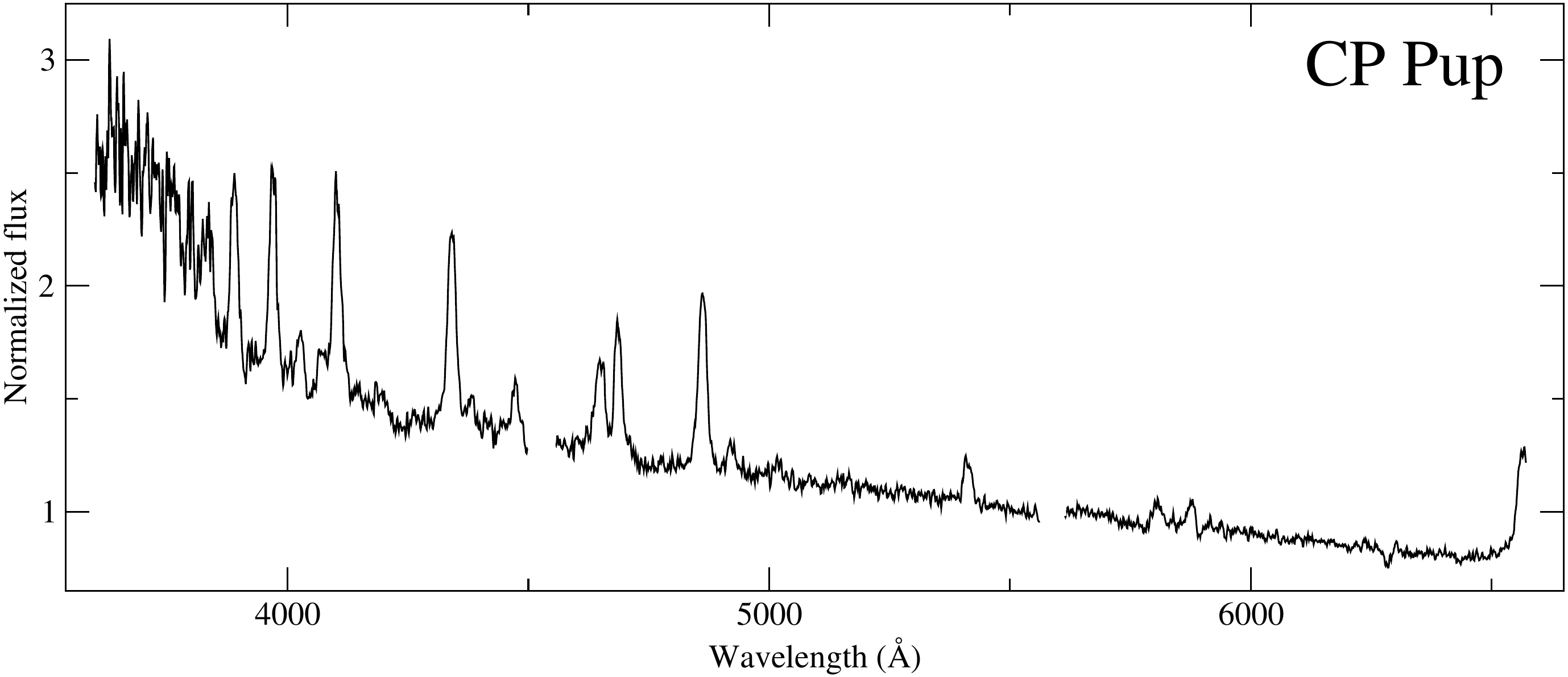} \\
\vspace{2mm}
\includegraphics[width=0.3\textwidth]{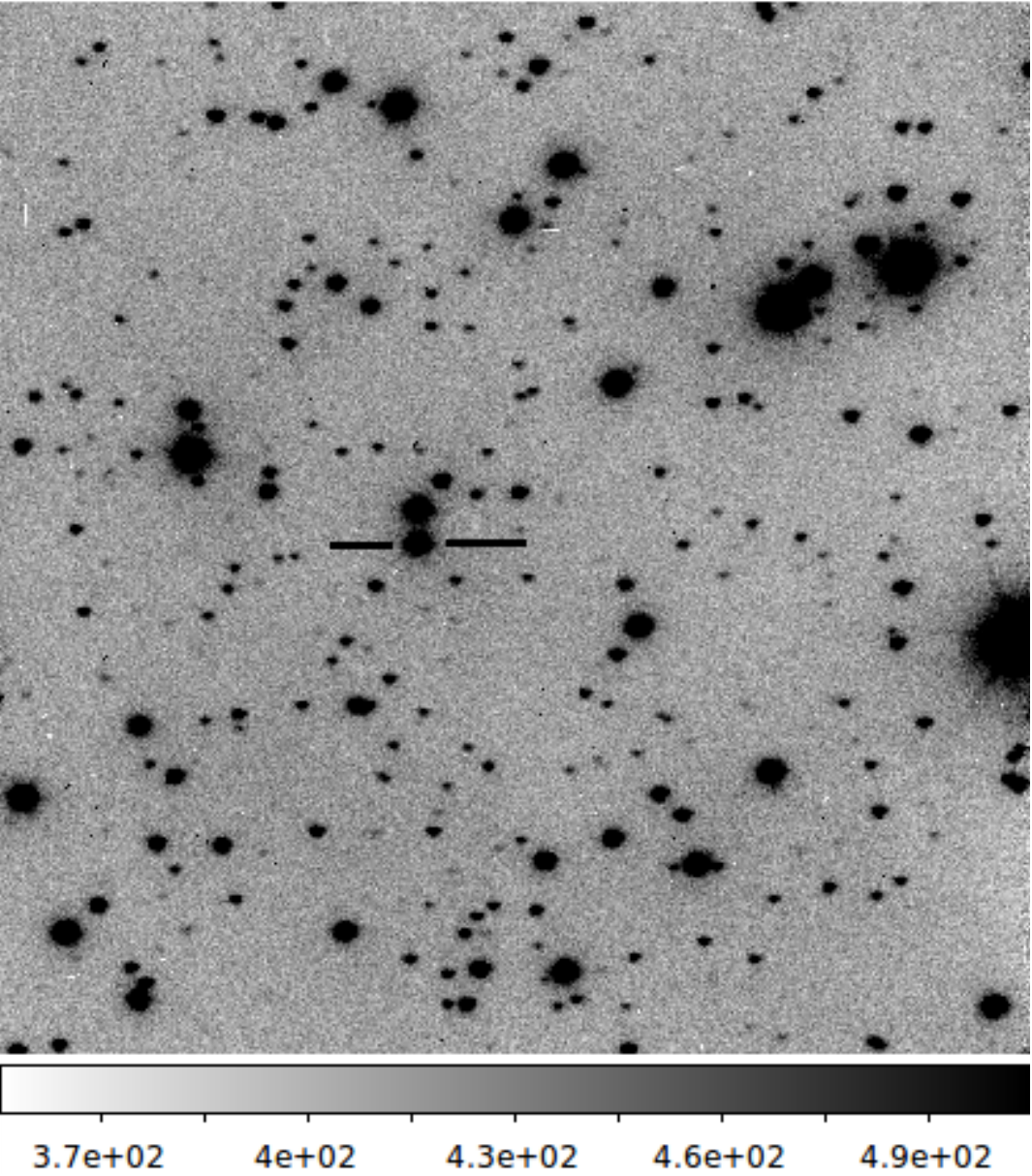}
  \hspace*{1mm}
\includegraphics[width=0.65\textwidth]{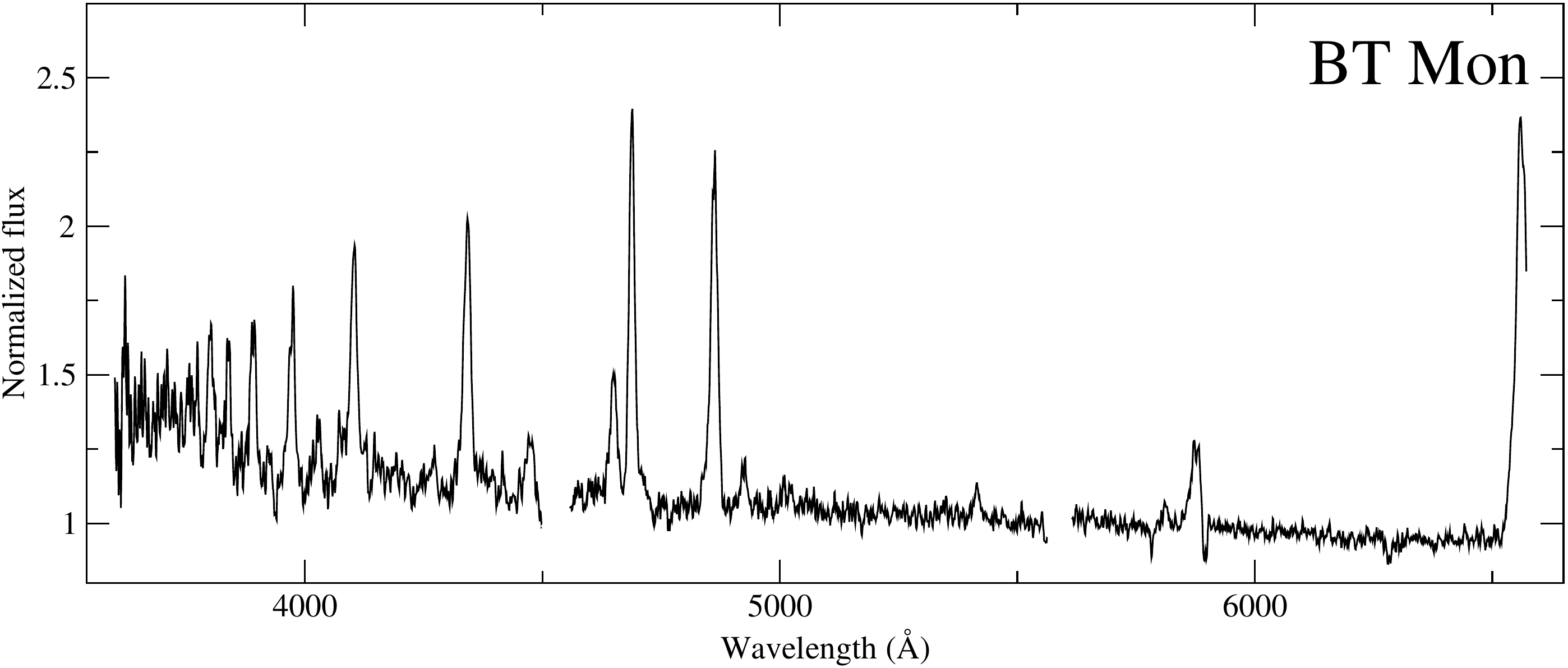} 
\caption{Continued}
\end{figure*}

\addtocounter{figure}{-1}
\begin{figure*}
 \centering
 \includegraphics[width=0.3\textwidth]{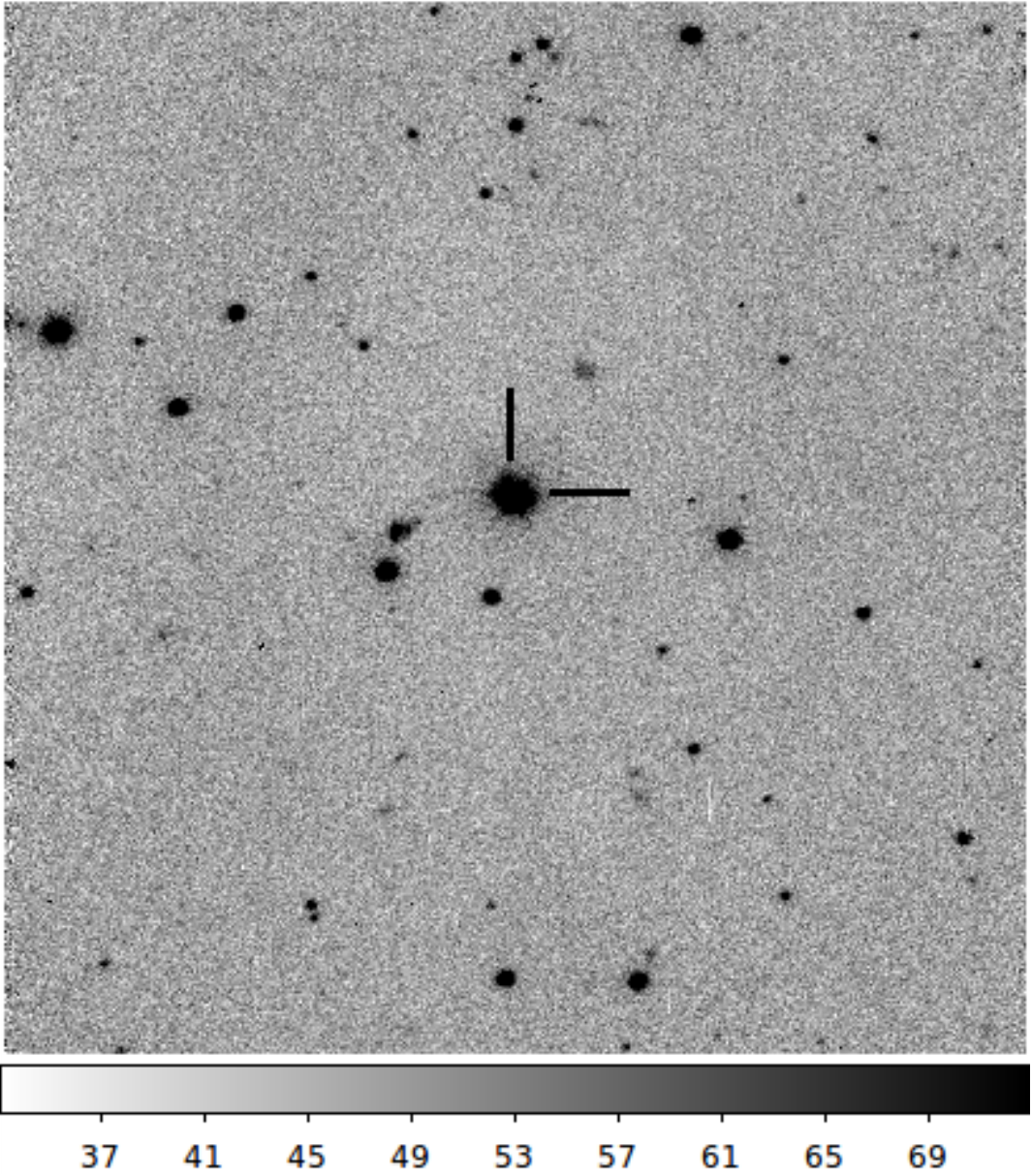}
  \hspace*{1mm}
 \includegraphics[width=0.65\textwidth]{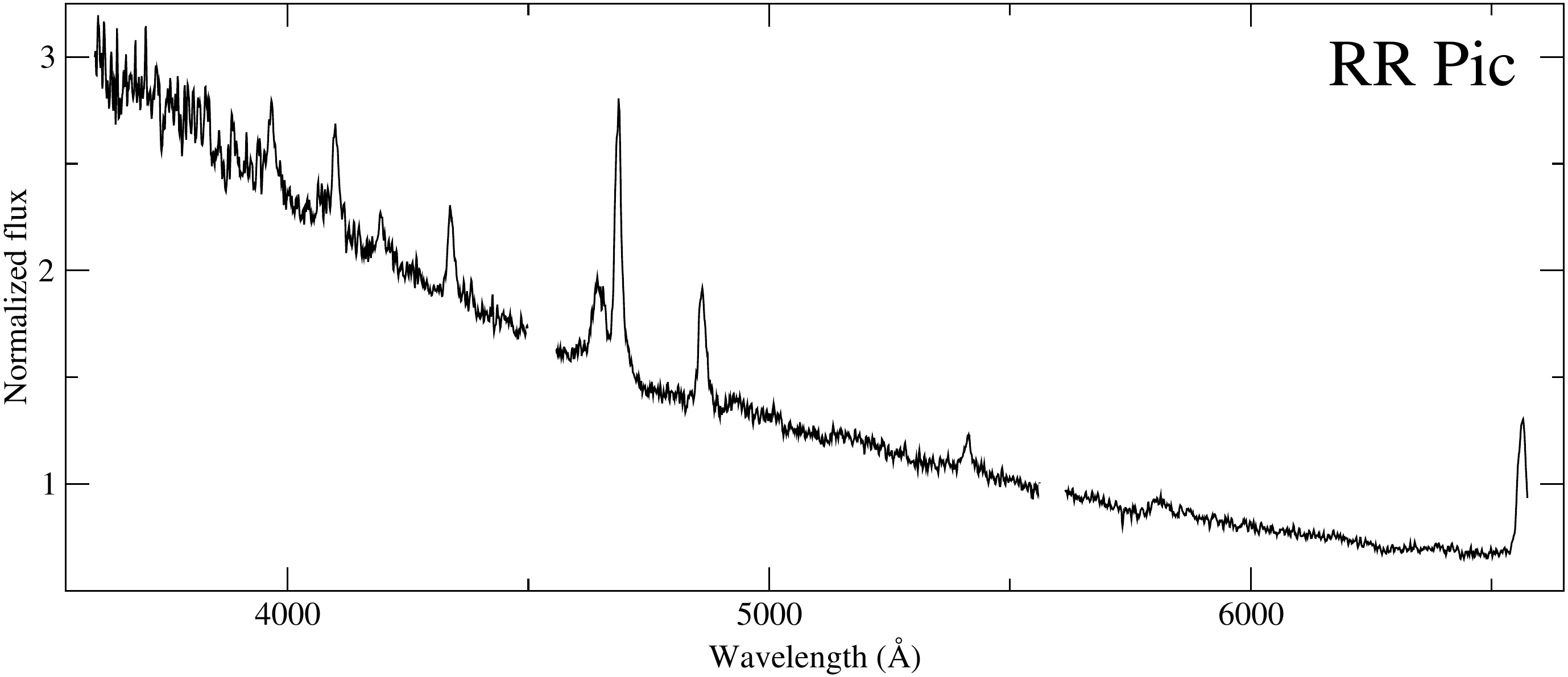} \\
\vspace{2mm}
\includegraphics[width=0.3\textwidth]{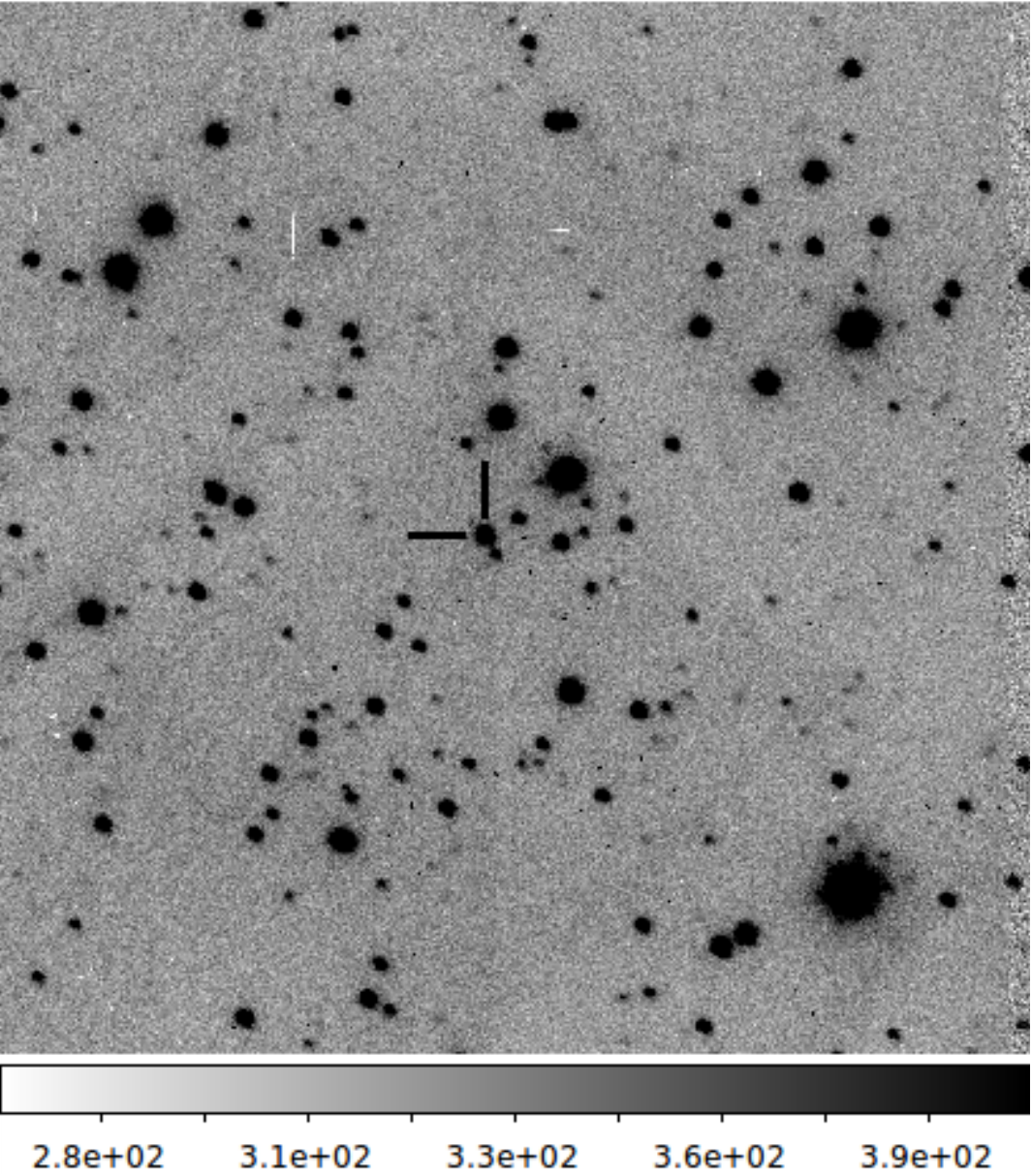}
  \hspace*{1mm}
\includegraphics[width=0.65\textwidth]{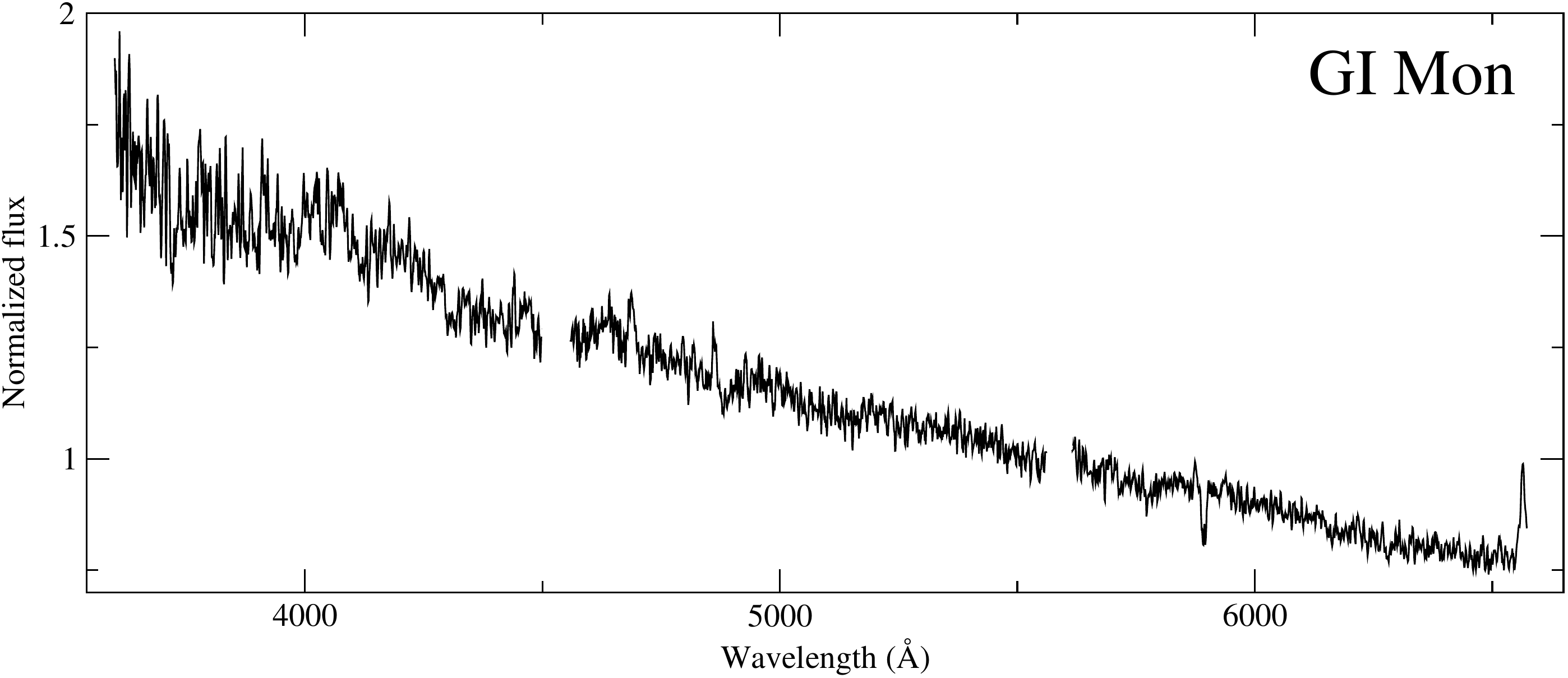} \\
\vspace{2mm}
\includegraphics[width=0.3\textwidth]{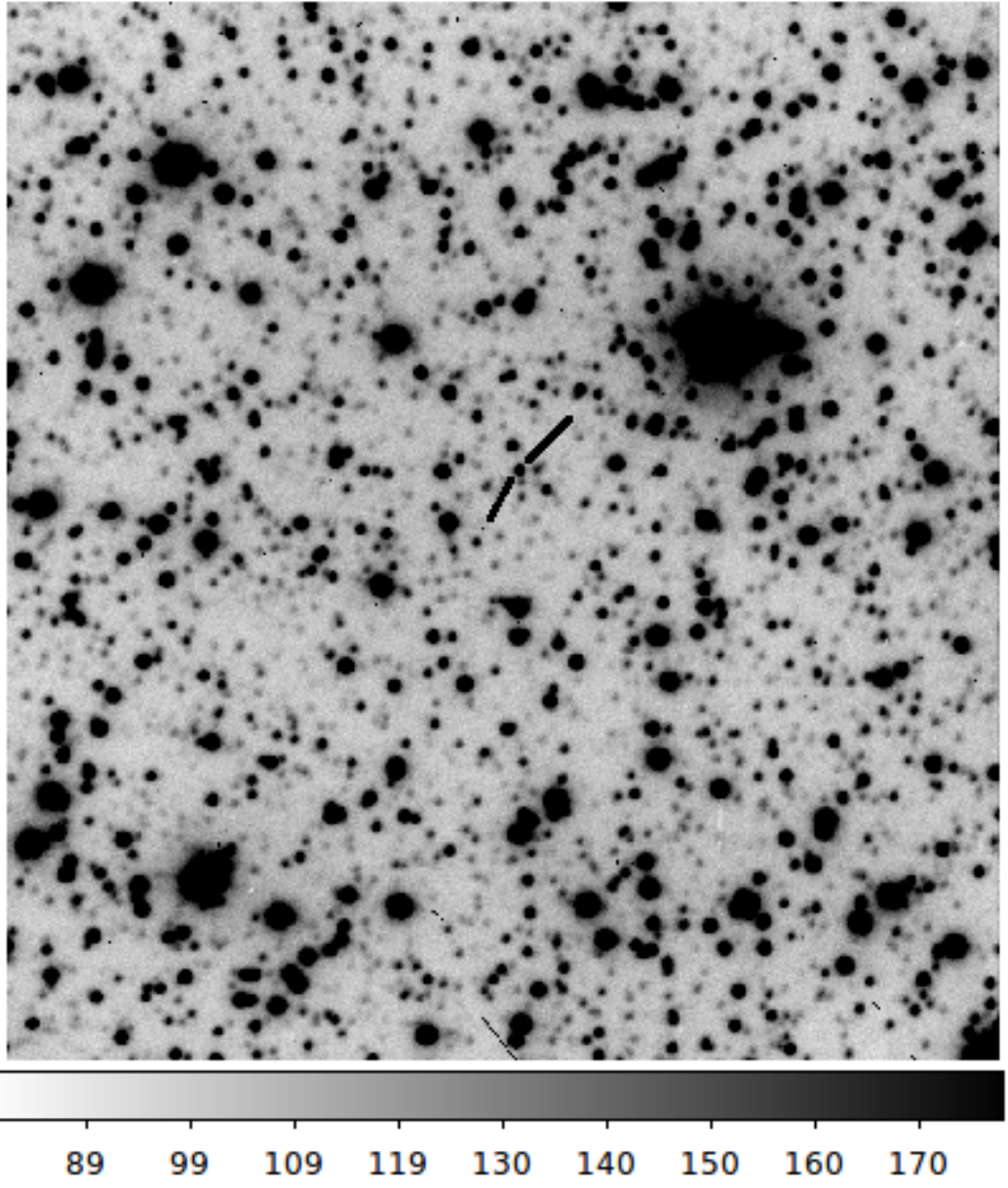}
  \hspace*{1mm}
 \includegraphics[width=0.65\textwidth]{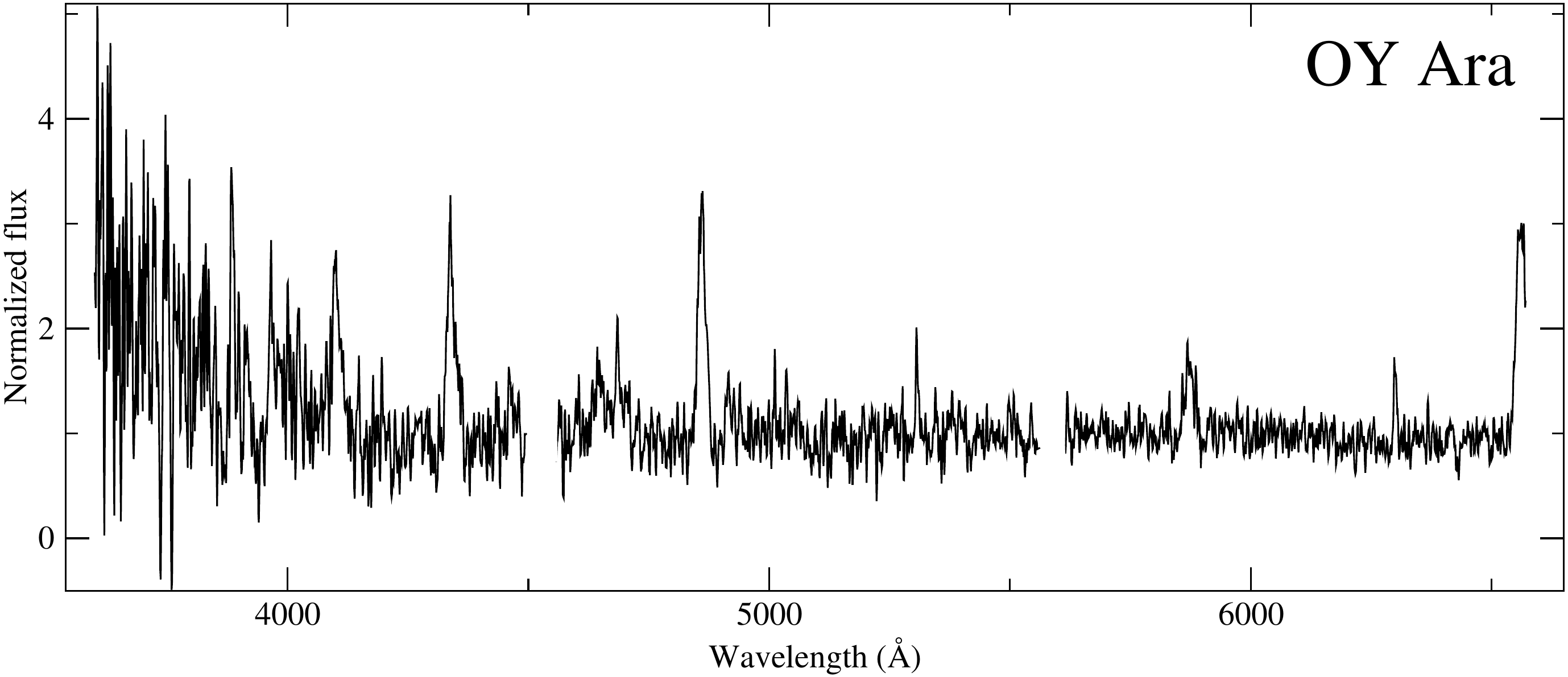} \\
\vspace{2mm}
\includegraphics[width=0.3\textwidth]{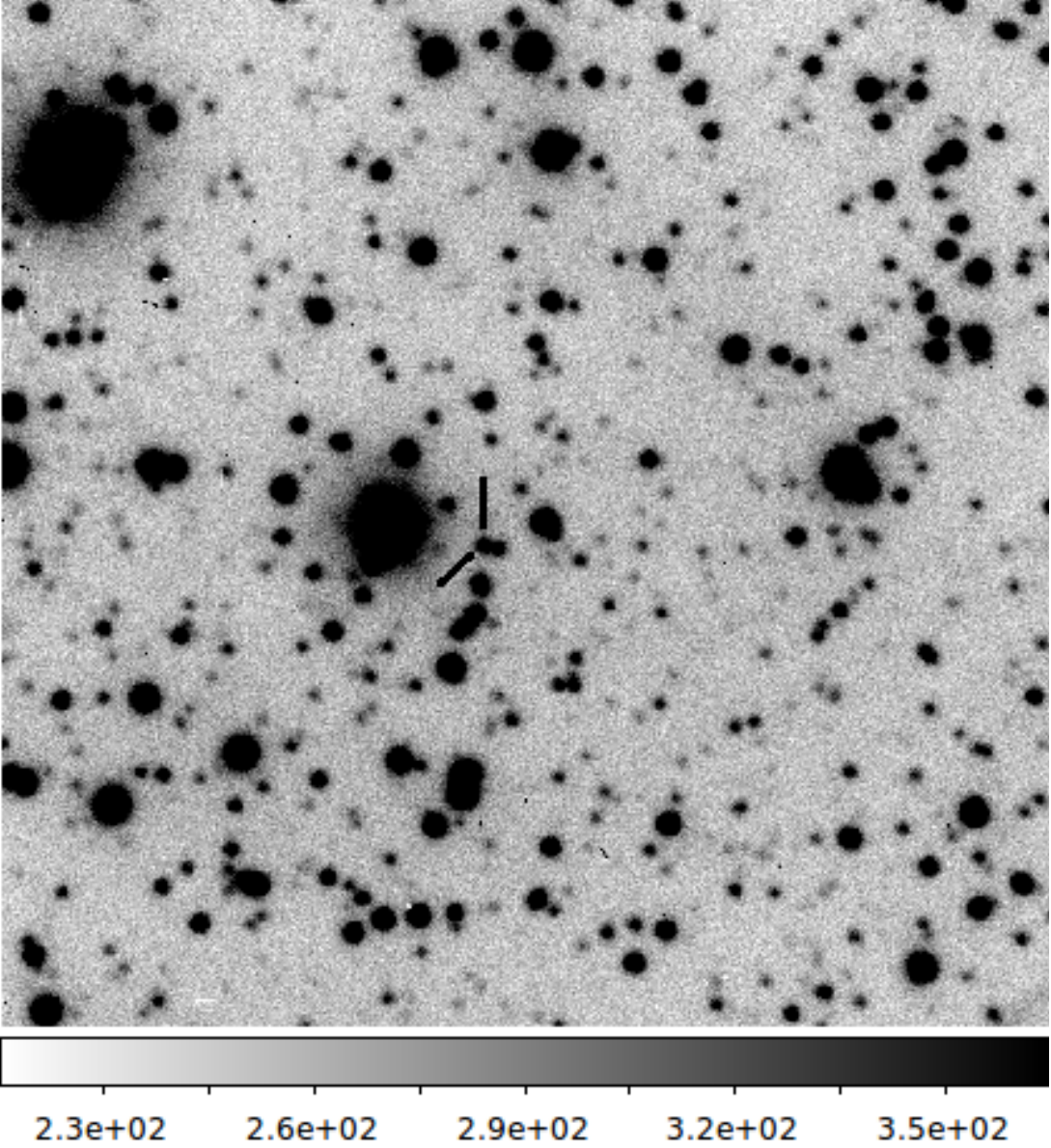}
  \hspace*{1mm}
\includegraphics[width=0.65\textwidth]{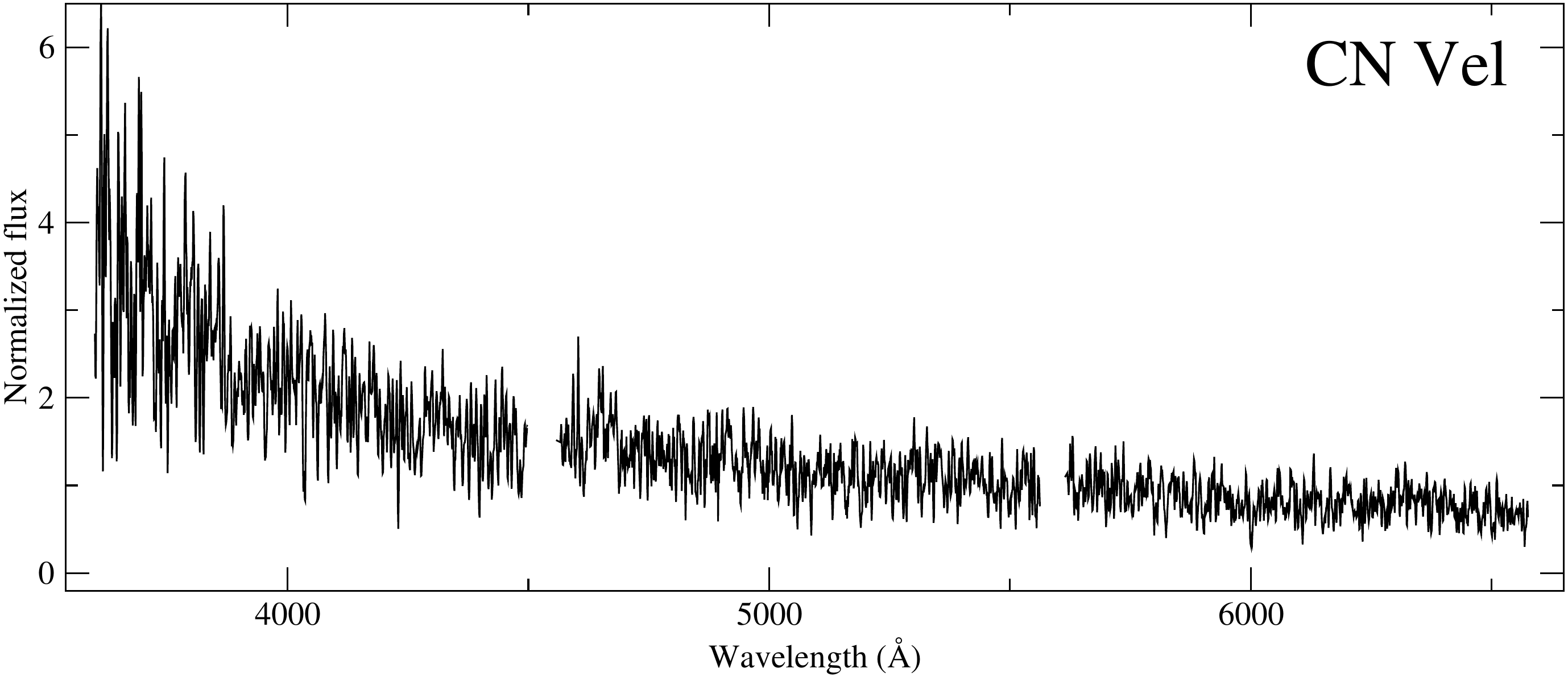} 
\caption{Continued}
\end{figure*}

\addtocounter{figure}{-1}
\begin{figure*}
 \centering
 \includegraphics[width=0.3\textwidth]{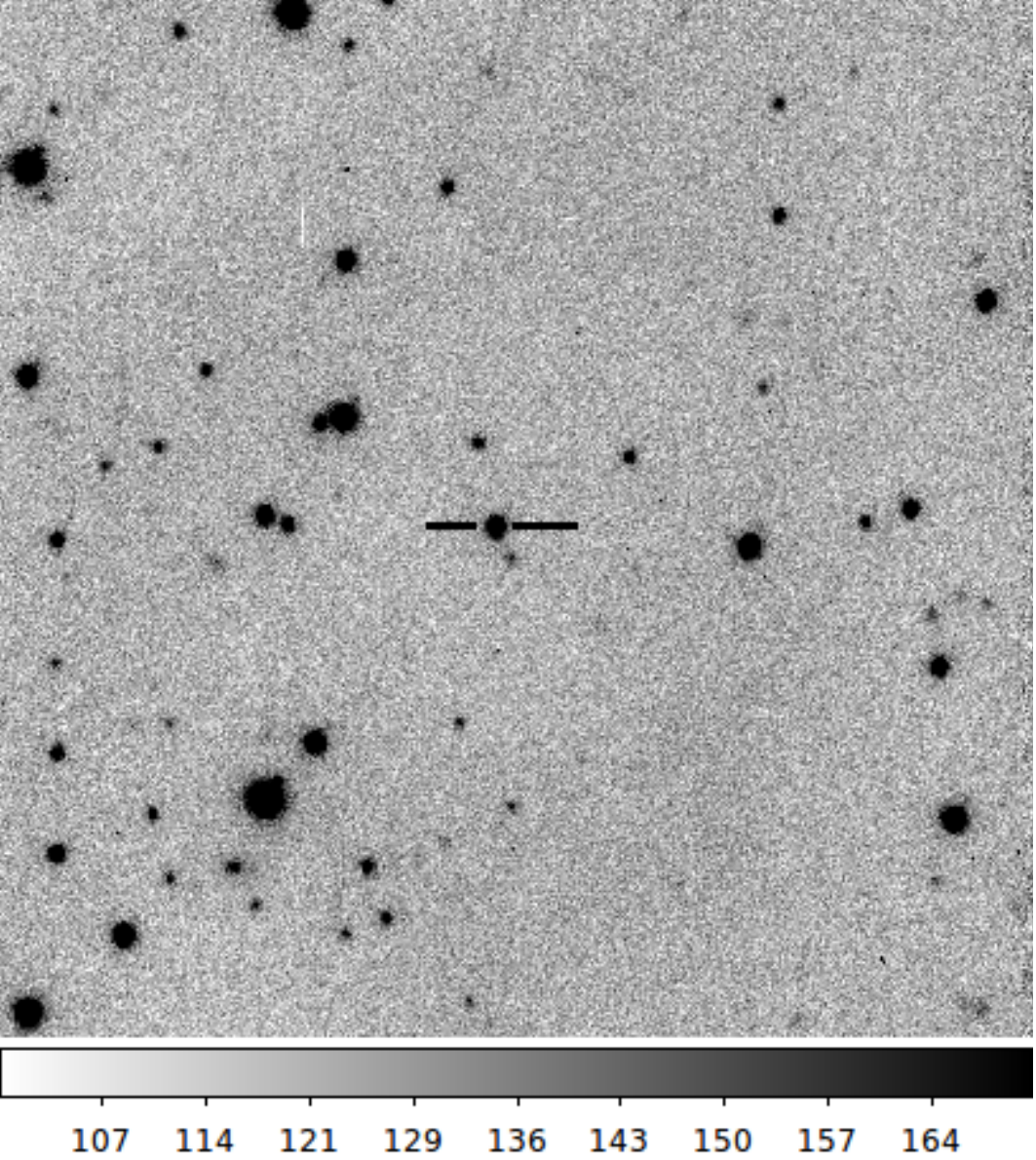}
  \hspace*{1mm}
 \includegraphics[width=0.65\textwidth]{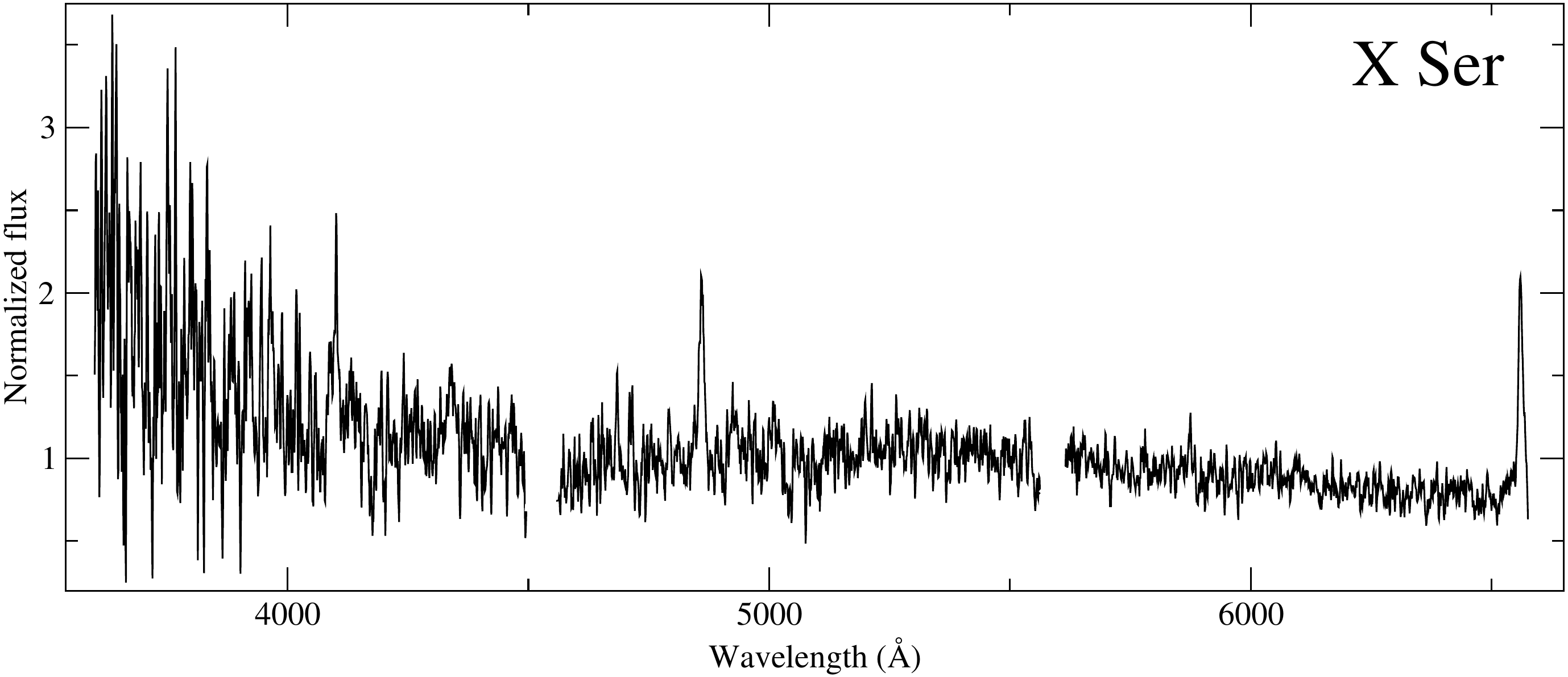} \\
\vspace{2mm}
\includegraphics[width=0.3\textwidth]{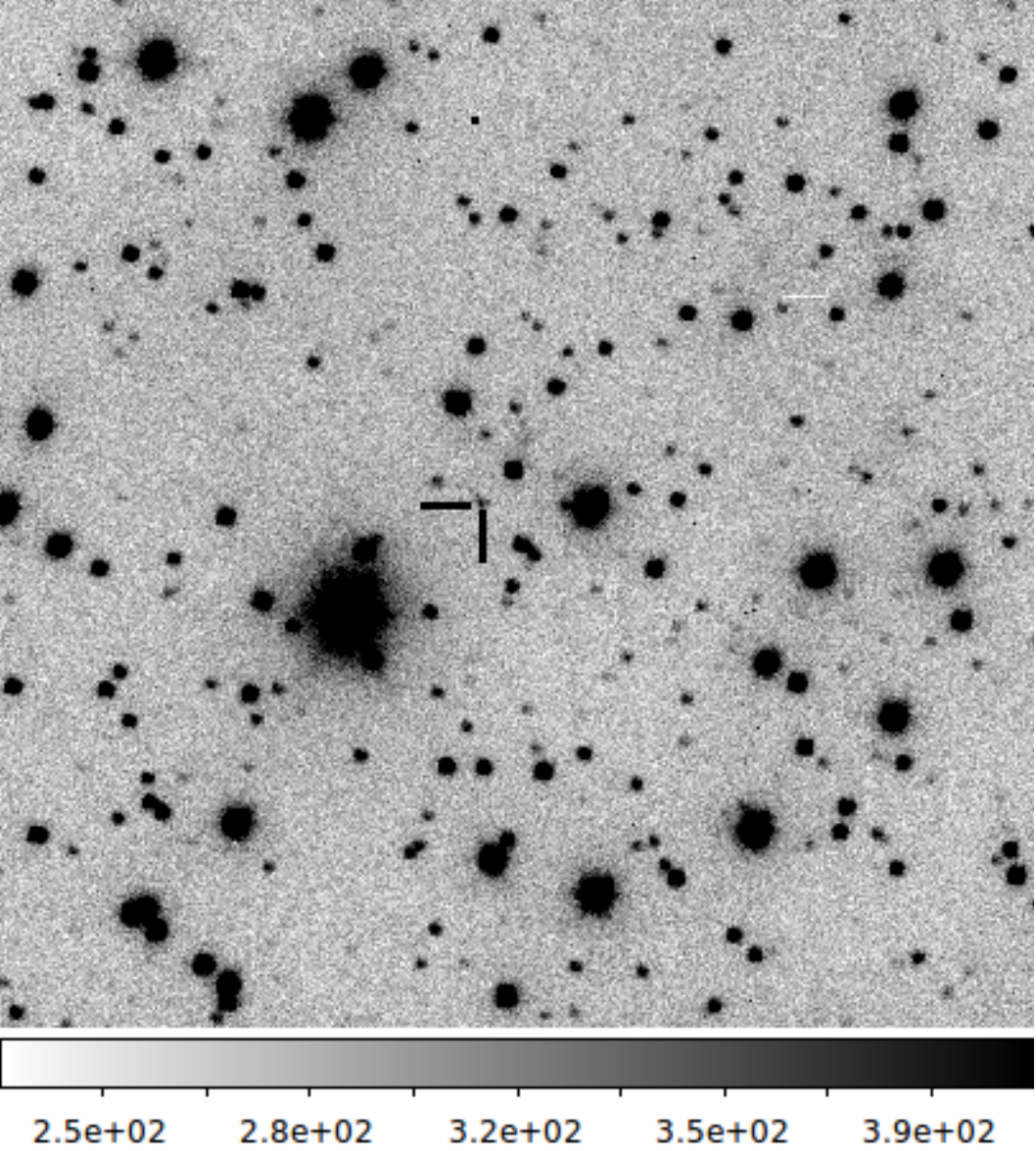}
  \hspace*{1mm}
\includegraphics[width=0.65\textwidth]{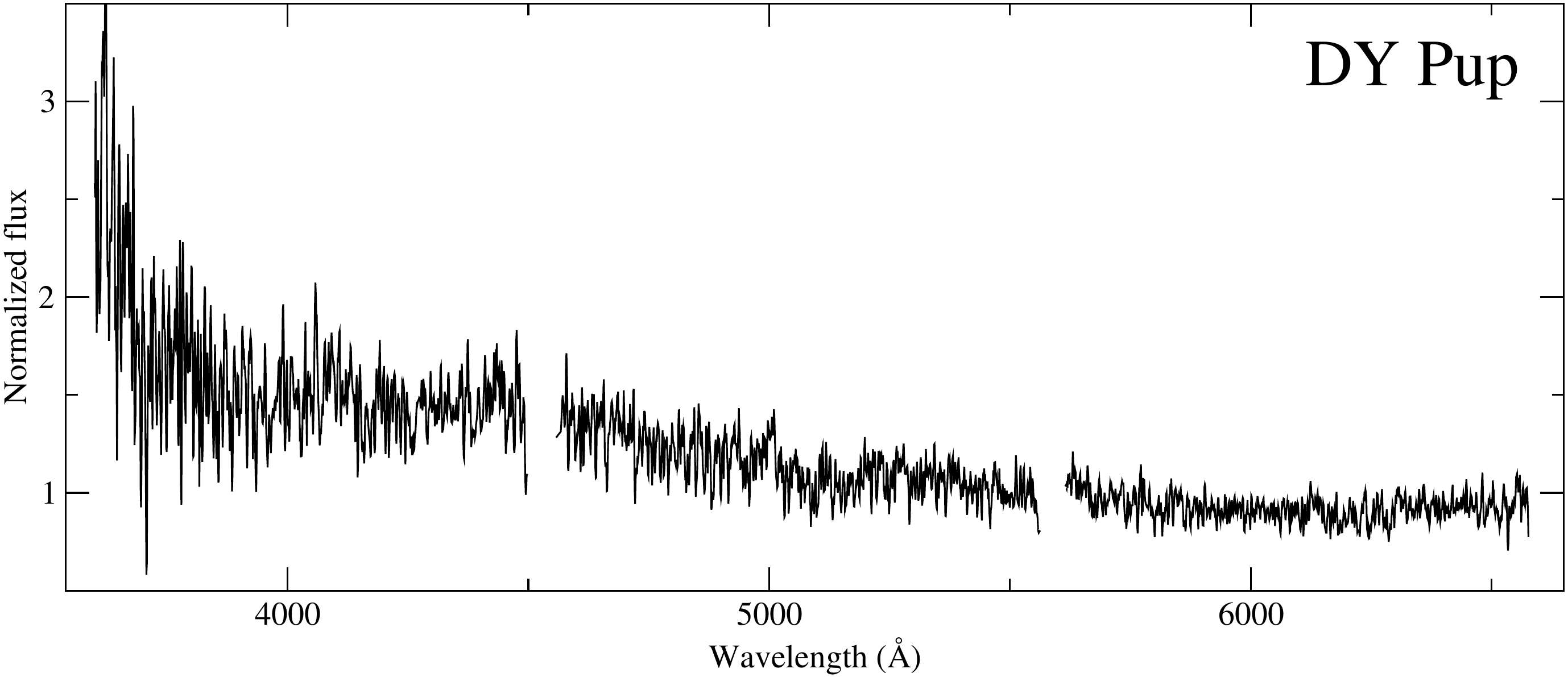} \\
\caption{Continued}
\end{figure*}

\begin{figure*}
\centering
\resizebox{\hsize}{!}{\includegraphics{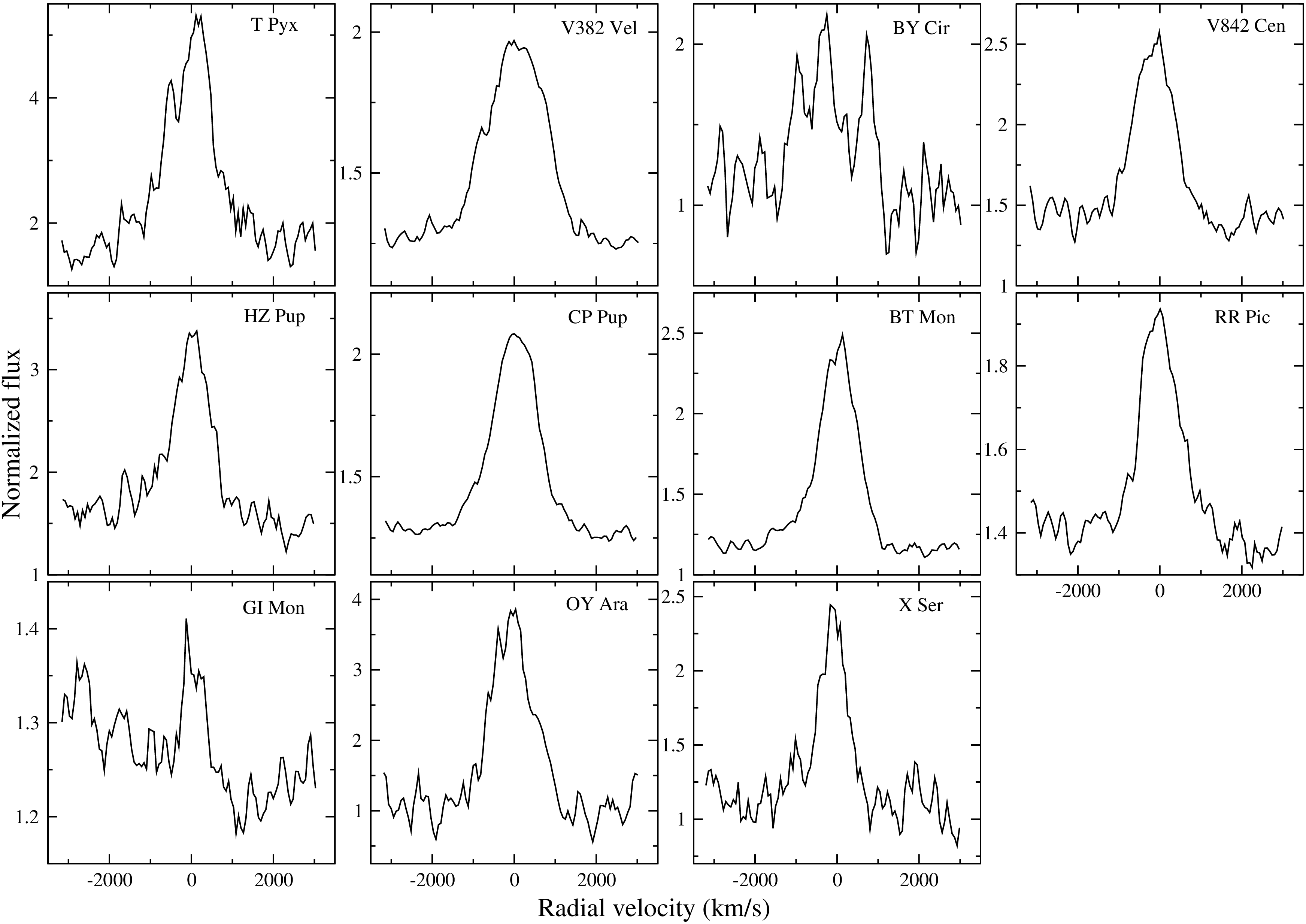}}
\caption{H$\beta$ profiles.}
\label{hb_prof} 
\bigskip
\resizebox{\hsize}{!}{\includegraphics{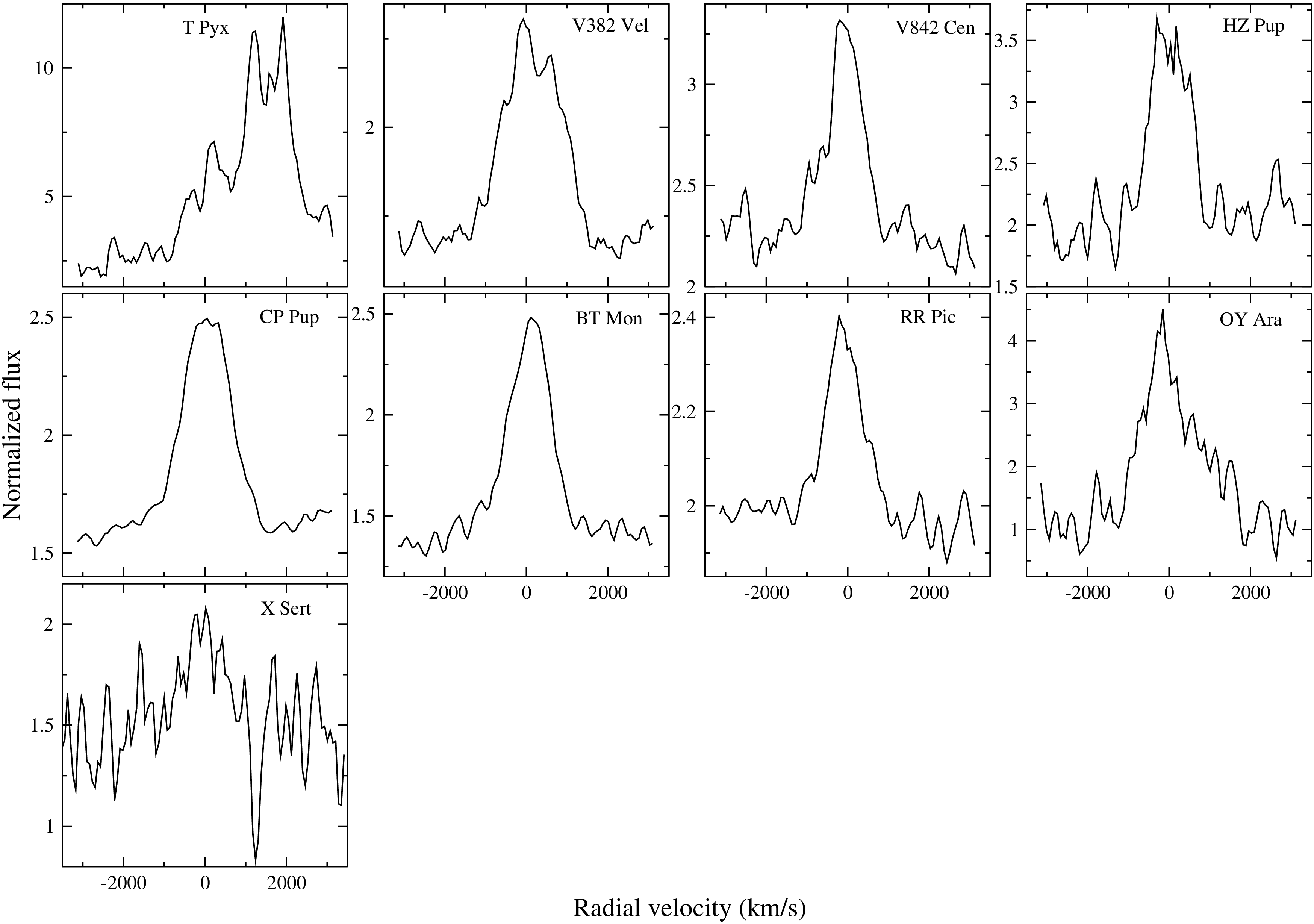}}
\caption{H$\gamma$ profiles.}
\label{hg_prof}
\end{figure*}

\begin{figure*}
\centering
\resizebox{\hsize}{!}{\includegraphics{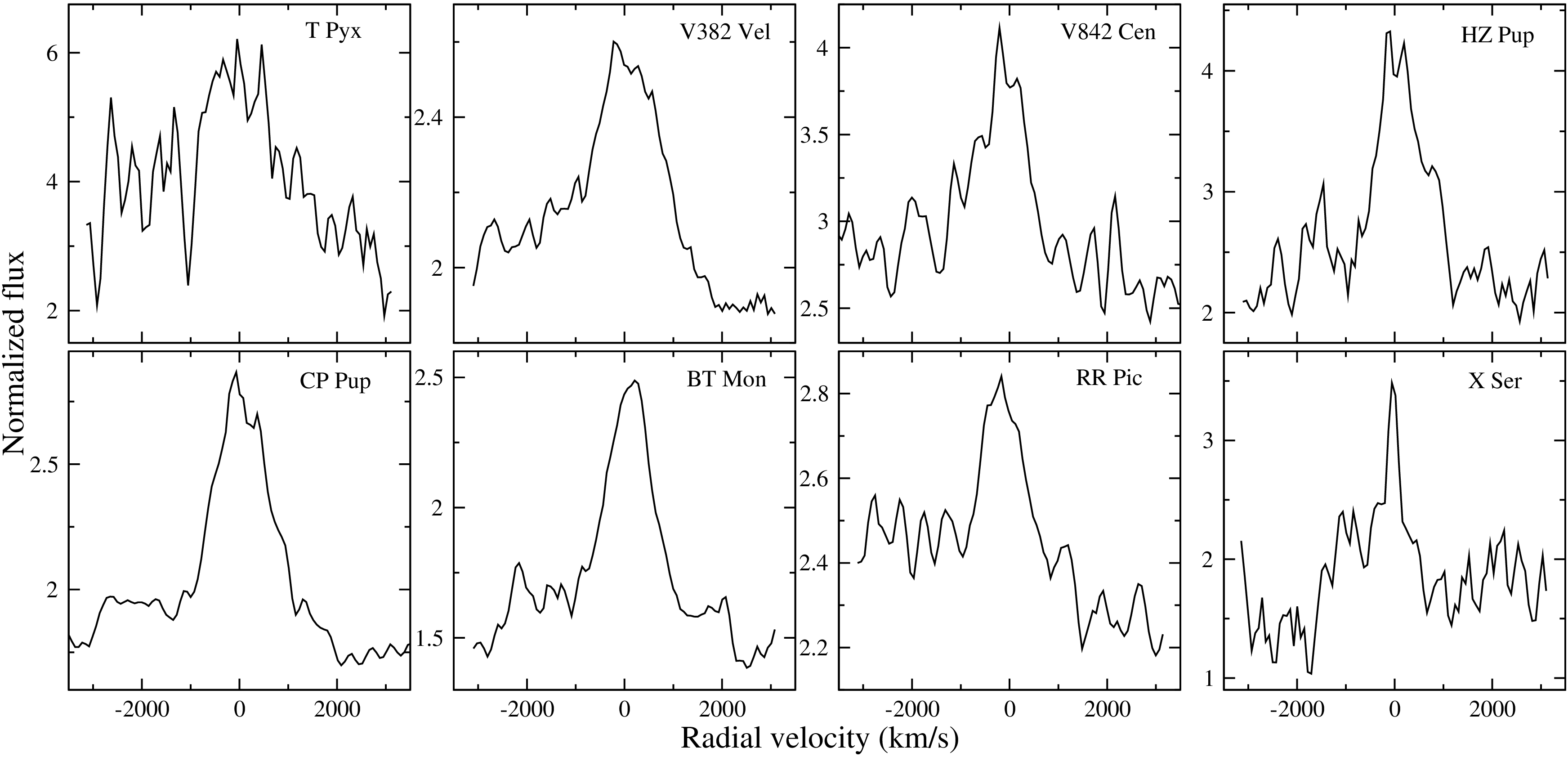}}
\caption{H$\delta$ profiles.}
\label{hd_prof} 
\bigskip
\resizebox{\hsize}{!}{\includegraphics{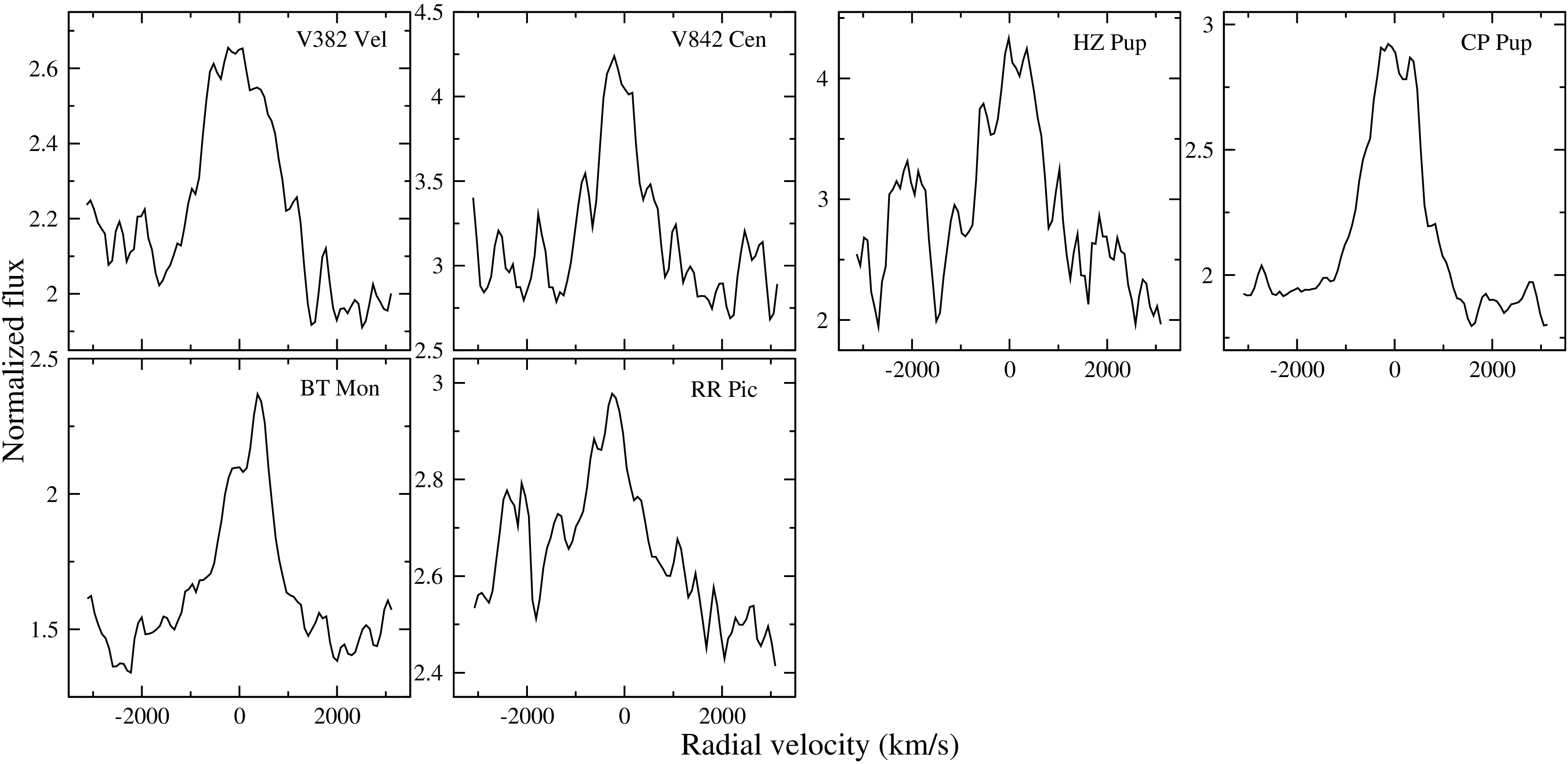}}
\caption{H$\varepsilon$ profiles.}
\label{he_prof}
\bigskip
\resizebox{\hsize}{!}{\includegraphics{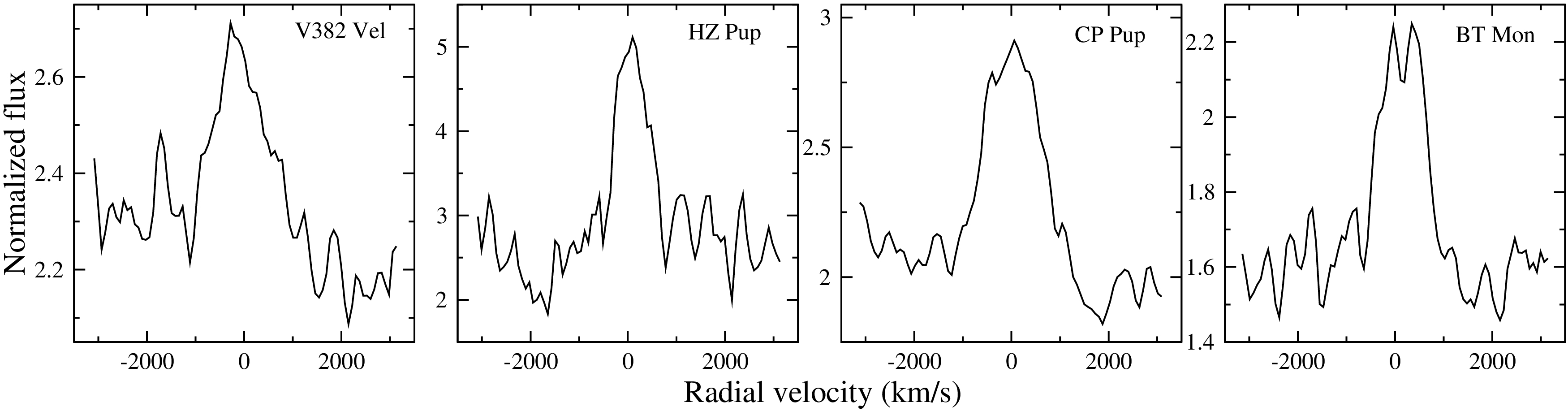}}
\caption{H8 profiles.}
\label{h8_prof} 
\end{figure*}

\begin{figure*}
\centering
\resizebox{\hsize}{!}{\includegraphics{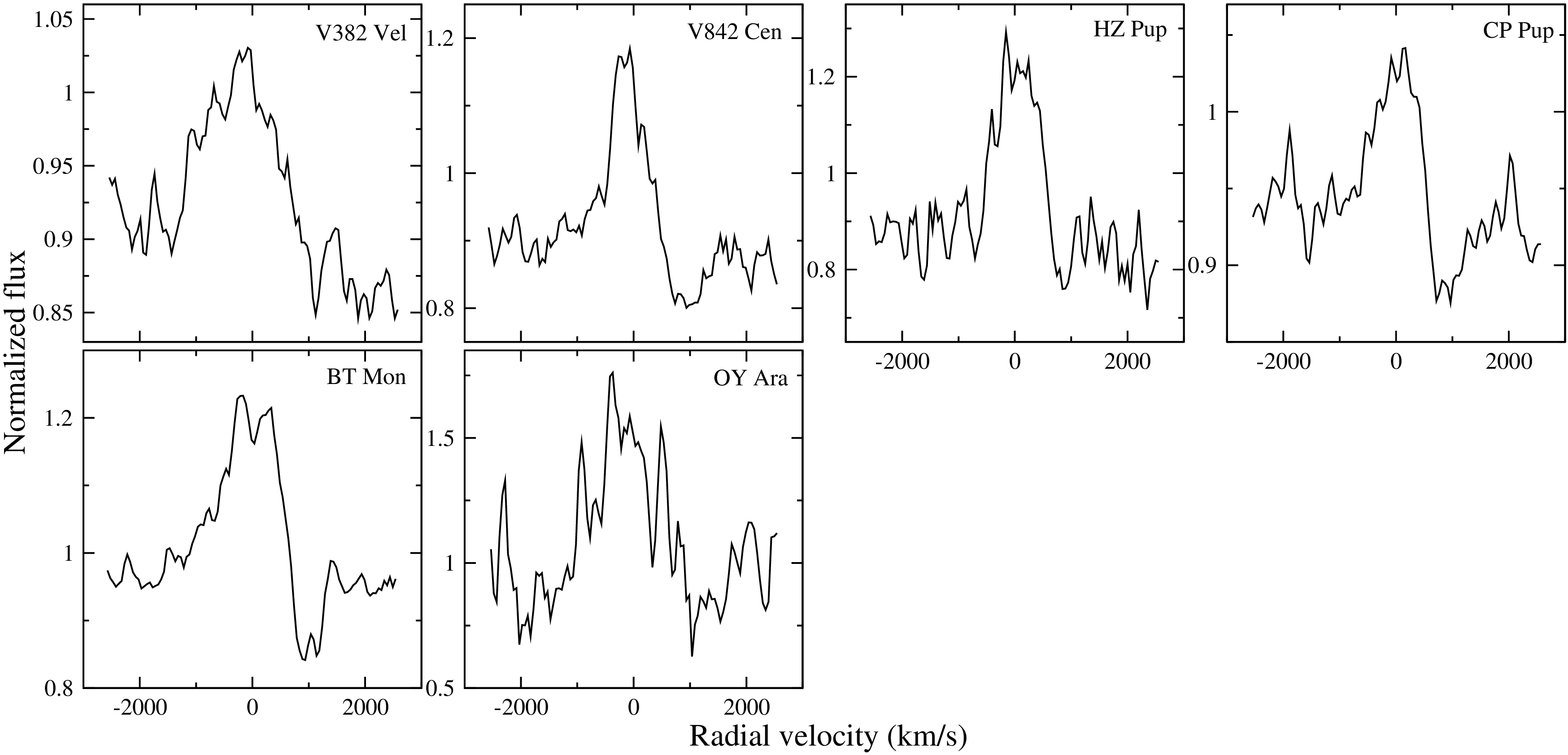}}
\caption{HeI 5876\,\AA\ profiles.}
\label{5876_prof}
\bigskip
\resizebox{\hsize}{!}{\includegraphics{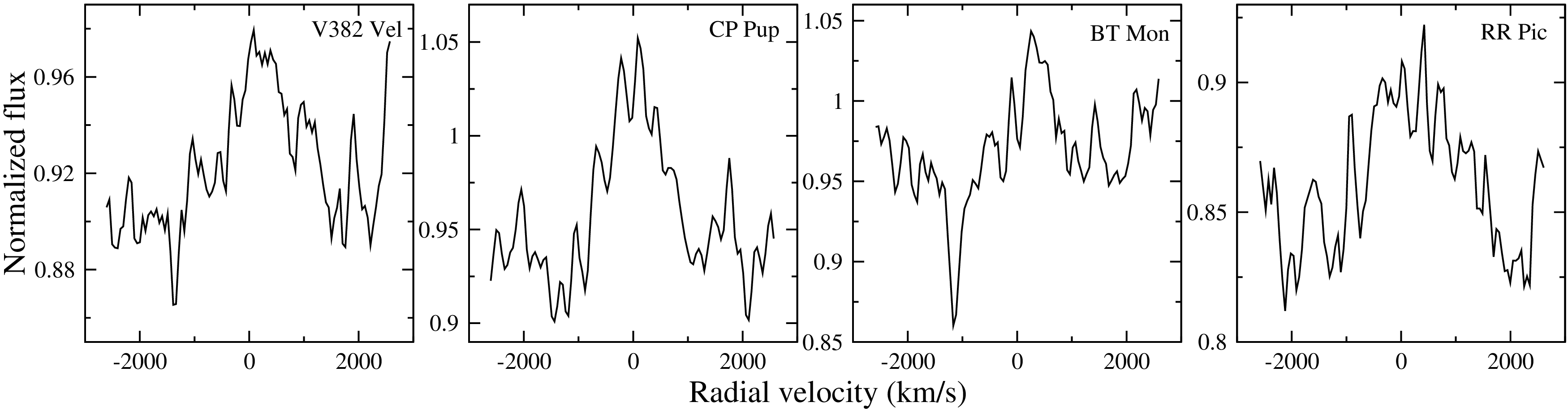}}
\caption{CIV 5805\,\AA\ profiles.}
\label{5805_prof} 
\bigskip
\flushleft
\resizebox{.75\hsize}{!}{\includegraphics{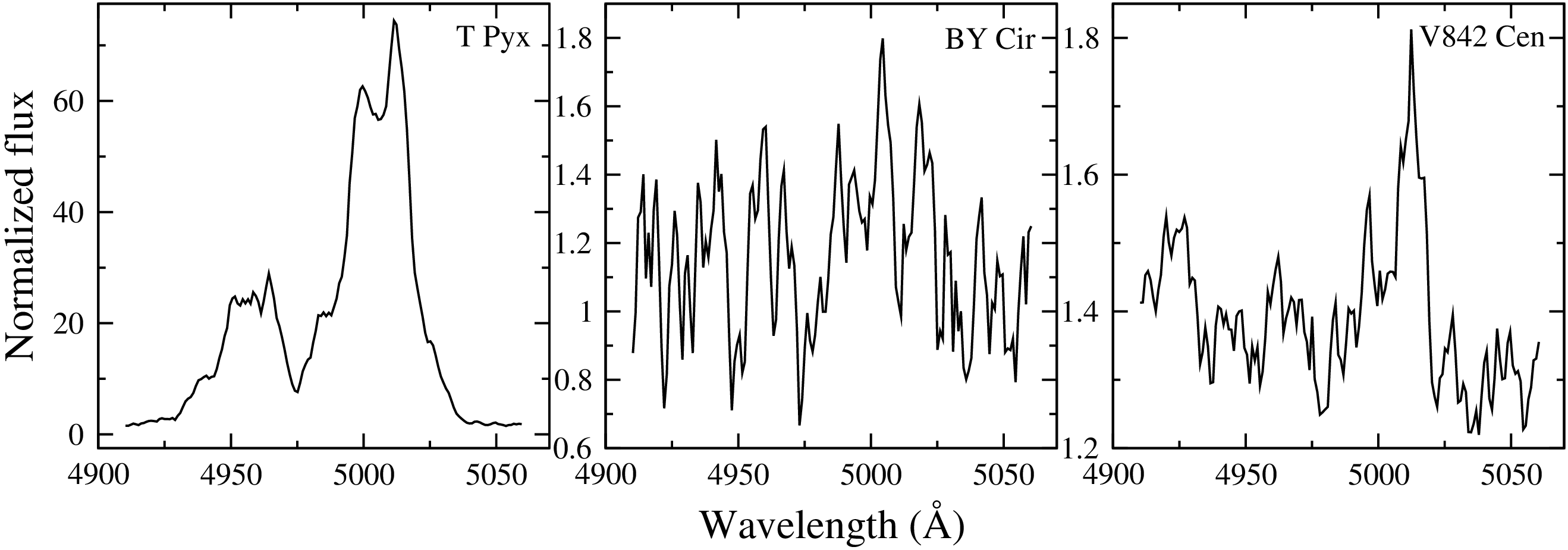}}
\caption{[OIII] 4959 and 5007\,\AA\ profiles.}
\label{oiii_prof}
\end{figure*}

\begin{figure*}
\centering
\resizebox{\hsize}{!}{\includegraphics{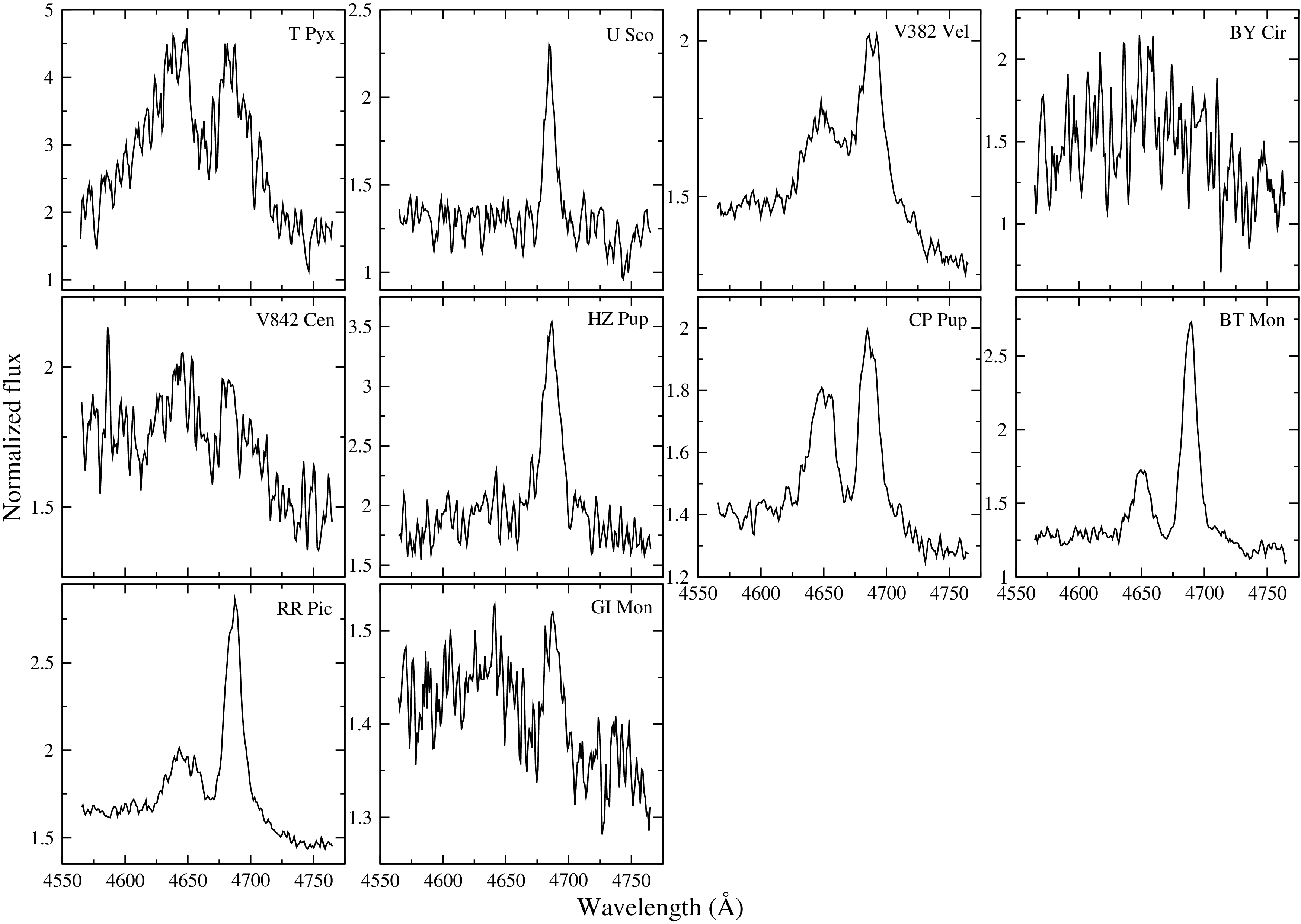}}
\caption{HeII and Bowen blend profiles.}
\label{bow_prof} 
\end{figure*}

}

\begin{acknowledgements}
All of the observations reported in this paper were obtained with the Southern African Large Telescope (SALT). We are grateful to the referee Linda Schmidtobreick and the editor Steven Shore, for the valuable comments and suggestions.
This research made use of the SIMBAD database, operated at the CDS, Strasbourg, France, and NASA's Astrophysics Data System Bibliographic Services.
\end{acknowledgements}

\bibliographystyle{aa}
\bibliography{tomov_salt_novae}


\Online

\end{document}